\documentclass[aps,reprint,onecolumn, longbibliography]{revtex4-1}

\usepackage{tikz}
\usepackage{bm}
\usepackage{amssymb,amsfonts}
\usepackage{amsmath}
\usetikzlibrary{decorations.pathmorphing}
\usetikzlibrary{patterns}
\usepackage{caption,subcaption}
\usepackage{float}

\newcommand{\Rey}{\operatorname{\mathit{R\kern-.15em e}}}
\newcommand{\Weber}{\operatorname{\mathit{W\kern-.30em e}}}
\newcommand{\Capil}{\mbox{\textit{C}}}

\draft % marks overfull lines with a black rule on the right

\DeclareFontFamily{OT1}{pzc}{}
\DeclareFontShape{OT1}{pzc}{m}{it}{<-> s * [1.10] pzcmi7t}{}
\DeclareMathAlphabet{\mathpzc}{OT1}{pzc}{m}{it}

\begin{document}

%\preprint{}

\title{Instability and dripping of electrified liquid films flowing down inverted substrates} %Title of paper



\author{R. J. Tomlin}
\email[]{ruben.tomlin11@imperial.ac.uk}
\affiliation{$\;$Department of Mathematics, Imperial College London$\;$}
%\thanks{}
\author{R. Cimpeanu}
\affiliation{Mathematical Institute, University of Oxford}
\affiliation{Department of Mathematics, Imperial College London}

\author{D. T. Papageorgiou}
\affiliation{Department of Mathematics, Imperial College London}


\date{\today}

\begin{abstract}

We consider the gravity-driven flow of a perfect dielectric, viscous, thin liquid film, wetting a flat substrate inclined at a non-zero angle to the horizontal. The dynamics of the thin film is influenced by an electric field which is set up parallel to the substrate surface -- this nonlocal physical mechanism has a linearly stabilizing effect on the interfacial dynamics. Our particular interest is in fluid films that are hanging from the underside of the substrate; these films may drip depending on physical parameters, and we investigate whether a sufficiently strong electric field can suppress such nonlinear phenomena. For a non-electrified flow, it was observed by Brun et al.~(Phys.~Fluids 27, 084107, 2015) that the thresholds of linear absolute instability and dripping are reasonably close. In the present study, we incorporate an electric field and analyse the absolute/convective instabilities of a hierarchy of reduced-order models to predict the dripping limit in parameter space. The spatial stability results for the reduced-order models are verified by performing an impulse--response analysis with direct numerical simulations (DNS) of the Navier--Stokes equations coupled to the appropriate electrical equations. Guided by the results of the linear theory, we perform DNS on extended domains with inflow/outflow conditions (mimicking an experimental set-up) to investigate the dripping limit for both non-electrified and electrified liquid films. For the latter, we find that the absolute instability threshold provides an order-of-magnitude estimate for the electric field strength required to suppress dripping; the linear theory may thus be used to determine the feasibility of dripping suppression given a set of geometrical, fluid and electrical parameters.

\end{abstract}


\maketitle %\maketitle must follow title, authors, abstract and \pacs

\section{Introduction}
\label{SecIntro}

{\color{black}

Thin liquid films are encountered in numerous industrial applications such as coating processes, lab-on-a-chip systems \cite{doi:10.1146/annurev.fluid.36.050802.122124}, and liquid film cooling \cite{Miyara1999,serifi2004transient}, as well as geological and geomorphological applications 
\cite{Shorts_et_al, Camporeale}.
Generally such flows can become unstable to inertial and gravitational instabilities and can support complicated wave structures including
spatiotemporal chaotic dynamics; it is of interest to consider ways to control such phenomena.
%However, despite the stabilising action of surface tension, it may be impossible to achieve a stable or flat film within the range of desired parameters (such as film thickness and inclination angle), given a restrictive range of real fluid parameters (such as density and viscosity) due to inertial and gravitational instabilities.
In this work we investigate the possible flow control of gravity-driven perfect dielectric liquid films by imposing an electric field parallel to the substrate on which the fluid lies (and hence
parallel to the undisturbed film surface). With this orientation, the electric field is stabilizing and we show that it can be utilized quite dramatically
in films on inverted substrates, for example, to arrest dripping or even damp interfacial waves completely.

The non-electrified falling film problem has been studied extensively since the pioneering experiments of Kapitza and Kapitza \cite{kapitza1949wave}. 
For an overlying film flow with a sufficiently shallow inclination angle, a flat interface with semi-parabolic velocity profile is an exact and stable solution -- the Nusselt solution \cite{nusselt1}. Linear stability analyses by Yih \cite{yih1955proceedings,Yih1} and Benjamin \cite{FLM:367246} find 
a critical Reynolds number, dependent on the inclination angle, above which the Nusselt solution first becomes unstable. 
Two-dimensional (2D) solitary waves form \cite{liu1994solitary}, comprising a humped wave front with leading capillary ripples, which transition to 3D waves due to secondary instabilities \cite{kharlamov2015transition}. Inlet forcing can be used to induce these 2D waves (as opposed to allowing the instability to develop from noise alone), and can also be employed as a flow control to preserve the 2D phase of the dynamics \cite{alekseenko1994wave,park2003three}.
%In agreement with the prediction from the linear stability analysis, two-dimensional (2D) solitary waves form \citep{liu1994solitary}, , which transition to 3D waves due to secondary instabilities \citep{alekseenko1994wave,liu1995three,park2003three,kofman2014three,kharlamov2015transition}. A shadow image of such a transition observed in experiments by \cite{park2003three} is shown in Figure \ref{2to3dparketal2003}(a). The waves found in experiments are long compared to the film thickness. Inlet forcing can be used as a control to preserve the 2D phase of the dynamics -- see Figure \ref{2to3dparketal2003}(b). 
 For hanging film flows on inverted substrates, the situation is more complicated and there are fewer experimental and theoretical studies. The cross-stream component of gravity now destabilizes the film interface, aiding the inertial (Kapitza) instability in the streamwise direction, and inducing the classical Rayleigh--Taylor instability in the spanwise direction. The latter gives rise to rivulet structures as observed in experiments \cite{rothrock1968study,markidesexperimental,PhysRevFluids.3.114002}, with waves on the crests which may drip depending on parameters \cite{indeikina1997drop,Brun1}. A recent numerical study of both overlying and hanging liquid films in the 2D setting was performed by Rohlfs et al.~\cite{Rohlfs}, in which the onsets of flow reversal and circulating waves are investigated.

}

Rothrock \cite{rothrock1968study} was the first to perform a careful experimental study of liquid films on inverted substrates, 
additionally carrying out a linear stability analysis to obtain onset conditions for waves on pendant rivulets -- these conditions agreed with experiments. 
Film dewetting was observed, as the rivulets which formed near the inlet fed into a wedge-shaped fluid film which ended with a single pendant rivulet. 
Regimes of dripping and no dripping were observed. Drop pinch-off from pendant rivulets was studied by Indeikina et al.~\cite{indeikina1997drop}, with further experiments and the consideration of an inertialess long-wave lubrication equation coupled with a matched asymptotics procedure. 
The authors identified two distinct mechanisms for dripping depending on the static contact angle of the rivulet and the substrate inclination angle: a jet mechanism at high flow rates due to failure of axial curvature to counteract gravity, and another due to the failure of azimuthal curvature. Charogiannis and Markides \cite{markidesexperimental} and Charogiannis et al.~\cite{PhysRevFluids.3.114002} performed experiments with fully wetting films on the underside of a flat substrate inclined at $15^\circ, 30^\circ$, and $45^\circ$ beyond vertical, for which no dripping was reported. The interface shapes observed were strongly 3D, with clear rivulet structures in the majority of cases with low to moderate Reynolds numbers. The pulses on the crests of the rivulets increased in amplitude as the Reynolds number increased. For larger Reynolds numbers, wavefronts formed across multiple rivulets, causing loss of fluid mass from the rivulets into the separating troughs. By approximating the transverse wavelength from experimental data, Charogiannis et al.~\cite{PhysRevFluids.3.114002} classified two types of rivulet formation depending on the inclination angle and the Kapitza number. They found that the wavelengths of the transverse rivulets for the more extreme inclinations (i.e.~$45^\circ$ beyond vertical) and/or lower Kapitza numbers, were as predicted by the linear stability of the flat film solution, matching the wavelength of the most unstable transverse mode arising from the competition between the cross-stream component of gravity and surface tension. However, for much smaller inclinations beyond vertical and/or larger Kapitza numbers, they found that the transverse wavelength matched that of the canonical Rayleigh--Taylor instability for a film hanging from a horizontal substrate (full vertical gravity versus surface tension). They suggested that in the former case, the primary instability was Rayleigh--Taylor, whereas in the latter case, the rivulet formation was due to a secondary instability of suspended 2D wavefronts. 
%This latter phenomenon is likely impossible to capture with reduced-order models.
%{\bf Confusing. Is the former not the more likely to suffer from Rayleigh--Taylor? Need to clearly define the geometry}

A hierarchy of reduced-order models may be constructed using a long-wave methodology to describe the dynamics of the fluid film; these simplify the problem both analytically and numerically while retaining the relevant physical effects. A so-called Benney equation \cite{Benney,gjevik1970occurrence} for the film thickness arises for Reynolds numbers close to critical. This highly nonlinear equation retains the effects of inertia, gravity, viscosity and surface tension. Analytical and numerical studies of the Benney equation were carried out by a number of authors \cite{pumir1983solitary,rosenau1992bounded,Salamon,oron2002nonlinear,gottlieb2004stability,Oron2004,scheid2005validity}, and finite time blow-up was observed in simulations (regions of parameter space where blow-up occurs coincide closely with parameters for which nonlinear traveling waves 
cease to exist \cite{pumir1983solitary,scheid2005validity}). Rosenau et al.~\cite{rosenau1992bounded} considered the Benney equation with a full-curvature regularisation, however this was not effective for all parameters for which finite-time blow-up occurred. 
Coupled systems of equations for the interface height and fluid flux may be derived using an averaging methodology, giving much better agreement 
with the true dynamics of liquid films for moderate Reynolds numbers. Such models were first constructed by Kapitza \cite{kapitza1948wave1,kapitza1948wave2} and Shkadov \cite{shkadov1967wave}, however their systems predicted an incorrect critical Reynolds number. This issue was corrected by the weighted integral boundary layer (WIBL) models developed by Ruyer-Quil and Manneville \cite{ruyer1998modeling,ruyer2000improved,ruyer2002further}; WIBL models show good agreement with DNS and experiment \cite{denner2018solitary}, and 
better matching with full linear theory (based on the Orr--Sommerfeld problem) than the Benney equation \cite{kalliadasis2011falling}.

The complete mechanisms for the dripping of a hanging film are not yet fully understood, but their relation to the {\color{black} absolute (or spatial) instability} of the Nusselt solution has been the subject of recent research. Overlying and vertical film flows exhibit convective instabilities, 
whereas films hanging from the underside of a horizontal substrate exhibit an absolute Rayleigh--Taylor instability. Thus, for given fluid parameters, a film flow transitions 
from convective to absolute instability as the inclination is increased to a critical angle beyond vertical. 
The connection between absolute instability and dripping of hanging films was investigated by Brun et al.~\cite{Brun1}. The authors derived an inertialess Benney equation for the 2D flow, and identified regions in parameter space for which the flat interface solution was convectively or absolutely unstable. They performed experiments of gravity-driven films hanging from inverted substrates, and found good agreement between the region of parameter space for which dripping was observed, and that in which the Benney model exhibited absolute linear instability. The experiments were conducted by pouring castor oil onto a flat substrate, and letting it spread until the fluid layer was roughly uniform with a given thickness. The substrate was then rotated to some angle beyond vertical, and the number of drips falling from a fixed region of the substrate was recorded. It is noticeable from their results that only a small amount of dripping was observed just beyond the absolute--convective (A/C) threshold, with much more intense dripping further into the absolute instability regime. Inertial effects and higher order terms were included by Scheid et al.~\cite{ScheidKofman1} in their study of the A/C transition for WIBL models. The authors reported a large discrepancy with 
the results of the inertialess Benney equation away from zero Reynolds numbers, and found a fluid-independent critical angle below which only convective instabilities exist for their models. Kofman et al.~\cite{kofman2018prediction} employed DNS for the 2D problem with
spatially periodic boundary conditions, finding that the dripping onset did not coincide closely with the A/C transition curve obtained from the WIBL models in \cite{ScheidKofman1}. They attribute dripping to a secondary instability of travelling wave solutions. We note that, the dripping thresholds computed by Kofman et al.~\cite{kofman2018prediction} improve in their agreement 
%{\bf Ruben, Radu: Is this their result or something you found?}
with the A/C threshold predictions as the length of the periodic domain increases (in the region where the A/C curve in terms of thickness and inclination angle is monotonic). We emphasize that due to nonlinear effects, dripping and absolute instability should not align exactly, however a predictor for dripping phenomena based on linear theory is useful. The consideration of 2D models to study the inherently 3D process of dripping is not entirely justified, in particular a 2D study
cannot capture the second mechanism identified by Indeikina et al.~\cite{indeikina1997drop}. 
However, for domains which are sufficiently small in the spanwise direction, i.e.~below the threshold of the spanwise Rayleigh--Taylor instability, 
a 2D flow assumption is appropriate. Furthermore, we believe that a 2D model is a reasonable approximation for the flow on the crest of a wide rivulet.

Lin et al.~\cite{lin2012thin} considered the dynamics of a 3D fluid front on the underside of an inclined plane 
and performed numerical simulations of a multidimensional Benney equation, regularizing the problem with the addition of
a thin precursor film. They found that the fluid fronts were unstable to a transverse fingering instability. Thin rivulets form with approximately equal width in the spanwise direction, and fast moving ``drop-like" waves appear on the rivulet crests as in the wetted case. Although the finger formation is not attributed to the Rayleigh--Taylor instability, the dynamics on their crests is of relevance to both the wetted and non-wetted cases. The effect of vertical electric fields 
and temperature gradients on the linear stability of such fluid fronts was considered by Conroy et al.~\cite{PhysRevFluids.4.034001}.

The use of horizontal electric fields to stabilize the Rayleigh--Taylor instability in stratified systems of dielectric fluids was considered by Cimpeanu et al.~\cite{cimpeanu2014control} and Anderson et al.~\cite{RaduAnder1}. The former work investigates an unbounded two-fluid system of viscous dielectric
fluids with a less dense lower fluid. The latter study considers the related problem of an
upper fluid layer and a hydrodynamically passive lower layer bounded above and below by solid dielectric substrates. 
Linear theory showed that the growth rates decrease as the field strength is increased, and the band of unstable wavenumbers shrinks
in extent. If the domain is finite in the horizontal direction, then complete linear stabilization of the flat interface solution is attained with a sufficiently strong field. 
DNS of the Navier--Stokes equations coupled with electrostatics was also carried out, 
and the parameters for which finger formation was suppressed was found to be in good agreement with the linear theory.
In addition, Anderson et al.~\cite{RaduAnder1} derived a nonlinear nonlocal evolution equation for the interfacial dynamics valid for sufficiently thin liquid layers.
Numerical solutions of the model accurately capture the primary collar and secondary lobe structures present in the early stages of finger formation 
(as validated with DNS). In both \cite{cimpeanu2014control} and \cite{RaduAnder1}, the authors 
demonstrate numerically the possibility of active control of the underlying Rayleigh--Taylor instability and production of sustained
nonlinear interfacial oscillations for arbitrarily long times.
Such oscillations can enhance mixing, for example, as in \cite{cimpeanu2015electrostatically}. 
Importantly, we note that such varying electric fields are required for this particular problem to give bounded nontrivial solutions at large times; 
for constant field strengths, either the flat state is stable or finger formation accompanied by film rupture occurs.
Of interest, therefore, is the recent study of Kord and Capecelatro \cite{Kord}, on optimal perturbations for controlling the Rayleigh--Taylor instability and the induced mixing.

The studies described above were conducted in the absence of a mean flow, which adds to the complexity of the physical system and further enriches the interplay between the competing mechanisms. In the present work, we consider the application of parallel electric fields to gravity-driven flows on inclined flat substrates. 
%The field is applied parallel to the substrate, and thus also the undisturbed dielectric liquid film surface.
%Electrodes are set-up at the fluid inflow and outflow so that the undisturbed electric field lines are parallel to the substrate (also parallel to the undisturbed flat film solution), with all phases assumed to be perfect dielectrics. 
We work with a 2D formulation of the problem, and allow the fluid to be either overlying or hanging. 
%The work considered here can be easily extended to the case of both layers being hydrodynamically active, i.e.~gravity-driven liquid--liquid stratification. 
A long-wave approximation allows us to construct a fully nonlinear Benney equation with nonlocal electric field effects. 
We also provide electrified versions of two WIBL models derived by Ruyer-Quil and Manneville \cite{ruyer1998modeling,ruyer2000improved,ruyer2002further}. Temporal linear stability analyses of all three models show the stabilizing effect of the electric field, and
complete stabilization of finite length systems (even in the absence of surface tension) as found in
\cite{cimpeanu2014control,RaduAnder1}.
The major effort focuses on
hanging films and the effect of the electric field on the A/C transition. 
Unsurprisingly, we find that an increased field strength restricts the range of parameters that yield absolutely unstable systems, i.e.~promotes convective instability (spatial stability) of the flat film solution. 
Using a WIBL model, we obtain a minimum critical angle depending on the electric field strength, below which only convective instabilities occur, 
extending the findings in \cite{ScheidKofman1}. Comparisons with results for the full stability problem at zero Reynolds number are also provided. {\color{black}A similar stability study of a nonlocal problem was performed for the Benney equation with (linearly destabilizing) normal electric field effects by Blyth et al.~\cite{blyth2018two}. They investigated the stability of solitary wave pulses by analysing absolute/convective instabilities in a reference frame travelling with the pulses. They found that increased field strengths promoted convective instability; their finding appears to be linked to the increased pulse speed as the electric field strength is raised, so that the pulses escape the expanding wave packets generated by localized disturbances. For large field strengths, the pulse solutions are hence much more stable and attracting structures, in agreement with full time-dependent computations. Although the results are not presented here, we found that normal electric fields promoted absolute instability of the flat film solution (in a fixed reference frame). }

Following \cite{Brun1,kofman2018prediction}, we investigate the relationship between absolute instability 
and dripping of hanging films, both with and without electric fields. The working fluid for the DNS is the same as 
used in experiments by Brun et al.~\cite{Brun1}, and we took care to faithfully reproduce the experiments {\it in silico}, consequently investigating
electric field effects through extensive computations that could be hard to sample experimentally. To this end, and in
contrast to \cite{kofman2018prediction} where spatially periodic boundary conditions are used, we initialize with a flat interface and excite the most unstable waves with time-periodic forcing at the inflow (as used in \cite{denner2018solitary} for the computation of solitary waves on overlying films). 
Hence, we both mimic experimental conditions and are able to produce solutions with a regular spacing of waves/drops. 
In the non-electrified case, we find a reasonably close agreement between the onset of dripping and the lower threshold of absolute instability. 
We believe that this is partly due to the low Reynolds numbers we consider, but also due to the use of long domains in our DNS 
coupled with appropriate inflow/outflow boundary conditions. In borderline cases, it can take many wavelengths from the inlet for a drip to develop. 
Our simulations indicate that (non-transient) dripping takes place as an instability of individual solitary pulses, agreeing with the discussions in \cite{kofman2018prediction}. 
For the electrified problem the agreement is less striking, but the A/C threshold still remains a good order-of-magnitude estimate for complete dripping suppression,
something that cannot be obtained from a temporal stability analysis. Increased electric field strengths delay the first dripping event, 
and induce temporally and spatially irregular dripping that does not appear to be transient (as in the non-electrified case). We believe that the 
nonlocality of the problem is an important factor here. We emphasize that, unlike the electric field stabilization of the classical 
Rayleigh--Taylor instability \cite{cimpeanu2014control,RaduAnder1}, the mean flow gives rise to nontrivial bounded solutions 
(wave trains that do not drip) below the threshold of stabilization of the flat interface solution. For this reason, we do not investigate active control strategies.

The structure of the rest of the paper is as follows. In Section \ref{sec:2},
we present the physical model and full set of governing equations for the 2D flow. 
In Section \ref{sec:Hierarchy}, we give the long-wave models 
for the electrified gravity-driven film flow, and perform a temporal linear stability analysis. In Section \ref{sec:absConv} we specialize to hanging film flows, and perform a spatial stability analysis of the long-wave models to determine parameter regimes of absolute or convective instabilities. We obtain a minimum critical angle, depending on the electric field strength, below which all instabilities are convective for all flow parameters. We also compare the linear 
results of the long-wave models at zero Reynolds number to those of the full Stokes flow problem. In Section \ref{sec:DNS}, we carry out
DNS of the full problem with small-amplitude pulse initial data to verify the regimes of absolute and convective instability predicted by the reduced-order models. Section \ref{sec:DNSdripping} provides DNS of the full problem to determine the dripping onset for both the non-electrified and electrified case, and investigates its relation to absolute instability. A discussion and conclusions are given in Section \ref{sec:Conclusions}.

\section{Physical Model and Governing Equations}\label{sec:2}

\begin{figure}
\includegraphics[scale=1]{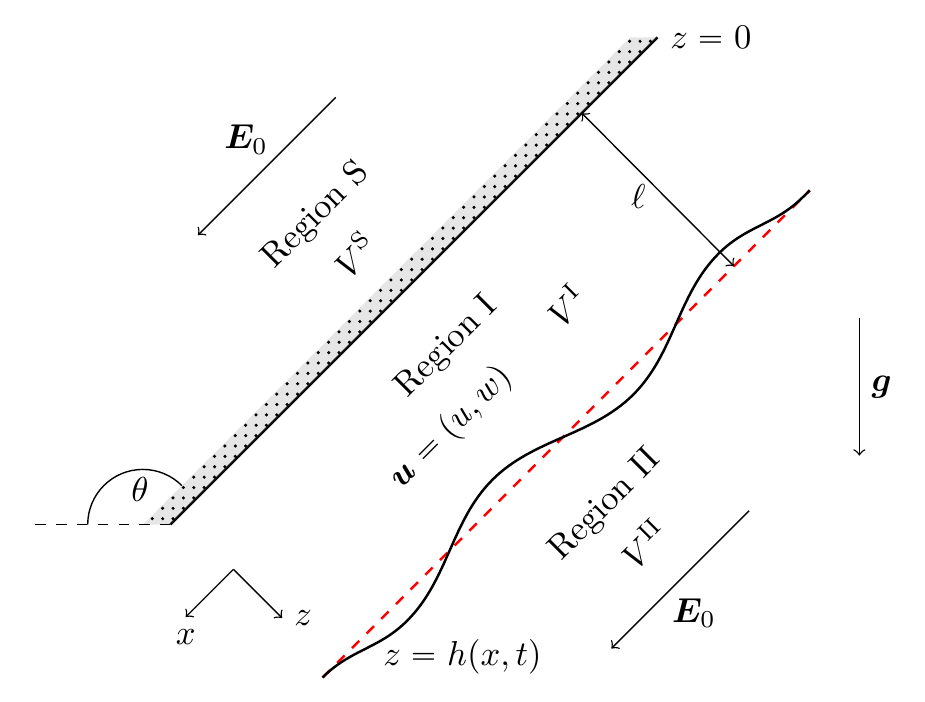}
\caption{Schematic for the problem of a liquid film influenced by a parallel electric field. As the inclination angle $\theta$ (measured from the horizontal) varies, the substrate and axis rotate, with overlying films found for $\theta < \pi/2$, and hanging films (as in the schematic) for $\theta > \pi/2$.} \label{Setupdiagram1}
\end{figure}

We consider a Newtonian fluid with constant density $\rho$ and dynamic viscosity $\mu$ (the kinematic viscosity is $\nu=\mu/\rho$), 
flowing under gravity on a flat substrate inclined at some angle $\theta$ to the horizontal. For the 2D problem considered here, we take coordinates $(x,z)$ with $x$ in the streamwise direction and $z$ perpendicular to the substrate surface -- see the schematic in figure \ref{Setupdiagram1}. We have overlying film flows for $\theta \in (0, \pi/2)$, a vertical film flow for $\theta = \pi/2$, and hanging flows for $\theta \in (\pi/2,\pi)$ as shown in figure \ref{Setupdiagram1}. The surface tension coefficient between the fluid and the surrounding medium is $\sigma$, and the 
gravitational force in our chosen coordinate system is $\bm{g} = g\sin\theta\, \bm{e}_x - g \cos\theta\,\bm{e}_z$, with
$\bm{e}_x$ and $\bm{e}_z$ denoting unit vectors in the $x$ and $z$ directions, respectively. 
The local film thickness is denoted by $h(x,t)$, a function of space and time, with unperturbed thickness $\ell$. The liquid layer is denoted by Region I, 
with Region II being the hydrodynamically passive medium defined by $z > h(x,t)$ (here, the pressure is constant and
denoted by $p_{\text{atm}}$). The fluid in Region I is governed by the usual Navier--Stokes equations
\begin{equation} \bm{u}_t + (\bm{u} \bm{\cdot} \bm{\nabla} )\bm{u} =  - \frac{1}{\rho} \bm{\nabla} p  + \nu {\nabla}^2 \bm{u} +  \bm{g} , \qquad \bm{\nabla}\bm{\cdot}\bm{u} =   0, 
\end{equation}
where $\bm{u}= (u,w)$ is the velocity field and $p$ is the pressure.
%, $\bm{\nabla}$ is the 2D gradient operator, and ${\nabla}^2 = \bm{\nabla} \bm{\cdot} \bm{\nabla}$ is the Laplacian operator. 
At the substrate surface, we have the no-slip and impermeability conditions, $\bm{u}|_{z=0} = \bm{0}$.

We assume that the substrate (Region $\textrm{S}$), fluid (Region $\textrm{I}$), and gas phase (Region $\textrm{II}$) are perfect dielectrics with dielectric
permittivities $\epsilon_0\epsilon^{\textrm{S}}$, $\epsilon_0\epsilon^{\textrm{I}}$, and $\epsilon_0\epsilon^{\textrm{II}}$, 
respectively, where $\epsilon_0$ is the permittivity of free space. We denote the voltage potentials in each region by
$V^{\textrm{S}}$, $V^{\textrm{I}}$, and $V^{\textrm{II}}$, each with corresponding electric fields
$\bm{E}^{\iota} = - \bm{\nabla} V^\iota$. Gauss' law provides harmonic problems for the voltages, i.e.~
\begin{equation}
\nabla^2 V^\iota = 0,\qquad
\iota= \textrm{S}, \textrm{I}, \textrm{II}.\label{eq:Laplace}
\end{equation}
These arguments follow from the electrostatic limit of Maxwell's equations appropriate to this study -- see
the review by Papageorgiou \cite{Papageorgiou2019} for details.
The imposed field is parallel to the solid substrate surface as shown in figure \ref{Setupdiagram1}, and hence the boundary conditions
far from the liquid phase in the normal direction are
\begin{equation}\label{farfieldV1} 
\bm{\nabla} V^{\textrm{S}}  \rightarrow -\bm{E}_0, \quad \textrm{as } z \rightarrow - \infty, \qquad  \bm{\nabla} V^{\textrm{II}} 
 \rightarrow - \bm{E}_0, \quad \textrm{as } z \rightarrow + \infty,
\end{equation}
where $\bm{E}_0 = E_0\,\bm{e}_x$ and $E_0=V_0/L_0$ measures the strength of the imposed field -- here $V_0$ is the voltage drop across a system of length $L_0$.
%If $V_0$ is the imposed potential difference and $L_0$ is the system length from the fluid inlet to outlet, then $E_0 = V_0/L_0$. 
%We will see that the orientation of the electrodes in unimportant, only the magnitude of the electric field strength. 
Making the usual assumption of zero impressed charges at interfaces, we have the following electrical boundary conditions at the substrate
%We assume that the surface charge density of the free charge at the substrate surface and film interface is zero, i.e.~the system is grounded before the potential difference is imposed. At the substrate surface we have the boundary condition for the voltages,
\begin{equation}\label{subsurfVconds1}
\left[\epsilon^{\iota} V^{\iota}_z \right]_{\textrm{I}}^{\textrm{S}}=0, \quad \left[V^{\iota} \right]_{\textrm{I}}^{\textrm{S}} = 0, \quad \textrm{at } z=0,
\end{equation} 
subscripts denote partial derivatives and
the jump notation $\left[\;\cdot\;\right]_{\textrm{I}}^{\textrm{S}}=(\;\cdot\;)_{\textrm{S}}-(\;\cdot\;)_{\textrm{I}}$ has been introduced. The first condition
corresponds to continuity of the displacement field and the
second to continuity of the voltage potentials -- see \cite{Papageorgiou2019}.

To calculate the boundary conditions at the gas--liquid interface, we first define the unit tangent and outward-pointing normal vectors there,
$\bm{t} = (1, h_x )^T/\sqrt{1 + h_x^2}$ and
$\bm{n} = (-h_x,1 )^T/\sqrt{1 + h_x^2}$. For the remainder of the section, all $z$-dependent expressions are evaluated at the interface $z=h(x,t)$. 
The kinematic condition reads
\begin{equation}\label{Kinematic1} 
w = h_t + u h_x, 
\end{equation} 
and the conditions on the voltage potentials, analogous to \eqref{subsurfVconds1}, are
\begin{equation}\label{surfacevoltcond1}
\left[\epsilon^{\iota} \nabla V^{\iota}\bm{\cdot} \bm{n} \right]_{\textrm{II}}^{\textrm{I}}=0, \quad \left[V^{\iota} \right]_{\textrm{II}}^{\textrm{I}} = 0.
\end{equation} 
The final two conditions are the continuity of normal and tangential stresses at the interface
\begin{equation}\label{eq:stresses}
\left[ \left(\bm{T}^{\iota} \bm{n}\right) \bm{\cdot} \bm{n}  \right]_{\textrm{II}}^{\textrm{I}} = \sigma  h_{xx} /(1 + h_x^2)^{3/2},
\qquad \left[\left( \bm{T}^{\iota}\bm{n} \right) \bm{\cdot} \bm{t}  \right]_{\textrm{II}}^{\textrm{I}}  = 0,
\end{equation}
where the stress tensors include hydrodynamic and/or Maxwell stresses as appropriate, and (employing the usual subscript notation) are given by 
\begin{subequations} \label{stressAB}
\begin{align}
\bm{T}^{\textrm{I}}_{jk} & = -  p\delta_{jk}+\mu \left( \frac{\partial u_k}{\partial x_j} + \frac{\partial u_j}{\partial x_k}\right)  
+ \epsilon_0 \epsilon^{\textrm{I}} \left( \frac{\partial V^{\textrm{I}}}{\partial x_j} \frac{\partial V^{\textrm{I}}}{\partial x_k} - 
\frac{1}{2} | \bm{\nabla}V^{\textrm{I}}|^2 \delta_{jk}\right), \label{stressA}\\
\bm{T}^{\textrm{II}}_{jk} & = - p_{\text{atm}} \delta_{jk}+\epsilon_0 \epsilon^{\textrm{II}} 
\left( \frac{\partial V^{\textrm{II}}}{\partial x_j} \frac{\partial V^{\textrm{II}}}{\partial x_k} - \frac{1}{2} | 
\bm{\nabla}V^{\textrm{II}}|^2 \delta_{jk}\right). \label{stressB}
\end{align}
\end{subequations}
%Note that the Maxwell stresses are divergence free since each phase is charge-free with constant permittivity and so
%there are no Lorentz forces in the Navier--Stokes equations \cite{pease2002linear}. 
The stress balances \eqref{eq:stresses} written out in full become
\begin{align}
2 \mu \left(u_x (h_x^2 - 1) - (u_z + w_x)h_x \right) +
\frac{\epsilon_0}{2} \left[  \epsilon^{\iota} \left( (h_x^2 - 1) ( (V^{\iota}_x)^2 - (V^{\iota}_z)^2 ) - 4 h_x V^{\iota}_x V^{\iota}_z \right)  
\right]_{\textrm{II}}^{\textrm{I}}\qquad & \nonumber \\
 + (p_{\text{atm}} - p)(1+h_x^2) & = \sigma \frac{ h_{xx} }{(1 + h_x^2)^{1/2}},\label{eq:NSBfull}\\
(1 -h_x^2)(u_z + w_x) + 4 w_z h_x  & = 0, \label{tangstressbal1}
\end{align}
where \eqref{surfacevoltcond1} is used to simplify the tangential stress balance \eqref{tangstressbal1}. Electrohydrodynamic coupling is present through the normal stress balance alone as expected for interfaces between perfect dielectrics (and also
between a perfect dielectric and a perfect conductor \cite{papageorgiou2004generation}). For finite conductivities, consideration of the Taylor--Melcher leaky dielectric model is appropriate
\cite{MelcherTaylor,doi:10.1146/annurev.fluid.29.1.27,pease2002linear, Papageorgiou2019}.

\subsection{Exact solution and non-dimensional equations}\label{subsecExactSoln1}

An exact solution to the full formulation (extending the Nusselt solution to the electrified problem) is
\begin{equation}\label{basestatesdimensional} \left.
\begin{array}{c} {\displaystyle \overline{h} = \ell, \qquad \overline{u} = \frac{g \sin\theta}{2\nu} (2\ell z - z^2), \qquad \overline{w} = 0,} \\[8pt]
{\displaystyle\overline{p} = p_{\textrm{atm}} - \frac{\epsilon_0 E_0^2}{2} ( \epsilon^{\textrm{I}} - \epsilon^{\textrm{II}}) - \rho g (z-\ell) \cos\theta,\quad \overline{V}^{\iota} = - E_0 x \quad \textrm{for $\iota=$ S, I, II}.} 
\end{array}   \right\}
\end{equation}
The velocity profile is semi-parabolic in $z$, and the voltage potential is linear in $x$. We non-dimensionalize velocities with the base velocity at the free surface, $U_0 = \overline{u}|_{z = \ell} = g \ell^2\sin\theta/2\nu$, and make use of the non-dimensional parameters
\begin{equation}\label{dimensionlessparameters} {\Rey} = \frac{U_0 \ell}{\nu} = \frac{g \ell^3\sin\theta}{2\nu^2},\quad {\Weber} = \frac{ \epsilon_0 E_0^2 \ell}{2 \mu U_0}= \frac{\epsilon_0 E_0^2}{\rho g \ell \sin\theta}, \quad {\Capil} = \frac{U_0 \mu}{\sigma} = \frac{ \rho g \ell^2\sin\theta}{2\sigma}.\end{equation}
Here, ${\Rey}$ is the Reynolds number measuring the ratio of inertial to viscous forces, 
${\Weber}$ is the electric Weber number measuring the ratio of electrical to fluid pressures, 
and ${\Capil}$ is the capillary number measuring the ratio of viscous to surface tension forces. 
To non-dimensionalize we write
\begin{equation} \label{nondimandshift1}
\left. \begin{array}{c} {\displaystyle x^{*} = \frac{1}{\ell} x, \quad z^{*} = \frac{1}{\ell} z, \quad  \bm{u}^{*} = \frac{1}{U_0} \bm{u}, \quad t^{*} = \frac{U_0}{\ell} t , \quad h^{*} = \frac{1}{\ell} h,} \\[8pt]
{\displaystyle p^{*} = \frac{\ell}{\mu U_0} \left(p - p_{\textrm{atm}} + \frac{\epsilon_0 E_0^2}{2}  ( \epsilon^{\textrm{I}} - \epsilon^{\textrm{II}}) + \rho g z \cos \theta \right), \quad (V^{\iota})^{*} = \frac{1}{E_0\ell} (V^{\iota} + E_0 x) \quad \textrm{for }\iota= \textrm{S}, \textrm{I}, \textrm{II},}
\end{array}\right\}  
\end{equation}
substitute into the governing equations and boundary conditions, and drop the stars. In Region I, the Navier--Stokes equations transform to
\begin{equation}\label{Navierstokesfinal1}
{\Rey} ( \bm{u}_t + (\bm{u} \bm{\cdot} \bm{\nabla} )\bm{u} ) = - \bm{\nabla} p  + {\nabla}^2 \bm{u} +  2\bm{e}_x ,\qquad \bm{\nabla}\bm{\cdot}\bm{u} =  0.
\end{equation}
The no-slip and impermeability conditions, and Laplace's equations for the voltage potentials are unchanged under the change of variables \eqref{nondimandshift1}, the far-field conditions \eqref{farfieldV1} become
\begin{equation}\label{farfieldV2nondim} \bm{\nabla} V^{\textrm{S}}  \rightarrow \bm{0} , \quad \textrm{as } z \rightarrow - \infty, \qquad \bm{\nabla} V^{\textrm{II}}  \rightarrow  \bm{0} , \quad \textrm{as } z \rightarrow + \infty,\end{equation}
while the conditions \eqref{subsurfVconds1} for the voltage potentials at the substrate surface are unchanged. For the interfacial conditions, the kinematic condition \eqref{Kinematic1}, continuity of voltage (\ref{surfacevoltcond1}b), and the tangential stress balance \eqref{tangstressbal1} are also unchanged. Equation (\ref{surfacevoltcond1}a) transforms to
\begin{equation}\label{surfacevoltcond1Aredo}\left[\epsilon^{\iota} (-h_x( V^{\iota}_x - 1) + V^{\iota}_z) \right]_{\textrm{II}}^{\textrm{I}}=0.\end{equation} 
Finally, the normal stress balance becomes
\begin{align} & u_x (h_x^2 - 1) - (u_z + w_x)h_x + h (1+h_x^2)\cot\theta -  \frac{1}{2 } p(1+h_x^2) \nonumber
\\ & \qquad +  \frac{{\Weber}}{2} \left[  \epsilon^{\iota} \left( (h_x^2 - 1) ( (V^{\iota}_x - 1)^2 - (V^{\iota}_z)^2 ) - 4 h_x (V^{\iota}_x - 1) V^{\iota}_z + (1 + h_x)^2 \right)  \right]_{\textrm{II}}^{\textrm{I}}  =  \frac{1}{2 {\Capil}} \frac{ h_{xx} }{(1 + h_x^2)^{1/2}}. \label{normalstressbal2}\end{align}
The exact Nusselt solution to the Navier--Stokes equations \eqref{Navierstokesfinal1} and the above boundary conditions in non-dimensional form is
\begin{equation}\label{basestatesdimensional2}  \overline{h} = 1, \qquad \overline{u} = 2 z - z^2,\qquad \overline{w} = 0, \qquad \overline{p} = 0,\qquad \overline{V}^{\iota} = 0 \quad \textrm{for $\iota=$ S, I, II}.
\end{equation}
In Appendix \ref{OrrSommerfeldAppendix}, we provide the Orr--Sommerfeld system for the linearization of the dynamics about \eqref{basestatesdimensional2}; the electrostatics component is 
analytically tractable and
the remaining problem is a modification of the usual Orr--Sommerfeld system for thin films \cite{kalliadasis2011falling}. 
We also consider the Stokes flow limit in Appendix \ref{Stokesflowappend1}, for which we obtain an exact dispersion relation and compare this with the linear behavior of the long-wave models given next.

\section{Hierarchy of nonlinear long-wave models}\label{sec:Hierarchy}

We utilize a hierarchy of long-wave models to analyse the nature of the linear instabilities present in the electrified flow -- the spatial stability of these models is investigated in Section \ref{sec:absConv}, and detailed comparisons with DNS are undertaken in Section \ref{sec:DNS}. The models considered are the Benney and two WIBL models. We omit details of their derivations since they appear in \cite{tseluiko2006wave,tomlin_papageorgiou_pavliotis_2017} for the Benney equation and \cite{ruyer1998modeling,ruyer2000improved,ruyer2002further} for the WIBL models (albeit using a different non-dimensionalisation), but give the calculations used to obtain the electric field contribution that enters through the normal stress balance \eqref{normalstressbal2}.

{\color{black}

To derive the models, we assume that (in the original dimensional variables) the typical wavelength of
interfacial deformations $\lambda$ is long in comparison to the undisturbed liquid height $\ell$, so that
$\delta = \ell/\lambda\ll 1$. We introduce the rescalings
\begin{equation}\label{eq:lubScalings}
(\partial_t,\partial_x,w) \rightarrow \delta (\partial_t,\partial_x,w),
\end{equation} 
and apply a systematic asymptotic procedure. 
For the derivations, it is also assumed that ${\Rey} = O(1)$, although this has been found to be an unnecessary restriction for the WIBL models.
The Benney equation arises from an asymptotically correct elimination of the flow field variables in the kinematic equation \eqref{Kinematic1}. For an approximation of the interface thickness $H$ (the first two terms of an asymptotic expansion of $h$), the Benney model with errors of $O(\delta^2)$ is
\begin{equation}\label{Benney1withCap2}H_t  + \left[ \frac{2}{3} H^3 + \delta \left[ \frac{8{\Rey}}{15} H^6 H_x - \frac{1}{3} H^3 P_x \right]  \right]_{x} = 0.
\end{equation}
Here, $P$ is the leading-order pressure at the interface, taken to be $O(1)$ so that it enters the dynamics. Equation \eqref{Benney1withCap2} is effective at modelling flows for Reynolds numbers just beyond critical; increasing $\Rey$ further, singular phenomena such as finite time blow-up are observed in numerical simulations \cite{pumir1983solitary,rosenau1992bounded}.
}

Even though 
WIBL models rely on closure assumptions rather than rational asymptotic approximations, they are 
far more accurate than Benney equations at describing thin films for Reynolds numbers away from critical; 
they correctly capture the dynamics beyond the drag--gravity regime to which Benney equations are restricted \cite{kalliadasis2011falling}. 
In order to obtain the WIBL models, we rewrite the kinematic condition \eqref{Kinematic1} as 
\begin{equation}\label{Kinematic2} 
h_t + f_x = 0, \qquad f(x,t) =  \int_{0}^{h(x,t)} u(x,z,t) \; \mathrm{d} z,
\end{equation}
where $f$ is the fluid flux through a slice of the film in the $z$-plane. Following a weighted residuals strategy, in which the flow field is expanded in polynomials of $z$, the WIBL2 (simplified second-order) model, which comprises an approximation of the kinematic condition \eqref{Kinematic2} coupled to an equation for the time evolution of $F$ (an approximation of $f$), is given by
\begin{align}H_t + F_x = & \; 0, \label{WIBL2firsteq} \\
 F + \delta\frac{2{\Rey}}{5} H^2 F_t = & \; \frac{2}{3} H^3 + \delta \left[ \frac{18{\Rey}}{35}  H_x F^2 - \frac{34{\Rey}}{35} H F F_x  - \frac{1}{3} H^3 P_x \right] \nonumber \\ 
& \;  + \delta^2 \left[\frac{8}{5}  H_x^2 F - \frac{9}{5} HH_x F_x - \frac{12}{5}  H H_{xx} F + \frac{9}{5} H^2 F_{xx}\right]. \label{WIBL2secondeq}
\end{align}
The WIBL1 (first-order) model is obtained by omitting the $O(\delta^2)$ term in \eqref{WIBL2secondeq}. The dependence on the flow field variables is not completely eliminated as in the Benney equation \eqref{Benney1withCap2}; the latter may be obtained from WIBL1 
by using the leading order relation $F = 2H^3/3$ in the $O(\delta)$ terms of \eqref{WIBL2secondeq}, and substituting the result into \eqref{WIBL2firsteq}.
%with elimination of $F$ in the $O(\delta)$ terms of \eqref{WIBL2secondeq} using the leading-order relation $F = 2H^3/3$. 
Furthermore, Benney and WIBL1 are identical at zero Reynolds numbers, and all models reduce to the same 
nonlinear hyperbolic equation for $\delta = 0$. Denner et al.~\cite{denner2018solitary} reported that simulations of WIBL2 correctly captured the main humps in solitary wave trains even up to ${\Rey} = 100$, agreeing with DNS and experiments, but overestimated the amplitude of the leading capillary ripples for ${\Rey} \gtrsim 10$ (note that our ${\Rey}$ is $1.5$ times greater than the Reynolds number used in \cite{denner2018solitary}). We consider Reynolds numbers up to these values in the linear theory to follow -- it is known that 
there is good agreement with the Orr--Sommerfeld linear theory for these values \cite{kalliadasis2011falling}. 
For all models it remains to compute $P$ in terms of $H$, and this is undertaken next to clarify the additional effect due to the electric field.

{\color{black}

The normal stress balance \eqref{normalstressbal2} under the lubrication scalings \eqref{eq:lubScalings} becomes
\begin{equation} h \cot\theta  -  \frac{1}{2 } p +  \frac{{\Weber}}{2} \left[  \epsilon^{\iota} \left( 2\delta V^{\iota}_x  + (V^{\iota}_z)^2 + 4 \delta h_x V^{\iota}_z \right)  \right]_{\textrm{II}}^{\textrm{I}}  =  \frac{\delta^2 }{2 {\Capil}} h_{xx} + O(\delta), \label{normalstressbal3delta_leadingorder}\end{equation}
where we have retained leading and next-order terms inside the bracket corresponding to the electric field effect. 
We expand the solutions as
%In order to calculate the contribution of the electric field to the interfacial pressure, we consider the asymptotic expansions 
\begin{equation}\label{expansionsinregion1} 
h = h_0 + \delta h_1 + \ldots,\qquad V^{\iota} = V^{\iota}_0 + \delta V^{\iota}_1  + \ldots,
\quad \iota = {\rm I, II, S}.
\end{equation}
We also introduce a stretched normal variable $z=\zeta/\delta$, $\zeta=O(1)$, 
in the non-slender Regions S and II, so that the Laplace equations in the three regions become
\begin{equation}\delta^2 V^{\textrm{I}}_{xx} + V^{\textrm{I}}_{zz} = 0, \qquad V^{\iota}_{xx} + V^{\iota}_{\zeta\zeta} = 0, \quad \textrm{for } \iota = \textrm{S}, \textrm{II}. \end{equation}
The solutions for $V^{\textrm{S}}$, $V^{\textrm{II}}$ can be found by taking a Fourier transform in $x$ and applying the far-field conditions \eqref{farfieldV2nondim}; these are
\begin{equation} 
\widehat{V}^{\textrm{S}} = \widehat{A}^{\textrm{S}}(\xi) e^{\operatorname{sign}(\xi_{\textrm{r}}) \xi \zeta}, \qquad \widehat{V}^{\textrm{II}} =  \widehat{B}^{\textrm{II}}(\xi) e^{ -\operatorname{sign}(\xi_{\textrm{r}})\xi \zeta }, 
\end{equation}
where the complex wavenumber $\xi =  \xi_{\textrm{r}} + i \xi_{\textrm{i}}$ has been introduced in anticipation of the spatial
linear stability analysis to follow.
The multipliers $\widehat{A}^{\textrm{S}}(\xi)$ and $\widehat{B}^{\textrm{II}}(\xi)$ have asymptotic expansions 
similar to \eqref{expansionsinregion1}. The solution in Region I has the asymptotic expansion
\begin{equation} \widehat{V}^{\textrm{I}} = \widehat{C}_0^{\textrm{I}}(\xi) z + \widehat{D}_0^{\textrm{I}}(\xi) + \delta \left( \widehat{C}_1^{\textrm{I}}(\xi) z + \widehat{D}_1^{\textrm{I}}(\xi) \right) + O(\delta^2).\end{equation}
The electric field boundary conditions at the substrate \eqref{subsurfVconds1} yield the relations
\begin{equation}  \widehat{A}_0^{\textrm{S}}(\xi)  = \widehat{D}_0^{\textrm{I}}(\xi), \quad \widehat{A}_1^{\textrm{S}}(\xi) = \widehat{D}_1^{\textrm{I}}(\xi), \quad \widehat{C}_0^{\textrm{I}}(\xi) = 0, \quad \epsilon^{\textrm{S}} \operatorname{sign}(\xi_{\textrm{r}}) \xi  \widehat{A}_0^{\textrm{S}}(\xi)  - \epsilon^{\textrm{I}} \widehat{C}_1^{\textrm{I}}(\xi) = 0.\end{equation}
From the interfacial boundary conditions for the voltage potentials (\ref{surfacevoltcond1}b,\ref{surfacevoltcond1Aredo}), we also obtain
\begin{equation}\widehat{B}_0^{\textrm{II}}(\xi)  = \widehat{D}_0^{\textrm{I}}(\xi), \qquad ( \epsilon^{\textrm{I}} - \epsilon^{\textrm{II}} ) i \xi \widehat{h_0}(\xi) +\epsilon^{\textrm{I}} \widehat{C}_1^{\textrm{I}}(\xi) + \epsilon^{\textrm{II}} \operatorname{sign}(\xi_{\textrm{r}}) \xi  \widehat{B}_0^{\textrm{II}}(\xi) = 0.    \end{equation}
These give the Fourier transform of the voltage potential at the interface to leading order as (c.f.~equation (40) in \cite{RaduAnder1} with 
$\epsilon_{B} = 1$, corresponding to the case of no lower bounding solid)
\begin{equation}\label{realsigndefn1}\widehat{V}^{\iota} =  \left( \frac{ \epsilon^{\textrm{II}} - \epsilon^{\textrm{I}} }{ \epsilon^{\textrm{S}} + \epsilon^{\textrm{II}}} \right) i   \operatorname{sign}(\xi_{\textrm{r}}) \widehat{h_0}(\xi) + O(\delta) , \quad \Rightarrow \quad V^{\iota} = \left( \frac{ \epsilon^{\textrm{I}} - \epsilon^{\textrm{II}} }{ \epsilon^{\textrm{S}} + \epsilon^{\textrm{II}}} \right) \mathcal{H}( h_0) + O(\delta),\end{equation}
for $\iota =$ I, II, where $\mathcal{H}$ denotes the Hilbert transform with Fourier symbol $\widehat{\mathcal{H}}(\xi) = -i \operatorname{sign}(\xi)$ for $\xi \in \mathbb{R}$. The expression for $\widehat{V}^{\iota}$ for complex wavenumbers given in (\ref{realsigndefn1}a) will be important for the spatial stability analysis in the next section. It follows from this result and \eqref{normalstressbal3delta_leadingorder} that
to retain electrical and capillary effects in the leading order evolution, we require
${\Weber} = O(\delta^{-1})$ and ${\Capil} = O(\delta^2)$. 
The normal stress balance \eqref{normalstressbal3delta_leadingorder} gives
\begin{equation}\label{leadingorderP1} P  =  2 \left[ H \cot \theta +  \delta\Weber'   \mathcal{H}(H_x)  - \frac{\delta^2}{2{\Capil}} H_{xx} \right], \end{equation}
where
\begin{equation} \label{rescaledweber1}{\Weber}' = \frac{( \epsilon^{\textrm{I}} - \epsilon^{\textrm{II}})^2}{ \epsilon^{\textrm{II}} + \epsilon^{\textrm{S}}} {\Weber}. \end{equation}
It follows that the field has no effect on the interfacial dynamics if the electrical permittivities in Regions I and II are equal as ${\Weber}'$ is always zero. The impact of the electric field is maximized for $\epsilon^{\textrm{I}}$ large, with $\epsilon^{\textrm{S}}$ and $\epsilon^{\textrm{II}}$ both small. 
%Finally, the $z$-momentum equation yields $p_z = O(\delta)$, so \eqref{leadingorderP1} gives the constant leading-order pressure across a fluid slice. 

}

With the rescaling $\delta (\partial_t,\partial_x) \rightarrow (\partial_t,\partial_x)$, we may return to the original time and space variables (as in the non-dimensional Navier--Stokes setting), formally setting $\delta = 1$ in the long-wave models (assumed from this point). As we will see in the following linear stability analysis, both surface tension and the electric field have a stabilizing effect on the interface dynamics; for $\Capil^{-1/2} \sim \Weber'$, both effects are relevant. For the results of the models to be valid, parameters should be chosen so that at least one of these effects is retained; if both surface tension and the electric field are negligible in the leading-order dynamics, the instability moves towards the short waves, invalidating the initial long-wave assumption. Although the small parameter is now scaled out of the equations, it remains in the problem implicitly, indicating the subset of parameter space which is compatible with the long-wave assumption. For non-zero field strengths, the models are well-posed in the limit of weak surface tension, i.e.~$\Capil^{-1/2} = o(\Weber')$, in the sense that high wavenumbers remain damped. The corresponding weakly nonlinear evolution for electrified flows with weak surface tension is described by the nonlocal Kuramoto--Sivashinsky type equation
\begin{equation}\label{KSnonlocal} \eta_t + \eta \eta_x + \eta_{xx} - \mathcal{H}(\eta_{xxx}) = 0, \end{equation}
which is also well-posed. A fourth-order term is included if surface tension effects are not negligible. We do not consider \eqref{KSnonlocal} in the current work.

\subsection{Temporal linear stability analysis}\label{sec:temporal}

The dispersion relation for the Benney model is obtained by substituting $H = 1 + \tilde{H} e^{i \xi x + \omega t}$, with $\xi$ real and complex $\omega=\omega_{\textrm{r}} + i \omega_{\textrm{i}}$,  
into \eqref{Benney1withCap2} and 
linearizing for small $\tilde{H}$. This gives
\begin{equation}\label{lindispBenney1} 
\omega = -2 i \xi + \left(\frac{8{\Rey}}{15}-\frac{2\cot\theta}{3}\right) \xi^2 -  \frac{2{\Weber}' }{ 3 } |\xi|^3 -  \frac{1}{3\Capil} \xi^4.
\end{equation}
The phase velocity is $-\omega_{\textrm{i}}/\xi = 2$, inertial forces are linearly destabilizing, with both 
electric and surface tension forces being linearly stabilizing. Gravitational forces are stabilizing for overlying films ($\theta < \pi/2$), and destabilizing for hanging films. The critical Reynolds number at which the flat film state destabilizes is recovered, ${\Rey}_{\textrm{c}} \equiv 5\cot\theta/4$, a quantity which is zero for vertical films. It follows 
from \eqref{lindispBenney1}, and also holds for the WIBL models and the full system, that if a flow is above critical, ${\Rey}>{\Rey}_{\textrm{c}}$ (which is always verified for hanging flows since $\Rey_{\textrm{c}} < 0 \leq \Rey$), then increasing surface tension and/or the strength of the parallel electric field cannot prevent linear instability in all wavenumbers; we have $\omega_{\textrm{r}} \sim \xi^2$ locally near $\xi = 0$, and thus, for any $\Weber'$ and $\Capil$, the flat state will be linearly unstable on a sufficiently large spatial domain.
%{\color{red}
%It is important to note that an inertial instability (and gravitational instability for $\theta > \pi/2$) in this model cannot be fully saturated by the higher order stabilising terms for all wavenumbers; for ${\Rey}>{\Rey}_{\textrm{c}}$ and any $\Weber'$ and ${\Capil}$, there is always a low wavenumber band of linearly unstable modes satisfying $\omega_{\textrm{r}}(\xi) > 0$ -- the flat state will be linearly unstable on a sufficiently large spatial domain.
%}
For the temporal linear stability of the WIBL models, we substitute $H = 1 + \tilde{H} e^{i \xi x + \omega t}$ and $F = 2/3+ \tilde{F} e^{i \xi x + \omega t}$ into (\ref{WIBL2firsteq},\ref{WIBL2secondeq}), and linearize to obtain a quadratic in $\omega$:
\begin{equation} \label{lindispWIBL1}\frac{2{\Rey}}{5} \omega^2 +\left(1 + \frac{68{\Rey}}{105} i \xi +  \underline{\underline{\frac{9}{5} \xi^2}} \right)\omega + 2 i \xi +  \left( \frac{2\cot\theta}{3} - \frac{8{\Rey}}{35} \right) \xi^2 + \frac{2\Weber' }{ 3} |\xi|^3 + \frac{1}{3\Capil} \xi^4  + \underline{\underline{ \frac{8}{5} i \xi^3}} = 0.\end{equation}
The two double-underlined terms in the above expression correspond to $O(\delta^2)$ terms in the WIBL2 model, and are dropped for the consideration of WIBL1. Both WIBL models have the same critical Reynolds number ${\Rey}_{\textrm{c}}$ as the Benney model and Orr--Sommerfeld theory.
%, improving on the IBL model of Shkadov \cite{shkadov1967wave} near critical. 

\begin{figure}
\begin{subfigure}{2.9in}
\caption{${\Rey} = 1$.} 
\includegraphics[width=2.8in]{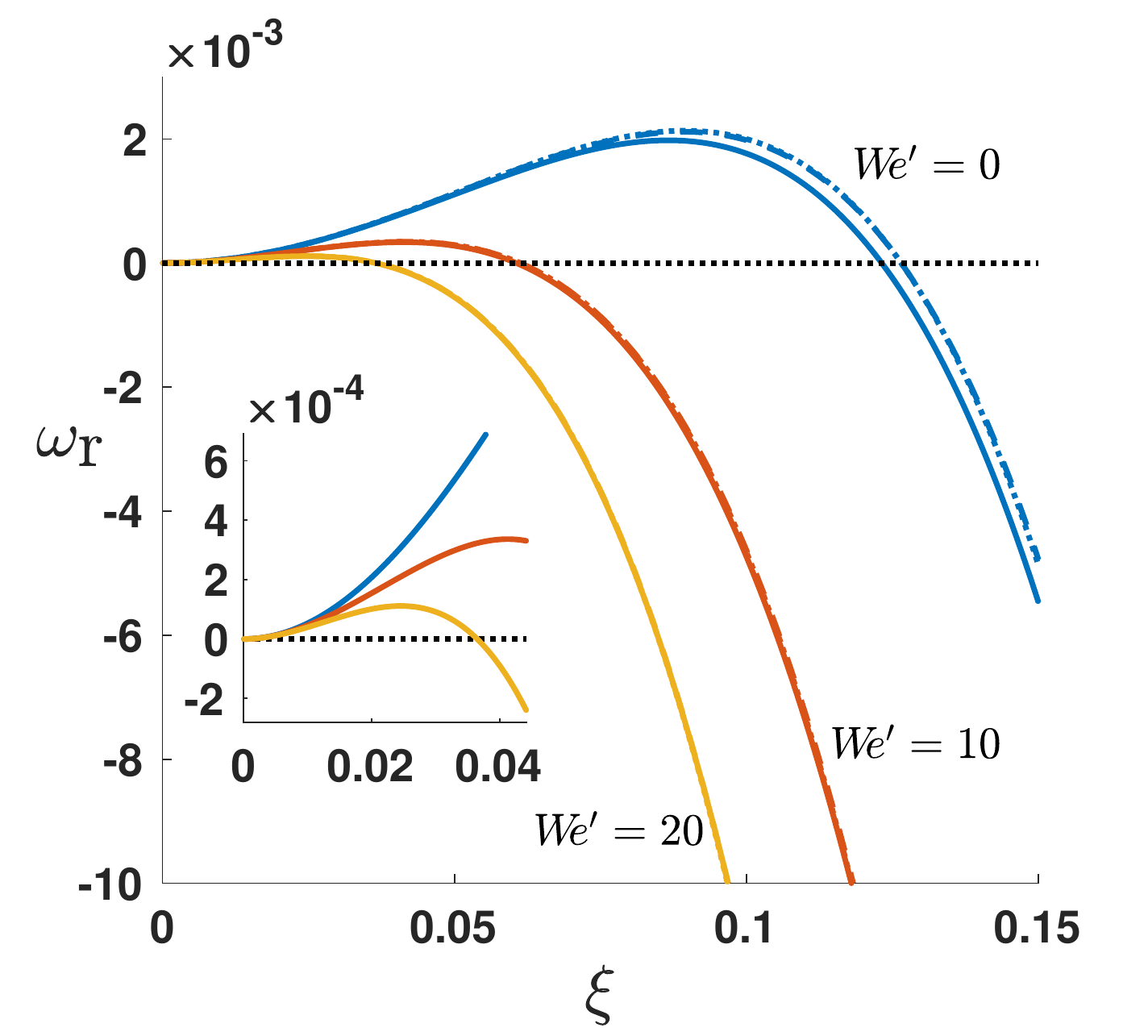}
\end{subfigure}
\begin{subfigure}{2.9in}
\caption{${\Rey} = 4$.} 
\includegraphics[width=2.8in]{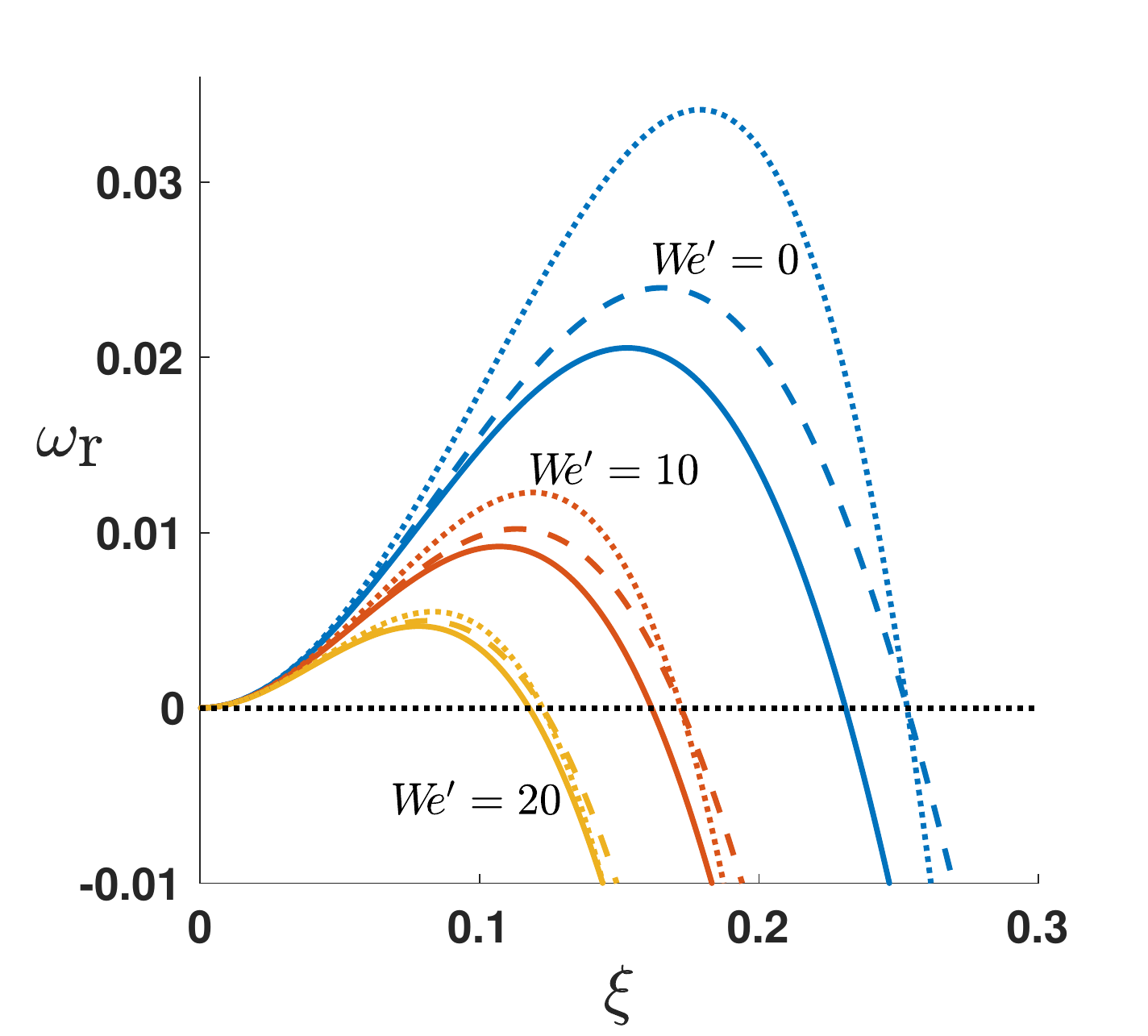}
\end{subfigure}
\caption{Linear growth rates as a function of the wavenumber for a range of electric Weber numbers, $\Weber' = 0, 10 , 20$. The parameters taken are $\theta = \pi/2$ ($\cot\theta = 0$) and ${\Capil} = 0.01$. The dotted/dashed/solid line corresponds to the Benney/WIBL1/WIBL2 model, respectively. Panels (a,b) take the cases ${\Rey} = 1, 4$, respectively. }\label{linearstabilityplot1}
\end{figure}

Figure \ref{linearstabilityplot1} plots growth rates for the three models in the vertical substrate case, $\theta = \pi /2$, with ${\Capil} = 0.01$ 
and $\Weber' = 0, 10 , 20$, for ${\Rey} = 1$ in panel (a) and ${\Rey} = 4$ in panel (b). 
The growth rates for the Benney model, WIBL1 and WIBL2 are plotted with dotted, dashed and solid curves, respectively. 
The results from the three models are in closer agreement for $\Rey=1$, which is nearer to
the critical Reynolds number ${\Rey}_{\textrm{c}} = 0$, with more visible discrepancies for $\Rey=4$ in panel (b). 
The stabilizing effect of the electric field is evident -- in panel (a), the band of modes with wavenumbers between approximately $0.06$ and $0.12$ become linearly stable as $\Weber'$ is increased from $0$ to $10$, and the maximum growth rate decreases.
\begin{figure}
\includegraphics[width=2.9in]{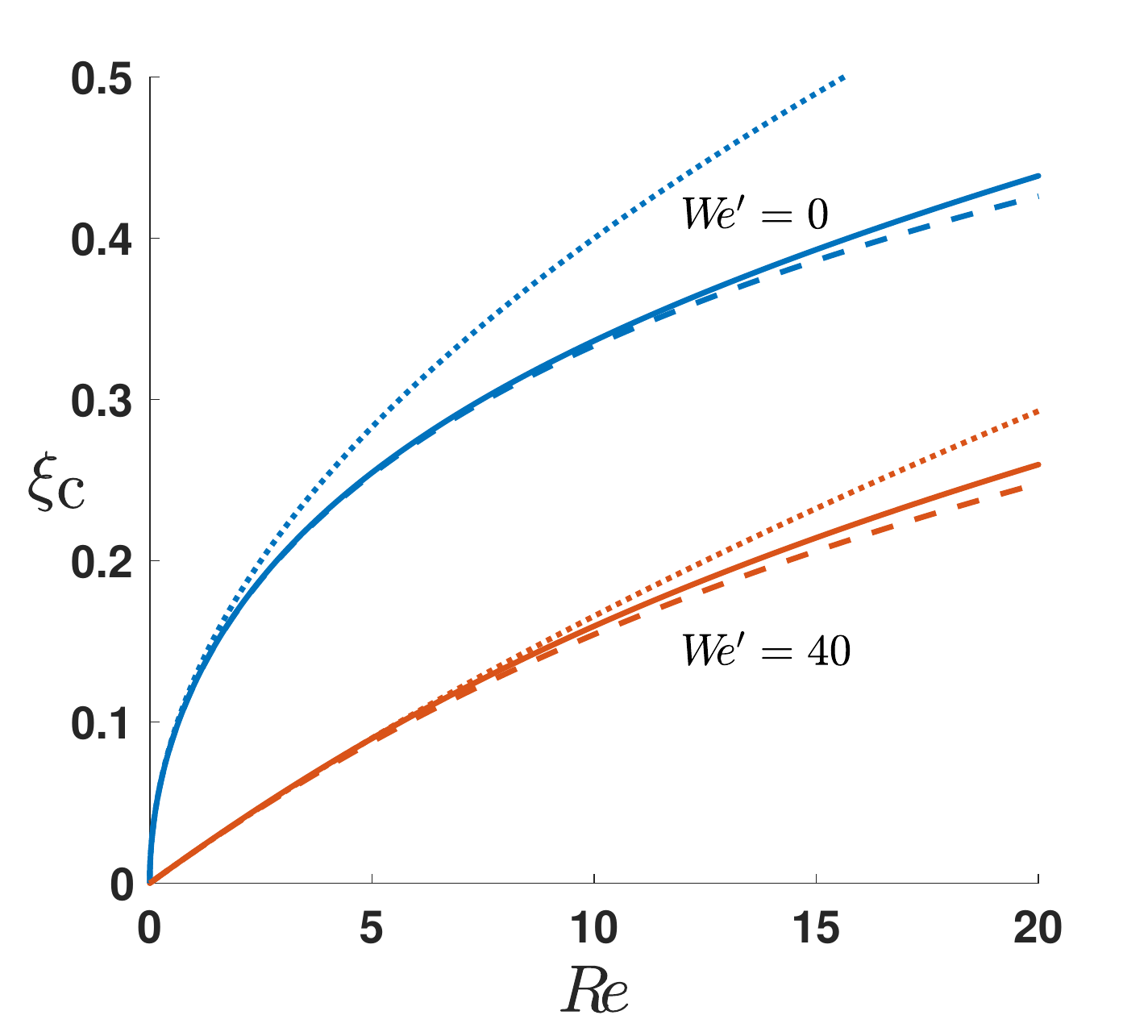}
\caption{Critical wavenumber dependence on $\Rey$ for $\theta = \pi/2$, ${\Capil} = 0.01$ and ${\Weber}' = 0, 40$. 
The dotted, dashed, and solid lines correspond to the results for the Benney/WIBL1, WIBL2, and Orr--Sommerfeld models, respectively. The electrical permittivities used in the Orr--Sommerfeld model are $\epsilon^{\textrm{S}} = 1.5$, $\epsilon^{\textrm{I}} = 2$, $\epsilon^{\textrm{II}} = 1$.}
\label{OSplot1}
\end{figure}
The band of unstable modes with positive wavenumbers $(0, \xi_{\textrm{c}})$ is the same for the Benney and WIBL1 models, 
where $\xi_{\textrm{c}}$ is the critical wavenumber satisfying $\omega_{\textrm{r}} = 0$,
with linear stability obtained if the non-dimensional system length $L = L_0/\ell$ satisfies $L < 2\pi /\xi_{\textrm{c}}$.

{\color{black}

In figure \ref{OSplot1} we plot the variation of the critical wavenumber $\xi_{\textrm{c}}$ against the Reynolds number for a vertical film flow. For a given $\Rey$, instability is found for wavenumbers below the curves. The long-wave models are compared to the full stability results based on the Orr--Sommerfeld equation discussed in Appendix \ref{OrrSommerfeldAppendix}, taking ${\Capil} = 0.01$ and ${\Weber}'= 0,\,40$. We note that the Benney and WIBL1 models give the same $\xi_{\textrm{c}}$, and thus are represented in the figure with the same (dotted) line. Unlike the long-wave models, dependence of the Orr--Sommerfeld system on the dielectric constants is not completely absorbed into the non-dimensional parameter ${\Weber}'$ -- see expression \eqref{EfieldAppenContribute} in Appendix \ref{OrrSommerfeldAppendix}. Thus, we must also prescribe electrical permittivities; we took $\epsilon^{\textrm{S}} = 1.5$, $\epsilon^{\textrm{I}} = 2$, and $\epsilon^{\textrm{II}} = 1$ to obtain the solid curve in figure \ref{OSplot1}. It is clear that WIBL2 agrees with the Orr--Sommerfeld theory much more closely than Benney/WIBL1 in both the non-electrified (as reported in \cite{kalliadasis2011falling}; see their Fig.~4.2, Ch.~4, p.~72) and electrified case.
%We see that WIBL2 does better than Benney/WIBL1 for non-zero ${\Weber}^\prime$ as well as ${\Weber}^\prime=0$ as reported
%in \cite{kalliadasis2011falling} -- see their Fig.~4.2, Ch.~4, p.~72. 
%Note from \eqref{EfieldAppenContribute} that the dispersion relation for the Orr--Sommerfeld system depends on both $\Weber'$ and the dielectric constants explicitly (unlike the long-wave models where all dependence on the dielectric constants is absorbed into $\Weber'$). 
% with our Reynolds number and Serafims Kapitza (not his Reynolds) we have $\Capil = {\Rey}^{2/3}/2^{1/3}\Gamma$.

The critical electric field strength beyond which the flat film solution is linearly stabilized may also be computed from the above dispersion relations; of particular interest is its dependence on the system length. For example, taking the Benney model on a spatial domain of non-dimensional length $L = L_0/\ell$, we require
\begin{equation}    {\Weber}'  >   \left(  \frac{4{\Rey}}{5} - \cot\theta \right)\frac{L_0}{2\pi \ell} - \frac{\pi \ell }{ \Capil L_0} \end{equation}
for linear stability. This implies that for $L_0 \gg 1$, the critical $E_0$ required to stabilize the flat film solution scales with $L_0^{1/2}$. Nonlinearly, we expect that linear stability will imply that the flat film solution is reached after large times, as found by Anderson et al.~\cite{RaduAnder1} in the related problem without a mean flow. Thus, a sufficiently strong electric field can fully stabilize an overlying or hanging film on a finite spatial domain (at least in 2D). We do not provide nonlinear simulations of the long-wave models here as this is not the focus of the present study. We expect that the field strength required for dripping suppression should be independent of the system length (for $L_0 \gg 1$), and thus the temporal stability results cannot be used as a predictor of this nonlinear phenomena.

%It may also be computed from our dispersion relations that the critical $E_0$ required to stabilise the flat film solution scales like $L_0^{1/2}$. We can show this using the Benney dispersion relation as follows: 
%Returning to dimensional variables, $(x,\xi) \rightarrow (x\ell ,\xi/\ell)$, the flat solution of the Benney equation on a system of physical length $L_0$ is linearly stable if
%%\begin{equation}    {\Weber}'  >  \left(  \frac{4{\Rey}}{5} - \cot\theta \right)|\xi|^{-1} - \frac{|\xi|}{2\Capil}   ,\quad \textrm{for } \xi = \frac{2\pi\ell}{L_0}. \end{equation}
%\begin{equation}    {\Weber}'  >  \frac{1}{\ell} \left(  \frac{4{\Rey}}{5} - \cot\theta \right)|\xi|^{-1} - \frac{\ell|\xi|}{2\Capil}   ,\quad \textrm{for } \xi = \frac{2\pi}{L_0}. \end{equation}
%For large $L_0$, the electric Weber number $\Weber'$ that satisfies the above inequality is $O(L_0)$, and thus $E_0 \sim L_0^{1/2}$ for linear stabilisation of the flat interface. 

%This implies that the required potential difference between the inlet and outlet behaves like $L_0^{3/2}$.
% We note that, linear stability usually ensures nonlinear stability of the flat film solution, even with the addition of an electric field \citep{RaduAnder1}. Thus, the above results indicate that a sufficiently strong electric field can provide nonlinear stability to the the flat interface solution on a finite spatial domain in 2D, even for hanging arrangements.

}

\section{Absolute and convective instabilities}\label{sec:absConv}

In this section we investigate the absolute or convective nature of the linear instabilities for the various long-wave models. We compare these results with (analytical) full linear computations in the case of
zero Reynolds number. We also perform DNS with pulse initial conditions in section \ref{sec:DNS} to validate the linear results, and to investigate the nonlinear development of the instabilities.

Briefly, a system is classified as convectively unstable if its response to a small localized
perturbation is convected away from the location of the initial disturbance,
whereas absolute instabilities lead to growth both upstream and downstream of the disturbance location -- see \cite{huerre1990local} for details. 
Such instabilities are of interest in thin film flows and it is reasonable to expect that
an absolutely unstable hanging film is more susceptible to dripping than a convectively unstable one. 
However, the evolution of liquid films, and in particular the dripping process, is very nonlinear and our approach in addressing such
questions is a combination of linear theory and allied DNS.
Moreover, linear theory is not in a position to predict whether a convectively unstable film 
will drip far downstream from the inlet (assuming inlet forcing is driving the system for example), or converge to a bounded wave-train of 
nonlinearly saturated pulses. Brun et al.~\cite{Brun1} considered this scenario in the case of a 2D flow with small Reynolds numbers 
and used an inertialess Benney equation given in terms of our non-dimensional parameters as \eqref{Benney1withCap2} with ${\Rey} = {\Weber}' = 0$. 
They determined regions of absolute and convective instability analytically in terms of the inclination angle and the ratio of mean film thickness to capillary length (this can be rewritten as a relationship between $\theta$ and ${\Capil}$). They found reasonable agreement with experiments, 
where films in the convectively unstable regime (of their model) were found not to drip at all, or yielded relatively few drips compared to the 
absolutely unstable regime. An increased number of drips emerged deeper into the absolute instability regime. The linear theory was extended to more complex WIBL models by Scheid et al.~\cite{ScheidKofman1} without the low Reynolds number restriction. They calculated a fluid-dependent critical angle in terms of the Kapitza number (fixed for a given fluid) above which only convective instabilities may exist for any Reynolds number; the minimum of these is a fluid-independent critical angle (corresponding to the limit of zero Kapitza number). In what follows, we show that an electric field may be used to increase these critical angles towards the horizontal arrangement, i.e.~increase the parameter space of convectively unstable systems. 
%{\bf{ a phenomenon
%that could be useful in experimental investigations.}}

%The simple recipe utilised in \cite{Brun1,ScheidKofman1} for determining absolute and convective instabilities is immediately applicable for problems with dispersion relations which are analytically extensible to the complex plane. In such cases, it can be shown that the flow transitions for parameters when pinching of the zero contours of $\omega_{\textrm{r}}$ (the real part of the dispersion relation) occurs -- this is known as the Briggs--Bers criterion. This line in parameter space can be formulated as the solution to $\omega_{\textrm{r}} = 0$ with the zero group velocity condition $\partial_{\xi} \omega = 0$. For nonlocal problems, however, such as those considered here, care must be taken to apply this method (see \cite{huerre1990local} for a discussion of such issues). Despite this, we will show that the Briggs--Bers pinching criterion is applicable through a symmetry of the dispersion relations. In order to apply our analysis, the dispersion relations obtained in the previous section must be extended correctly to the complex plane. From (\ref{realsigndefn1}a), we see that $|\xi|^3$ must be extended as $\operatorname{sign}(\xi_{\textrm{r}}) \xi^3$; note that this term is not analytic on the imaginary wavenumber axis $\xi_{\textrm{r}} = 0$, yet is symmetric about it.

We apply the theory of Fokas and Papageorgiou \cite{fokas2005absolute} for the linearized model equations at hand. 
It is useful to extend the dispersion relation $\omega(\xi)$ to complex wavenumbers $\xi = \xi_{\textrm{r}} + i \xi_{\textrm{i}} \in \mathbb{C}$. 
%Generically, for real wavenumbers, instability occurs in a wavenumber band 
%$(-\xi_{\textrm{c}}, \xi_{\textrm{c}})$, where $\xi_{\textrm{c}} \in \mathbb{R}$ is the critical wavenumber.
For the long-wave models considered here, $\omega(\xi)$ is given by either \eqref{lindispBenney1} or 
\eqref{lindispWIBL1} with $|\xi|^3$ replaced by its correct generalisation to complex wavenumbers, $\operatorname{sign}(\xi_{\textrm{r}}) \xi^3$. 
Note also the conjugate symmetry $\omega(-\xi_{\textrm{r}}) = \overline{\omega(\xi_{\textrm{r}})}$, where the bar denotes the complex conjugate of a quantity, since the solution of the PDE is real-valued. Given an initial condition $H(x,0)=H_0(x)$, the solution is given by the Fourier transform pair
%Consider a general linear PDE for a real-valued function $H$ with dispersion relation $\omega(\xi)$ extended to complex wavenumbers $\xi = \xi_{\textrm{r}} + i \xi_{\textrm{i}} \in \mathbb{C}$. For the restriction of $\omega$ to real wavenumbers, we assume that all unstable frequencies are contained in the interval 
%$(-\xi_{\textrm{c}}, \xi_{\textrm{c}})$, where $\xi_{\textrm{c}} \in \mathbb{R}$ is the critical wavenumber. We also assume conjugate symmetry of the dispersion relation, $\omega(-\xi_{\textrm{r}}) = \overline{\omega(\xi_{\textrm{r}})}$, a condition that is satisfied by the dispersion relations under consideration since the solution is real-valued. Given an initial condition $H_0(x)$, the solution of the linear PDE may be expressed as
\begin{equation}\label{HFourierintegral1} 
H(x,t) = \frac{1}{2\pi} \int_{\mathbb{R}} \widehat{H_0}(\xi) e^{i \xi x + \omega(\xi) t} \; \mathrm{d}\xi, \quad \quad  \widehat{H_0}(\xi) = \int_{\mathbb{R}} H_0(x) e^{- i x \xi} \; \mathrm{d}x,
\end{equation}
where $\widehat{H_0}(\xi)$ is also conjugate symmetric since $H_0(x)$ is real. The first integral in \eqref{HFourierintegral1} is split into two parts, one integral over the stable frequencies, $\mathbb{R}\backslash(-\xi_{\textrm{c}}, \xi_{\textrm{c}})$, where $\xi_{\textrm{c}} \in \mathbb{R}$ is the critical wavenumber, and another integral over the interval $(-\xi_{\textrm{c}}, \xi_{\textrm{c}})$ which is denoted $J(x,t)$. 
The former integral decays to zero as $t\rightarrow \infty$ while keeping $x = O(t)$, thus $J(x,t)$ alone contains information
regarding the absolute or convective nature of the instability. 
Using the conjugate symmetry of $\omega$ and $\widehat{H_0}(\xi)$, we have
%\begin{equation} J(x,t) = \frac{1}{2\pi} \int_{0}^{\xi_{\textrm{c}}} \widehat{H_0}(\xi) e^{\omega(\xi) t + i x \xi} + \widehat{H_0}(-\xi) e^{\omega(-\xi) t - i x \xi}\; \mathrm{d}\xi.
%\end{equation}
\begin{equation} 
J(x,t) = \frac{1}{\pi} \operatorname{Real}\left[ \int_{0}^{\xi_{\textrm{c}}} \widehat{H_0}(\xi) e^{i \xi x + \omega(\xi) t} \; \mathrm{d}\xi \right].
\end{equation}
As shown in \cite{fokas2005absolute}, the type of instability is determined by the topology of the curves $\omega_{\textrm{r}}(\xi)$ in the right half-plane. Cauchy's theorem is applicable since the dispersion relation is analytic and bounded in this region, 
where $\operatorname{sign}(\xi_{\textrm{r}}) \xi^3 = \xi^3$. If the integral over $(0,\xi_{\textrm{c}})$ can be deformed 
to part of the curve $\omega_{\textrm{r}}(\xi) = 0$ in the right half-plane, then $J(x,t)$ decays for $t$ fixed with $|x| \rightarrow \infty$, 
and for $x$ fixed with $t \rightarrow \infty$, thus the instability is convective. 
In the case that the curves of $\omega_{\textrm{r}}(\xi) = 0$ do not connect the points $0$ and $\xi_{\textrm{c}}$ on $\mathbb{R}_+$, then the instability is absolute. The Briggs--Bers pinching criterion can then be applied to our problem -- see \cite{Brun1,ScheidKofman1} for its application to the non-electrified flow. 
Solving $\omega_{\textrm{r}} = 0$ and $\partial_{\xi} \omega = 0$ for complex wavenumbers in the right half-plane is an algebraic problem, the solutions of which separate the regions between absolute and convective instability.  In order to find the transition curve in parameter space, we proceed with numerical continuation of solutions to the algebraic problem using the continuation software $\textsc{AUTO-07P}$ \cite{doedel2007auto}.

The Benney and WIBL1 models are identical for ${\Rey} =0$, and without electric fields they yield the equation studied by Brun et al.~\cite{Brun1}. 
We have a unique non-trivial exact solution to the Briggs--Bers criterion for this system with $\xi_{\textrm{r}} > 0$. The critical angle $\theta \in (\pi/2,\pi)$ for this exact solution is defined by $\cot^3\theta = -  243/4{\Capil} (17 + 7^{3/2}),$
with wavenumber $\xi$ given by
\begin{equation}\xi_{\textrm{i}} = \frac{2^{4/3}Z^{2/3} + 3^{1/3} {\Capil} \cot\theta }{2^{5/3} 3^{2/3} Z^{1/3} }, \quad \xi_{\textrm{r}} = (3\xi_{\textrm{i}}^2 - {\Capil} \cot\theta)^{1/2},\quad \textrm{where}\quad Z = \sqrt{ \frac{729 {\Capil}^2}{256 } - \frac{3{\Capil}^3 \cot^3\theta}{16}} - \frac{27{\Capil}}{16},
\end{equation}
and $\omega_{\textrm{i}} = - 3 \xi_{\textrm{r}}/2 - 2\cot(\theta) \xi_{\textrm{r}} \xi_{\textrm{i}}/3$. In the case of ${\Capil} = 0.01$, this exact solution is $\theta \approx 2.963$ (about $80^\circ$ beyond vertical),
%, $Z \approx 9.25 \times 10^{-4}$
$\xi_{\textrm{i}} \approx -0.0873$, $\xi_{\textrm{r}} \approx  0.280$ and $\omega_{\textrm{i}} \approx  -0.510$.

\begin{figure}
\centering
\begin{subfigure}{2.9in}
\caption{$\omega_\textrm{r} = 0$ curves for Benney model.} 
\includegraphics[width=2.8in]{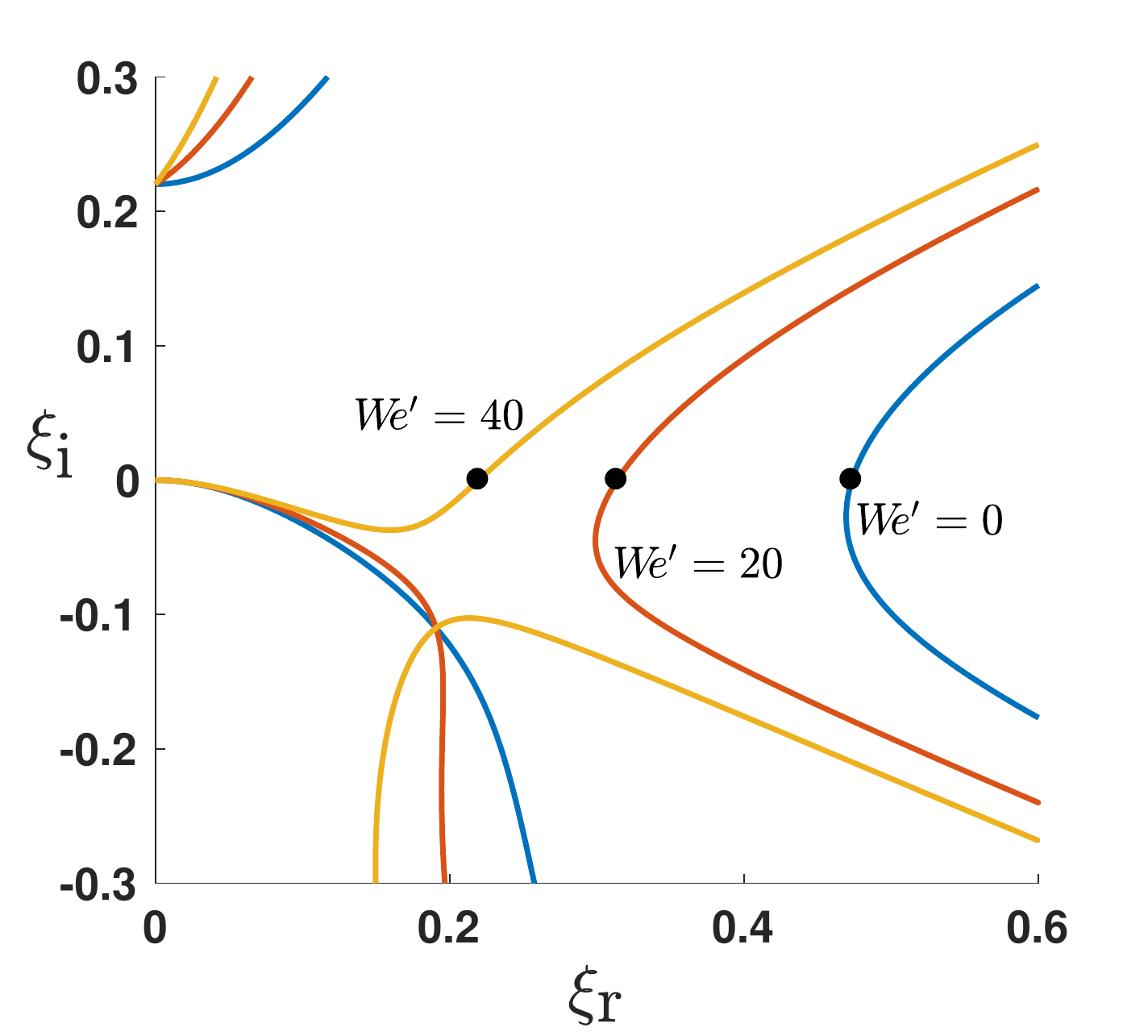}
\hspace{1.1cm}
\end{subfigure}
\begin{subfigure}{2.9in}
\caption{$\omega_\textrm{r} = 0$ curves for WIBL1 model.} 
\includegraphics[width=2.8in]{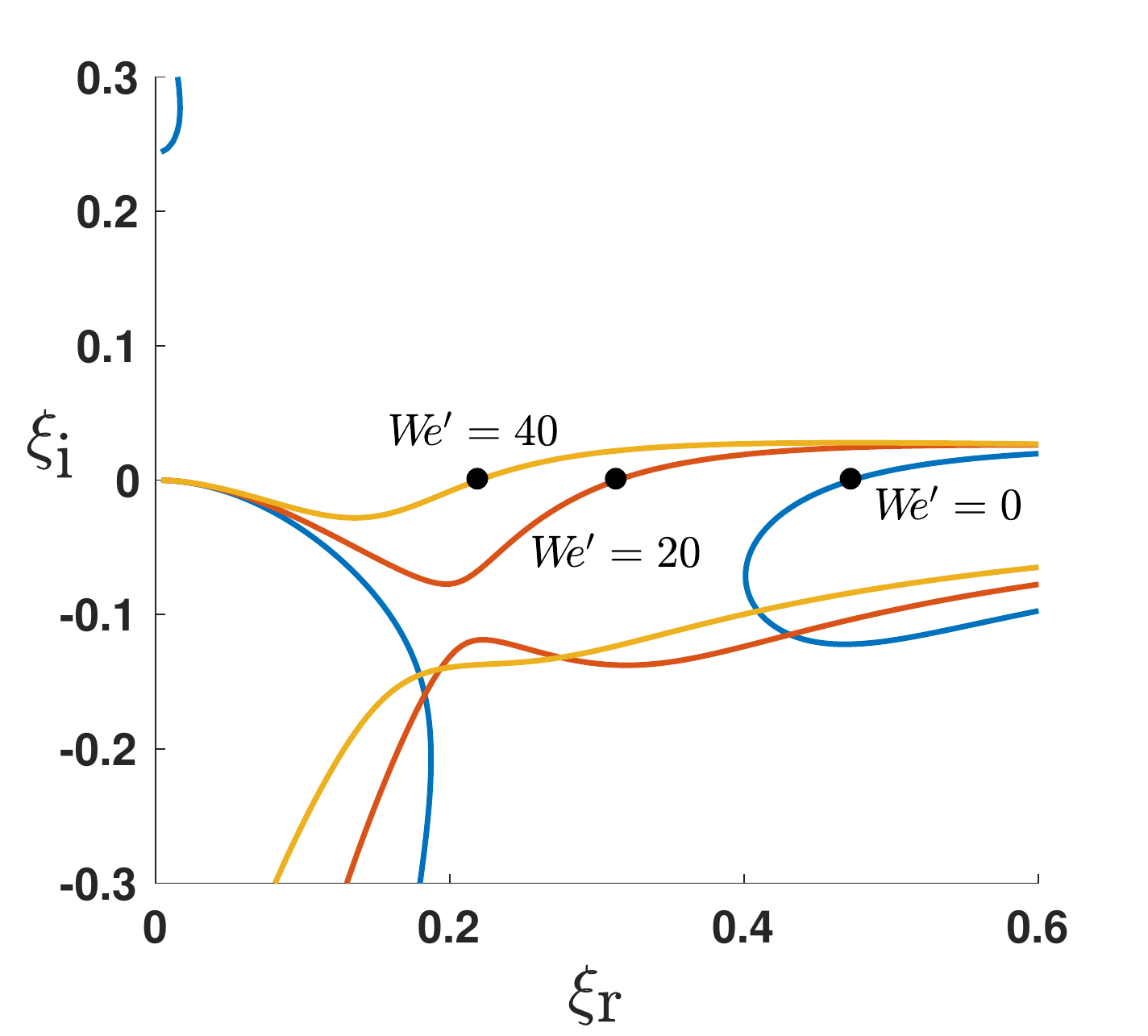}
\hspace{1.1cm}
\end{subfigure}
\begin{subfigure}{2.9in}
\caption{A/C regions for increasing $\Weber'$.} 
\includegraphics[width=2.8in]{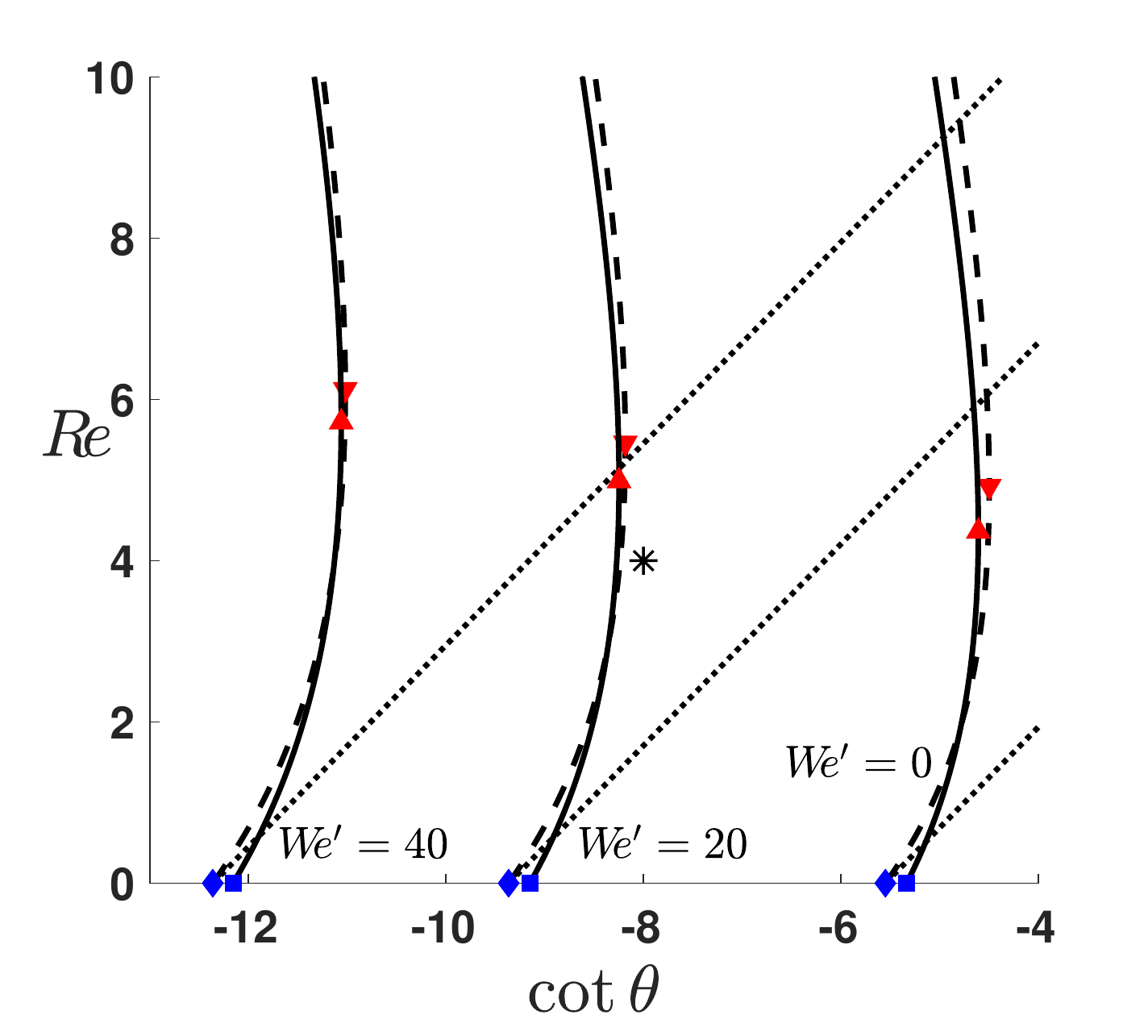}
\end{subfigure}
\begin{subfigure}{2.9in}
\caption{Zero Reynolds number and fold point continuation.} 
\includegraphics[width=2.8in]{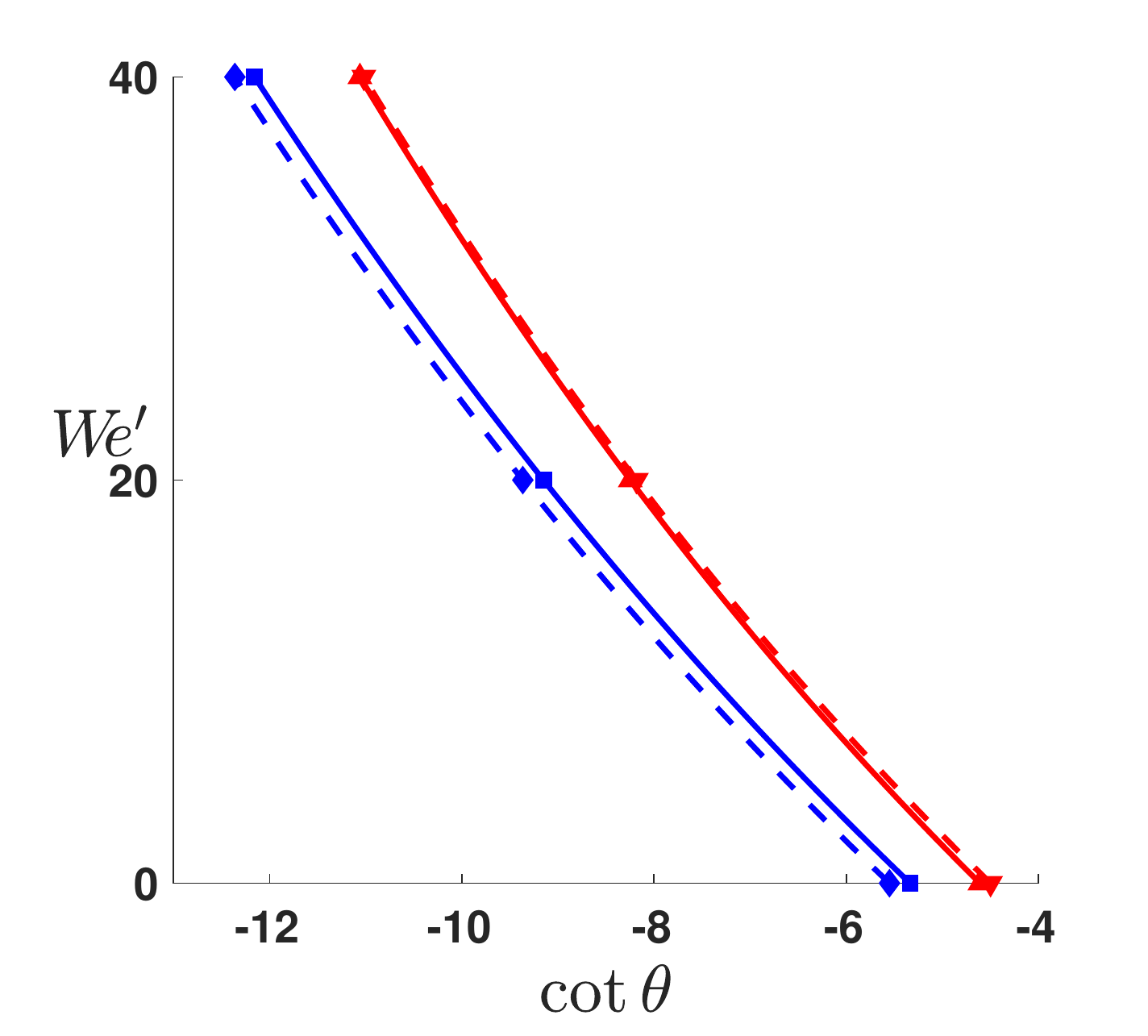}
\end{subfigure}
\caption{Curves of $\omega_\textrm{r} = 0$ in the right half-plane and A/C regions of the Benney and WIBL models. Panels (a,b) plot $\omega_\textrm{r} = 0$ for $\Capil = 0.01$, ${\Rey} = 4$, $\cot\theta = -8$, $\Weber' = 0, 20, 40$, for the Benney and WIBL1 models, respectively. The black dots correspond to the non-trivial real roots. Panel (c) displays the transition curves for the Benney (dotted line), WIBL1 (dashed line), and WIBL2 (solid line) models with ${\Capil} = 0.01$, $\Weber' = 0, 20, 40$, in terms of ${\Rey}$ and $\cot\theta$; convective instabilities occur for parameters on the right of the corresponding curve, with absolute on the left. The transition points at zero Reynolds number are marked with diamonds for the Benney and WIBL1 models, and a square for the WIBL2 model. The fold point for the WIBL1/WIBL2 model is marked with a down/up-pointing triangle. The star in panel (c) corresponds to the parameters used in panels (a) and (b). Panel (d) shows continuation of the ${\Rey} = 0$ transition point and the fold point for the WIBL1 and WIBL2 models.}\label{ACplot1}
\end{figure}

Given a capillary number ${\Capil}$, the curve in parameter space separating convective from absolute instability for either the Benney, WIBL1 
or WIBL2 models can be continued from this exact solution.
% as ${\Rey}$ increases from zero. 
%This solution can also be continued to solutions of the Briggs--Bers criterion for WIBL2 by introducing an additional parameter in front of the $O(\delta^2)$ viscous extensional stress terms in \eqref{WIBL2secondeq} which varies from $0$ (corresponding to WIBL1) to $1$ (corresponding to WIBL2). 
These transition curves can be validated from the topology of $\omega_{\textrm{r}} = 0$ for the given parameters. In figure \ref{ACplot1}(a,b), we give examples of the $\omega_{\textrm{r}} = 0$ curves in the right half-plane (the diagram is symmetric about the imaginary axis) for the Benney and WIBL1 models, respectively, and ${\Capil} = 0.01$, ${\Rey} = 4$, $\cot\theta = -8$, with ${\Weber}' = 0, 20, 40$. For the Benney model in panel (a), the non-trivial real root is connected to the origin when ${\Weber}' = 40$ implying convective
instability, but for ${\Weber}' = 0, 20$ we predict absolute instability.
The ``pinching" of the roots corresponding to the Briggs--Bers criterion occurs between ${\Weber}' = 20$ and ${\Weber}' = 40$ for the Benney model. 
The results for WIBL1 are different as shown in panel (b), with absolute instability only for ${\Weber}' = 0$ 
out of the three cases depicted
(WIBL2 with these parameters gives
very similar results). 
%The non-analyticity of the dispersion relation is apparent in panel (a), where the curves are not 
%smooth at the non-trivial purely imaginary roots for ${\Weber}' > 0$. 
In figure \ref{ACplot1}(c) we collect the A/C transition curves for the different models as a function of ${\Rey}$ and $\cot\theta$, for 
fixed ${\Capil} = 0.01$ and ${\Weber}' = 0,20,40$. The dotted, dashed, and solid curves correspond to the Benney, WIBL1, and
WIBL2 models, respectively. Absolute instability is found to the left of the curves with convective instability to their right. The results show that an increase in the applied electric field strength promotes convective instability, with the absolute instability threshold found for larger negative values of $\cot\theta$, i.e.~for substrate inclinations tending to the horizontal hanging film configuration -- note that $\cot\theta\to-\infty$ as $\theta\to\pi_-$ when the inclination becomes horizontal and the film is hanging vertically.
%In figure \ref{ACplot1}(c), we plot the A/C transition curves obtained from numerical continuation of the Briggs--Bers criterion for fixed ${\Capil} = 0.01$ and ${\Weber}' = 0,20,40$, as ${\Rey}$ and $\cot\theta$ vary; these lines separate the absolutely unstable parameter region (on the left) from the convectively unstable region (on the right) for the Benny and WIBL models. 
The Benney model has a linear transition curve due to its particular dispersion relation, and incorrectly predicts 
that a vertical film flow ($\cot\theta=0$) undergoes the transition from convective to absolute instability at ${\Rey} = 6.939$, and that overlying film flows may become absolutely unstable for higher Reynolds numbers.

In contrast, the more accurate WIBL1 and WIBL2 models 
yield fold points, marked with triangles in the figure, and implying the existence of a fluid-dependent critical angle 
above which all instabilities are convective irrespective of Reynolds number. 
The predicted critical angle corresponds to a hanging film, so that 
WIBL models for overlying films can only be convectively unstable. 
%Note that the curves for the inertialess 
%Benney would be vertical lines emanating from the square markers on the ${\Rey = 0}$ axis, making them much closer to the 
%WIBL curves than the inertial Benney equation for most parameters (this is purely coincidental, and does not attest to the accuracy of the model). 
%All models agree that an increase in ${\Weber}^\prime$ is stabilizing in the sense that the A/C transition curve moves to the left. 
The star in panel (c) corresponds to $\Rey=4$, $\cot\theta=-8$ used in panels (a,b), and verifies the A/C conclusions of the root plots. Figure \ref{ACplot1}(d) plots the zero Reynolds number transition lines and fold points as the electric Weber number increases; it is evident that there is good agreement between the long-wave models at ${\Rey} =0$, and that WIBL1 and WIBL2 agree closely on the location of the critical angle fold point for these parameters; increasing $\Weber'$ from $0$ to $40$ moves the ${\Rey} = 0$ transition and fold points by over $5^{\circ}$. Our parameters are related to those in Scheid et al.~\cite{ScheidKofman1} (denoted with a superscript Sch) by $\delta^{\textrm{Sch}} = 2^{4/3}  {\Capil}^{1/3} {\Rey}$, $\zeta^{\textrm{Sch}} = 2^{1/3} {\Capil}^{1/3} \cot \theta$, $\eta^{\textrm{Sch}} = 2^{2/3}  {\Capil}^{2/3}$. These are the reduced Reynolds number, the reduced inclination number, and the viscous extensional number, respectively. For the WIBL1 (and Benney) model, viscous extensional stresses are ignored, i.e.~$\eta^{\textrm{Sch}} = 0$ or dropping the $O(\delta^2$) terms in \eqref{WIBL2secondeq}, but for the computations for WIBL2 shown in figure \ref{ACplot1}(c,d) we have $\eta^{\textrm{Sch}} = 0.0737$ -- see figure 3 in \cite{ScheidKofman1}. From the results shown in figure \ref{ACplot1}, we recover their critical values of $\zeta_A^{\textrm{Sch}} \approx -1.507$ corresponding to the transition point for zero Reynolds numbers, and $\zeta_C^{\textrm{Sch}} \approx -1.221$, corresponding to the fold point for the WIBL1 transition curve; the latter of these is the maximum value of $ \zeta^{\textrm{Sch}}$ even for non-zero extensional stresses.

The results in figure \ref{ACplot1} are computed for a fixed capillary number ${\Capil}$, and varying it 
gives a 2D surface in $(\theta,\Rey,\Capil)$-space, separating the regions of absolute and convective instability; this surface changes with the electric field strength through variations in $\Weber'$. 
In the non-electrified case, Scheid et al.~\cite{ScheidKofman1} showed the existence of a fluid-independent critical angle $\theta_{\textrm{c}} = 147.4^{\circ}$ ($\theta_{\textrm{c}} - 90^{\circ} = 57.4^{\circ}$ beyond vertical) for 
the full WIBL2 model; below $\theta_c$ absolute instabilities cannot exist (see figures 4, 5 and 6 in \cite{ScheidKofman1}), providing 
a lower bound on the A/C transition surface in terms of the inclination angle. For ${\Weber}'=0$, this lower bound 
is attained at $\Rey=0$. We recover the minimum critical angle with the simplified WIBL2 system, but not WIBL1 -- inclusion of viscous extensional stresses is vital as absolute instability can be found for WIBL1 at any angle beyond vertical. 
%Such a minimum critical angle can be obtained by continuation of the fold points (up and down facing triangles) in figure \ref{ACplot1}(c) in the capillary number.

\begin{figure}
\centering
\begin{subfigure}{2.9in}
\caption{A/C transition surface for ${\Weber}' =0$.} 
\includegraphics[width=2.8in]{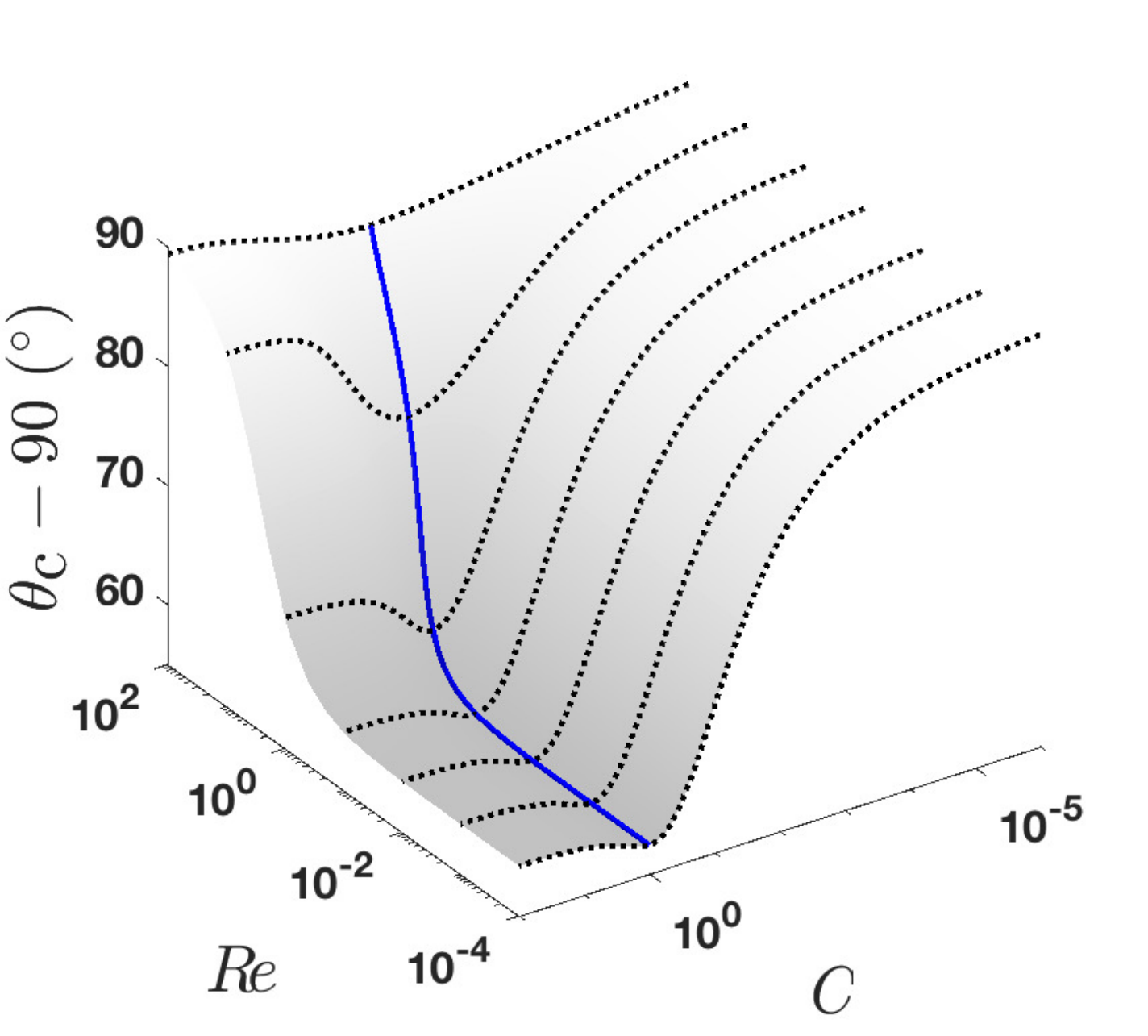}
\hspace{1.1cm}
\end{subfigure}
\begin{subfigure}{2.9in}
\caption{A/C transition surface for ${\Weber}' =5$.} 
\includegraphics[width=2.8in]{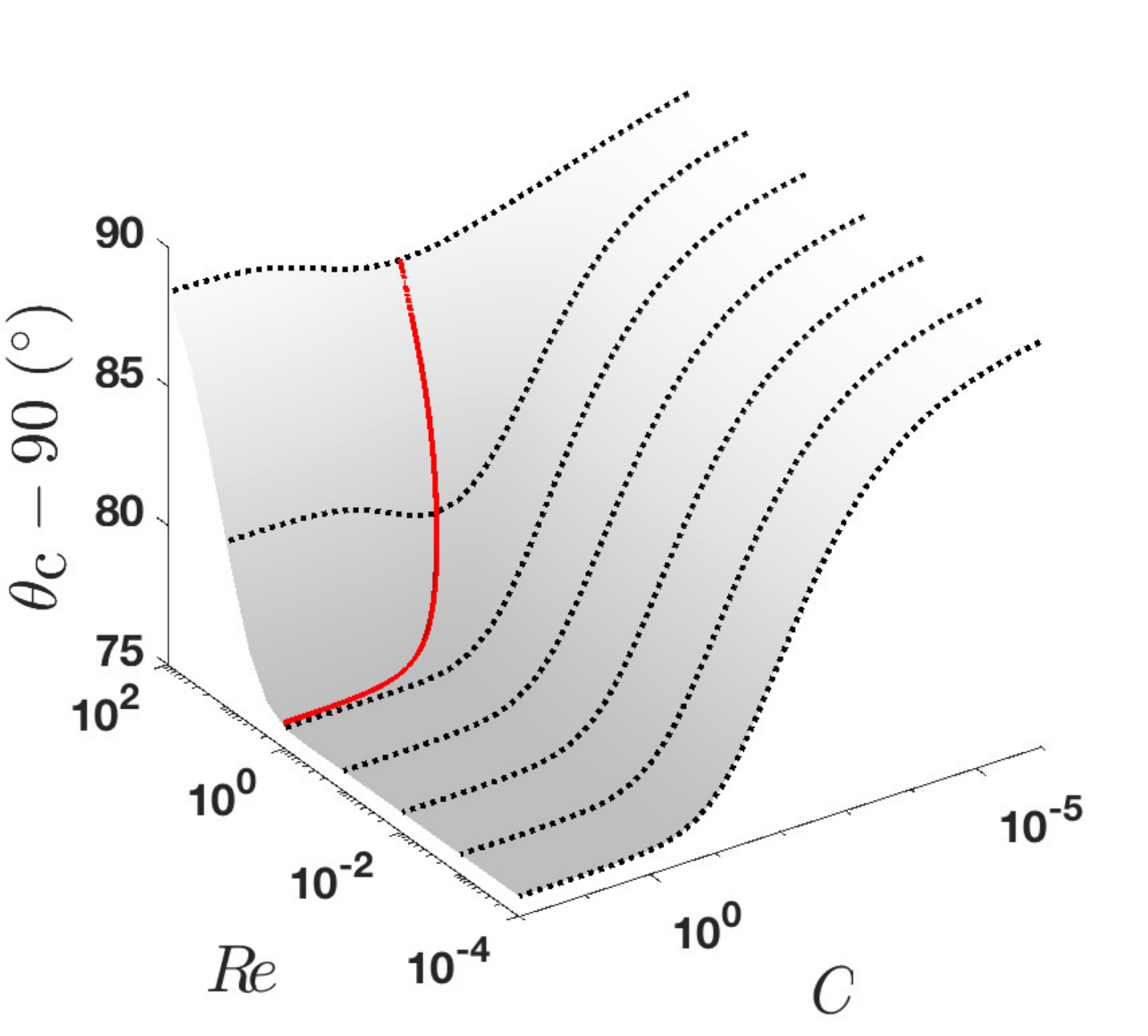}
\hspace{1.1cm}
\end{subfigure}
\begin{subfigure}{2.9in}
\caption{Critical angle minimized over ${\Capil}$ against ${\Rey}$.} 
\includegraphics[width=2.8in]{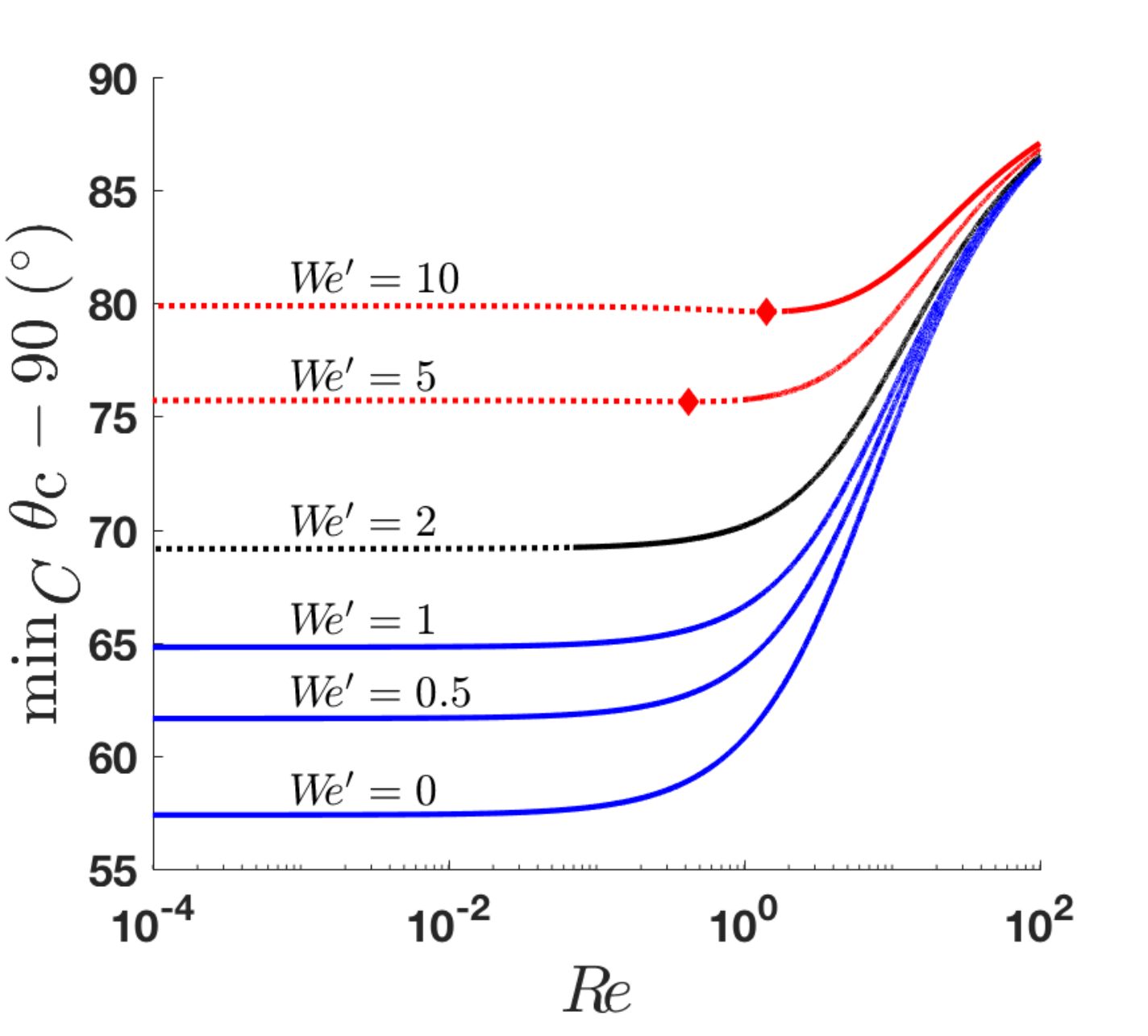}
\end{subfigure}
\begin{subfigure}{2.9in}
\caption{Minimum critical angle against ${\Weber}'\sin\theta$.} 
\includegraphics[width=2.8in]{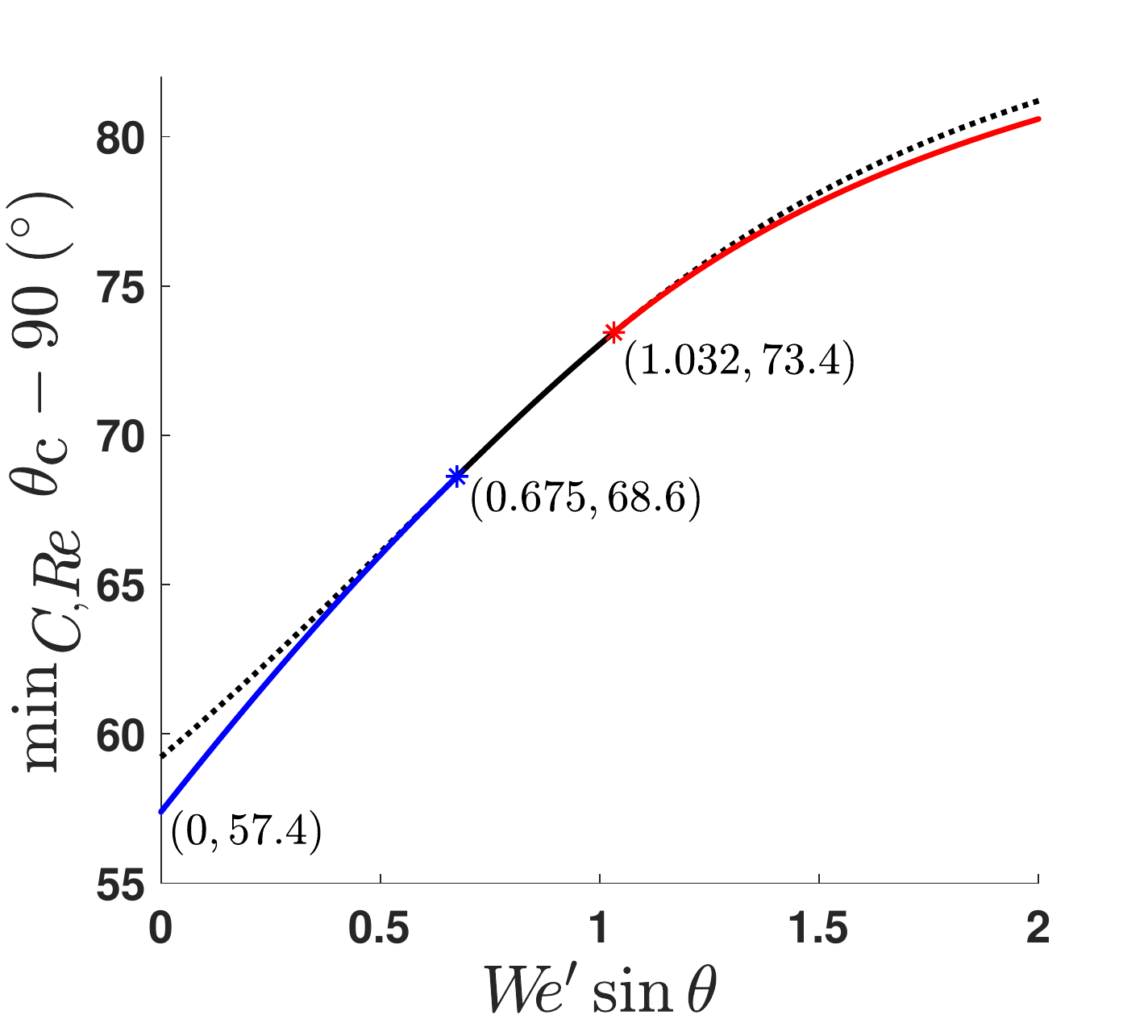}
\end{subfigure}
\caption{Effect of electric field strength on A/C transition surface for WIBL2. Panels (a,b) show the A/C transition surface for the choices of ${\Weber}' =0, 5$. Panel (c) shows the minima of these surfaces over $\Capil$ plotted against ${\Rey}$. The dotted lines signify that the minimum is attained in the ${\Capil} \rightarrow \infty$ limit, and the diamonds denote minima of curves where the minimum is found at ${\Rey} > 0$. Panel (d) plots the minima in the Reynolds number of the curves in panel (c) against ${\Weber}' \sin\theta$.}\label{ACplot2}
\end{figure}

Figure \ref{ACplot2}(a,b) plots the A/C transition surfaces of the WIBL2 model for ${\Weber}' = 0$ and ${\Weber}' =5$, respectively, with absolute
instability obtained above the surfaces (we give the critical inclination angle from the vertical, $\theta_{\textrm{c}} - 90^{\circ}$, expressed in degrees
for comparison with \cite{ScheidKofman1}).
The superimposed dotted curves depict the transition curve for fixed ${\Rey}$ as ${\Capil}$ varies, and the solid lines track the minima 
of $\theta_{\textrm{c}} - 90^{\circ}$ over all ${\Capil}$ as ${\Rey}$ is varied. 
As can be seen from the results in panel \ref{ACplot2}(a), there is a global minimum value of
$\theta_{\textrm{c}} - 90^{\circ}= 57.4^{\circ}$ at ${\Rey} = 0$ as found by Scheid et al.~\cite{ScheidKofman1} in their study of the non-electrified flow. 
Interestingly, the minima are found at $O(1)$ values of ${\Capil}$; this non-monotonic A/C threshold is not an artefact of long-wave modelling, as confirmed by DNS 
(see figure 7 in \cite{ScheidKofman1}). 
Unsurprisingly, the strong surface tension limit $\Capil\to 0$ yields critical angles which are very close to horizontal for any ${\Rey}$. We turn next to the inclusion of a field with 
${\Weber}' = 5$ shown in figure \ref{ACplot2}(b).
%The minimum $\theta_{\textrm{c}} - 90^{\circ}$
The solid curve tracing the minima diverges to ${\Capil} = \infty$ as the Reynolds number is decreased from $100$, corresponding to zero surface tension effects. The dotted curves for small ${\Rey}$ shown in panel (b) are monotonically decreasing in ${\Capil}$, and their minimum value is obtained in the limit ${\Capil} \rightarrow \infty$, which is qualitatively different to the non-electrified case. This is verified by performing continuation of the algebraic system corresponding to the Briggs--Bers criterion in the parameter $1/{\Capil}$, where it is found that the fold point passes to prohibited negative values of $1/{\Capil}$ as the Reynolds number is decreased.

Panel (c) gives the projection of the minimum curves in panels (a) and (b) with other choices of ${\Weber}'$ against ${\Rey}$. The region above a given line corresponds to either absolute or convective instability, whereas all parameters below the line give convectively unstable systems. We find three regimes depending on the value of ${\Weber}'$: For ${\Weber}' < 1.85$, the critical angle for each ${\Rey}$ is attained at finite ${\Capil}$, and the minimum critical angle is found in the zero Reynolds number limit. For ${\Weber}' \in (1.85,3.62)$, the critical angle for small Reynolds numbers is found in the limit of zero surface tension (indicated with a dotted line section), yet the minimum is still found at ${\Rey} =0$. The former is also true for ${\Weber}' >3.62$, except the minimum critical angle is found for ${\Rey} > 0$, shown with diamonds in panel (c). 
%The solid and dotted lines in the figure distinguish between cases of the minimum angle being attained at finite and infinite capillary numbers, respectively. 
%Furthermore, the diamonds denote the minima of the curves in the case that it is not found at ${\Rey} = 0$. 
The minimum critical angle (the minimum of the A/C transition surface over both ${\Capil}$ and ${\Rey}$) is plotted as a function of 
${\Weber}'\sin\theta$ (a $\theta$-independent quantity) in figure \ref{ACplot2}(d), where the three sections of the solid curve 
(separated with stars) correspond to the three cases in panel (c). The dotted lines are continuations of the middle section of the solid curve 
corresponding to ${\Rey} = 0$ and ${\Capil} \rightarrow \infty$.

%The first section of the solid line corresponds to ${\Weber}'\sin\theta<0.675$, with values of $\theta_{\textrm{c}} - 90^{\circ} \in (57.4^{\circ}, 68.6^{\circ})$, giving ${\Weber}'<1.85$. In this case, the minimum transition angle is found for finite capillary numbers and zero Reynolds number. The next section of the line corresponds to values of electric field strength for which the $\theta_{\textrm{c}} - 90^{\circ}$ is found for zero surface tension and Reynolds number; the line for this limiting case is continued beyond the stars with a dotted line. The final section of the line corresponds to the case where the minimum critical angle is found for ${\Capil} \rightarrow \infty$ and ${\Rey} > 0$; two such cases are marked with diamonds in panel (c). Here we have ${\Weber}'\sin\theta>1.032$ and values of $\theta_{\textrm{c}} - 90^{\circ} > 73.4^{\circ}$; in this case ${\Weber}'>3.62$.

\begin{figure}
\centering
\begin{subfigure}{2.9in}
\caption{A/C transition surface for ${\Capil} \rightarrow \infty$.} 
\includegraphics[width=2.8in]{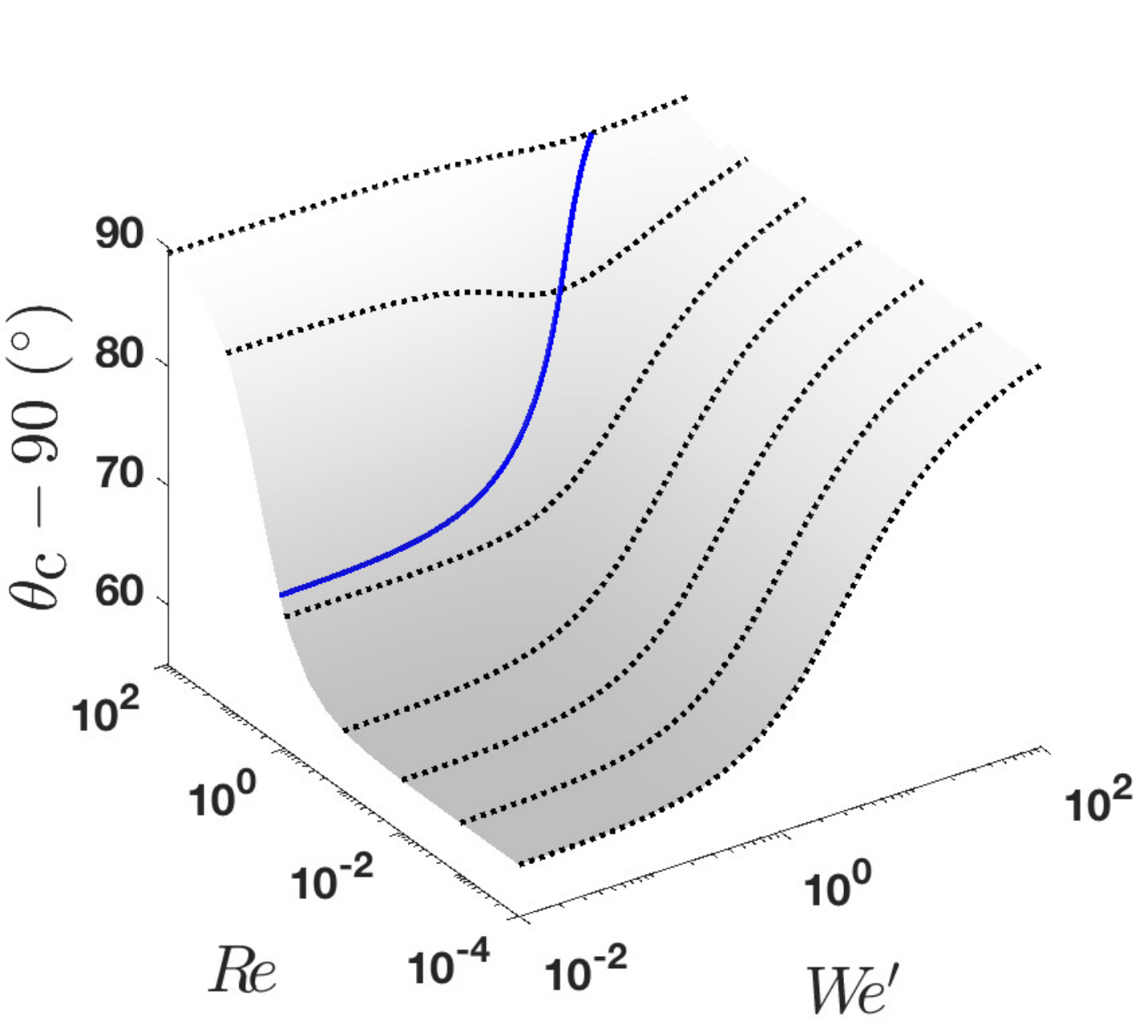}
\end{subfigure}
\begin{subfigure}{2.9in}
\caption{Critical angle minimized over ${\Weber}'$ against ${\Rey}$.} 
\includegraphics[width=2.8in]{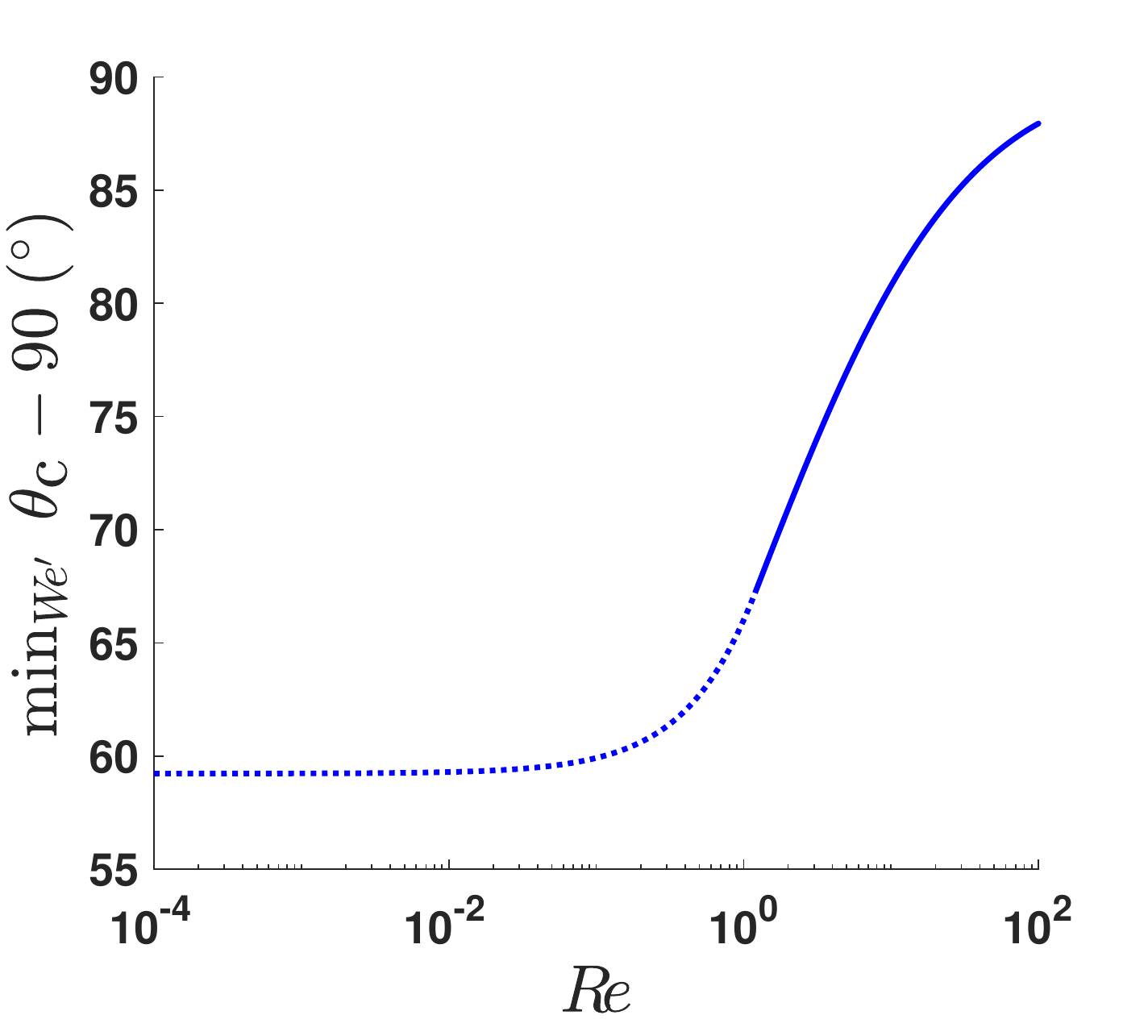}
\end{subfigure}
\caption{Effect of electric field strength on A/C transition for a flow with negligible surface tension. Panel (a) shows the A/C transition surface, and panel (b) gives the minimum of the surface over ${\Weber}'$ as a function of ${\Rey}$.}
\label{ACplot3}
\end{figure}

The A/C instability regions of WIBL2 in the weak surface tension limit are shown in figure \ref{ACplot3}. Panel (a) plots the transition surface in terms of ${\Rey}$ and ${\Weber}'$. The dotted lines are the transition curves for fixed ${\Rey}$ as ${\Weber}'$ varies, with the solid curve 
tracking their minima. We note that the minima are attained at ${\Weber}' = 0$ for $O(1)$ values of ${\Rey}$, but at non-zero ${\Weber}'$ for larger Reynolds numbers. It is interesting to note that figures \ref{ACplot2}(a) and \ref{ACplot3}(a) where the electric field or surface
tension is absent, respectively, are qualitatively different since either physical mechanism is
stabilizing albeit with a different spectral content.
It may be of theoretical interest to consider generalizations of the stabilizing terms in the interfacial pressure \eqref{leadingorderP1}, 
but this is beyond the scope of the current study. The projection of the solid curve in panel (a) onto the ${\Rey}$-direction is given in panel (b), where the dotted line indicates that the minimum is attained at ${\Weber}' = 0$. As before, parameters above the curve yield absolute or convective instability, with only convective instability below the curve. The minimum critical angle of $59.2^{\circ}$ below vertical is obtained for ${\Rey} = 0$ -- as shown above, inclusion of surface tension can increase or decrease this value. The minimum of the surface in figure \ref{ACplot3}(a) over ${\Rey}$ (this is not the projection of the solid curve onto the ${\Weber}'$-direction) can be obtained from figure \ref{ACplot2}(d); it comprises of the initial dotted curve for ${\Weber}' \sin \theta < 0.675$, and the two other solid curve segments -- these three sections correspond to minima attained in the zero surface tension limit.

\subsection{Comparison of long-wave models with Stokes flow results}
%\label{}

We now provide comparisons between the A/C regions for WIBL2 at ${\Rey} = 0$ with those of the corresponding Stokes flow problem. The regions of validity in parameter space of the long-wave models are restricted due to their asymptotic derivation, whereas for the full Stokes flow problem, there are no restrictions on ${\Capil}$, ${\Weber}$, or the electrical permittivities. 
From the full Orr--Sommerfeld system for the electrified problem given in Appendix \ref{OrrSommerfeldAppendix}, the exact dispersion relation 
at $\Rey=0$, computed in Appendix \ref{Stokesflowappend1}, reads
\begin{equation} \omega =  - \frac{  i \xi ( 1 + \xi^2 + \cosh^2(\xi) )  }{ \xi^2 + \cosh^2(\xi)} - \left( \cot\theta +  i {\Weber}  \xi [  \epsilon^{\iota}  \tilde{A}^{\iota} ]_{\textrm{II}}^{\textrm{I}} + (2{\Capil})^{-1} \xi^2 \right) \frac{\sinh(\xi ) \cosh(\xi ) - \xi }{\xi( \xi^2 + \cosh^2(\xi))} ,
  \end{equation}
where the electric field term is given by \eqref{EfieldAppenContribute}. As in the case of the long-wave models, 
we are able to take advantage of symmetries of the dispersion relation in order to apply the Briggs--Bers criterion in the right-half $\xi$-plane. 
We again utilize \textsc{AUTO-07P} to solve this algebraic problem and perform solution continuation. 
The expressions for the real and imaginary parts of $\omega$ and $\partial_{\xi} \omega$ are exceedingly long -- we used \textsc{Maple} to expand these expressions and convert them into a format which was appropriate for \textsc{AUTO-07P}.

\begin{figure}
\centering
\begin{subfigure}{2.9in}
\caption{A/C transition line for ${\Weber}' =0, 5$.} 
\includegraphics[width=2.8in]{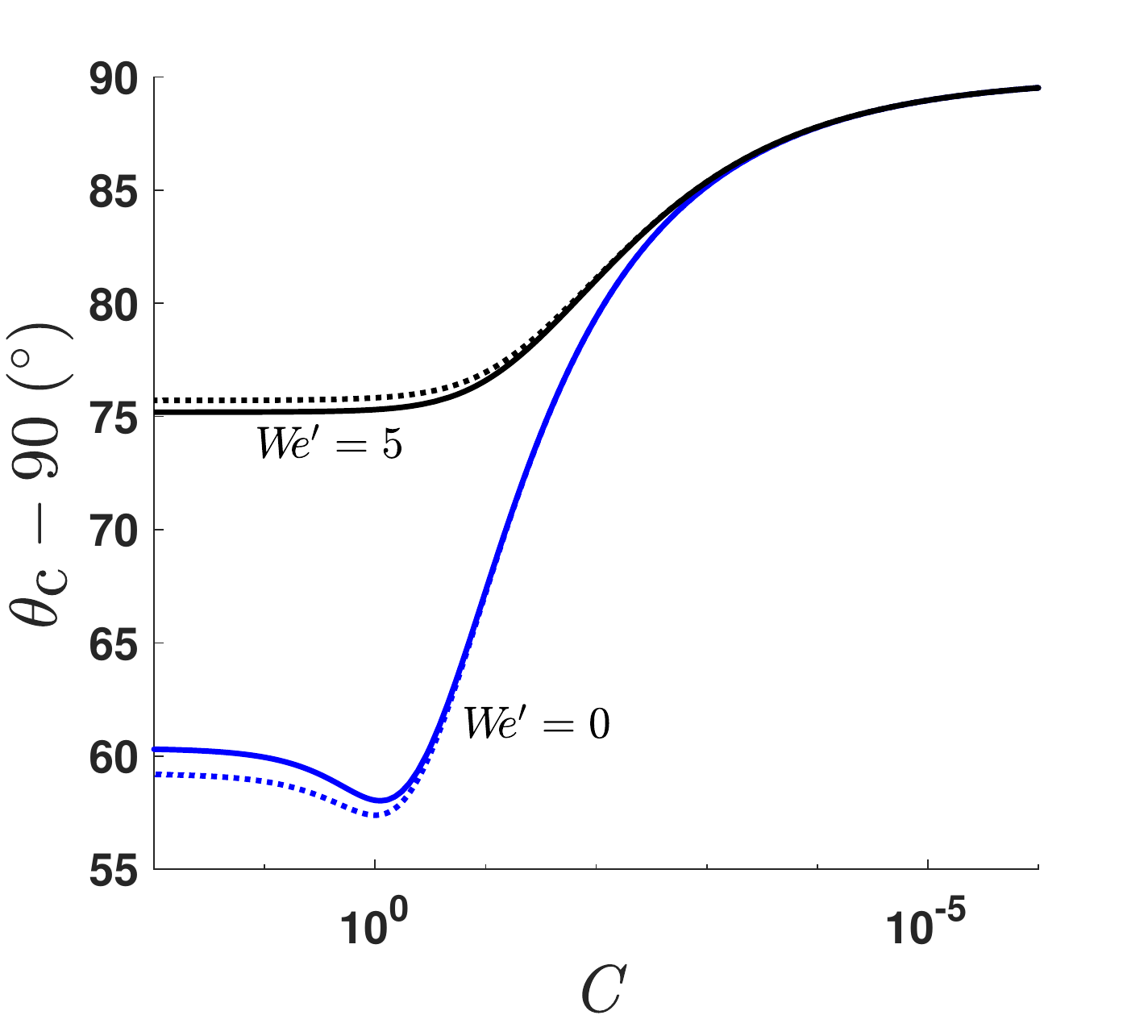}
\end{subfigure}
\begin{subfigure}{2.9in}
\caption{Minimum critical angle against ${\Weber}'\sin\theta$.} 
\includegraphics[width=2.8in]{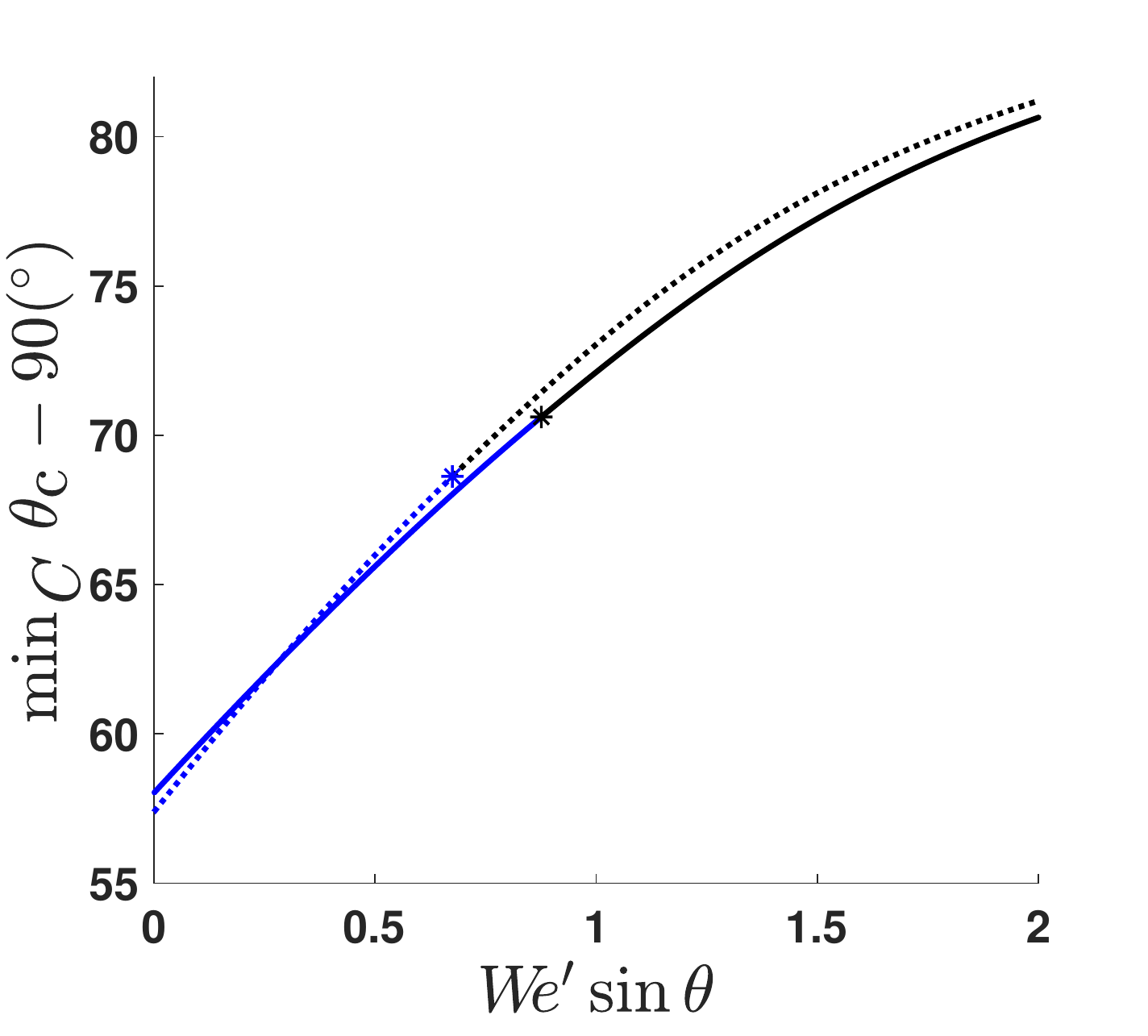}
\end{subfigure}
\caption{Comparison of A/C regions for Stokes flow (solid) with WIBL2 at zero Reynolds number (dotted). Panel (a) plots the transition curves for ${\Weber}' =0,5$. Panel (b) plots the minimum critical angle against ${\Weber}'\sin\theta$; the values to the right of the stars are attained in the weak surface tension limit, as in the ${\Weber}' =5$ case in panel (a). The permittivities used for the Stokes flow computations are $\epsilon^{\textrm{S}} = 1.5$, $\epsilon^{\textrm{I}} = 2$, and $\epsilon^{\textrm{II}} = 1$.}\label{ACplot4}
\end{figure}

Figure \ref{ACplot4} gives a comparison of the Stokes results with those of WIBL2 restricted to zero Reynolds number. 
Panel (a) plots the critical inclination angle for the Stokes problem (solid curve) and WIBL2 model (dotted curve) against 
${\Capil}$, for ${\Weber}' =0,5$. The curves agree closely in the case of strong surface tension, 
$\Capil \rightarrow 0$, with deviation for larger capillary numbers. This is expected as the system becomes unstable 
to shorter waves for these parameters, thus invalidating the long-wave model. In the non-electrified case, the minimum is again found at an $O(1)$ value of ${\Capil}$, although at a different value of $\theta_{\textrm{c}}$, further confirming that the non-monotonic A/C behavior is not a fabrication of the long-wave methodology. A minimum critical angle of $58.0^{\circ}$ below vertical is obtained, differing from the result of $57.4^{\circ}$ for WIBL2 (and the full second-order WIBL model). It may be the case that a higher-order long-wave model captures this value more accurately (WIBL3), but this is beyond the scope of the current study. This reveals a slight downfall in the use of long-wave models to predict the A/C behavior of the full problem. The results for ${\Weber}' =5$ also agree qualitatively, although with a discrepancy for negligible surface tension -- ${\Weber}'$ is not large enough to maintain the long-wave assumption in this limit. Panel (b) shows a continuation of the minimum critical angle against ${\Weber}'\sin\theta$. The curves intersect and do not converge together as the field strength increases; this is expected as the instability moves towards shorter waves, and the capillary number where the minima are attained increases (to the right of the stars, ${\Capil} \rightarrow \infty$). To compute the results for the Stokes flow problem, we used permittivities $\epsilon^{\textrm{S}} = 1.5$, $\epsilon^{\textrm{I}} = 2$, and $\epsilon^{\textrm{II}} = 1$ (as in figure \ref{OSplot1}).

\section{Direct Numerical Simulations: Pulse initialization}\label{sec:DNS}

In this section, we perform DNS of the full Navier--Stokes problem coupled with electrostatics, adding a small-amplitude pulse disturbance to the Nusselt solution as the initial condition. This is utilized to check the accuracy of the predictions of A/C instability for the full problem with that 
of the low-dimensional models. A similar numerical experiment was performed for the non-electrified case by Scheid et al.~\cite{ScheidKofman1}, who showed that the full second-order WIBL model captured the correct A/C behavior of the Navier--Stokes problem. This section provides a similar validation with the addition of the stabilizing electric field.

The hydrodynamical component of the results presented in the previous section can be recast in terms of the mean interface height $\ell$, inclination angle $\theta$, and a geometry-independent Kapitza number $\sigma/\rho g^{1/3} \nu^{4/3}$, which is fixed for a given fluid. Fixing the dielectric constants also means that $E_0$ alone parameterizes the electrical component of the problem. Thus, we may obtain a surface in $(\ell,\theta,E_0)$-space separating the regions of absolute and convective instability for each model. This surface is not monotonic (for a model including viscous extensional stresses) -- this is visible from the results for the non-electrified problem by taking a constant-Kapitza number slice of figure \ref{ACplot2}(a), as well as figure 7 in \cite{ScheidKofman1} and figure 3 in \cite{kofman2018prediction}. For brevity, we do not re-plot the results in terms of these parameters, but provide A/C thresholds in tables when comparing to DNS.

{\color{black}

We fix the working fluid to be the Castor Oil (H\"{a}nseler AG) used in experiments by Brun et al.~\cite{Brun1}. They measured the capillary and viscous lengths to be $1.91$ \emph{mm} and $4.4$ \emph{mm}, respectively, and also give the dynamic viscosity to be $865 \pm 5$ \emph{cP}. The fluid properties corresponding to these measurements are given in SI units in Table \ref{physproptable}, and we also fix $g = 9.81$ \emph{m}/\emph{s}$^2$ for all calculations. 
\begin{table}[H]
\begin{center}
\begin{tabular}{|c|c|c|c|}
  \hline
  $\rho$ &  $\nu$ & $\mu$ & $\sigma$ \\
    (\emph{kg}/\emph{m$^3$}) & (\emph{m$^2$}/\emph{s}) &  (\emph{kg}/\emph{m}$\cdot$\emph{s}) & (\emph{N}/\emph{m}) \\
  \hline
  $946$ & $9.14 \times 10^{-4}$  &  $0.865$ & $0.0339$  \\
  \hline
\end{tabular}
\end{center}
\caption{Physical parameters for Castor Oil (H\"{a}nseler AG).}\label{physproptable}
\end{table}
\noindent 
We assign to the fluid the dielectric constant $\epsilon^{\textrm{I}} = 5$, which is roughly the value found for castor oils. We note that, although castor oils typically have a very low electrical conductivity (less than $10^{-10}$ \emph{Siemens}/\emph{m}), they have been observed in certain experiments to behave as leaky rather than perfect dielectrics \cite{burcham2000electrohydrodynamic}. Setting $\epsilon^{\textrm{II}} = 1$, the value for air, we have $\epsilon^{\textrm{I}} > \epsilon^{\textrm{II}}$, and so the parallel field still has a linearly stabilizing effect on the interface even if the conductivity of the castor oil is accounted for \citep{papageorgiou2004generation,uguz2008electric}. We ignore conductivities for the purposes of the present study; if conductivities are included, our models become coupled to an equation for the interfacial charge, as in \cite{:/content/aip/journal/pof2/17/3/10.1063/1.1852459} -- the study of such models is warranted for comparison with experiments. By choosing the electrical permittivity of the solid substrate to match that of the liquid phase, $\epsilon^{\textrm{S}} =5$  -- a realistic value for Pyrex glass, we do not need to solve for $V^{\textrm{S}}$
(the numerical implementation is discussed later). For our calculations, we take the permittivity of free space to be $\epsilon_0 = 8.85 \times 10^{-12}$  \emph{A}$^2\cdot$\emph{s}$^4$/\emph{kg}$\cdot$\emph{m}$^{3}$. We note that the dielectric breakdown of air occurs for electric field strengths beyond approximately $E_0 = 3\times 10^6$ \emph{V}/\emph{m} \cite{tipler1987college}, and thus we do not consider values of $E_0$ beyond this.

}

For our numerical simulation of the full system with pulse initial conditions, we choose $\theta = 5\pi/6$ and $\ell = 2.5$ \emph{mm}, in which case all models (long-wave and Stokes) predict the non-electrified case to be absolutely unstable. However, this is sufficiently close to a region of convective instability as the field strength is increased, so that the required $E_0$ is feasible. The critical values of $E_0$ above which the models exhibit convective instablility are summarized in Table \ref{criticalE0}.
\begin{table}[H]
\begin{center}
\begin{tabular}{|c|c|c|c|c|c|}
  \hline
   Model & Inertialess Benney & Benney & WIBL1 & WIBL2 &  Stokes flow \\
  \hline
   $E_0$ (\emph{V}/\emph{m}) & $ 3.07 \times 10^5 $ & $ 3.45 \times 10^5$  &  $3.34 \times 10^5$ & $2.48 \times 10^5$  &  $2.10 \times 10^5$ \\
  \hline
\end{tabular}
\end{center}
\caption{Critical $E_0$ (\emph{V}/\emph{m}) for A/C transition with $\theta = 5\pi/6$ and $\ell = 2.5$ \emph{mm}, in
the case of Castor Oil (H\"{a}nseler AG).}\label{criticalE0}
\end{table}
\noindent The critical value of $E_0$ decreases with increasing accuracy of the inertial long wave models, with the inertialess Benney and Stokes flow dispersion relations also giving reasonable values. The field strengths in Table \ref{criticalE0} were produced by continuation of the dominant saddle point, and additionally verified with root plots as in figure \ref{ACplot1}(a,b).

\begin{figure}
\centering
\begin{subfigure}{2.9in}
\caption{Evolution of wavepacket for $E_0 = 0$ \emph{V}/\emph{m}.} 
\includegraphics[width=2.8in]{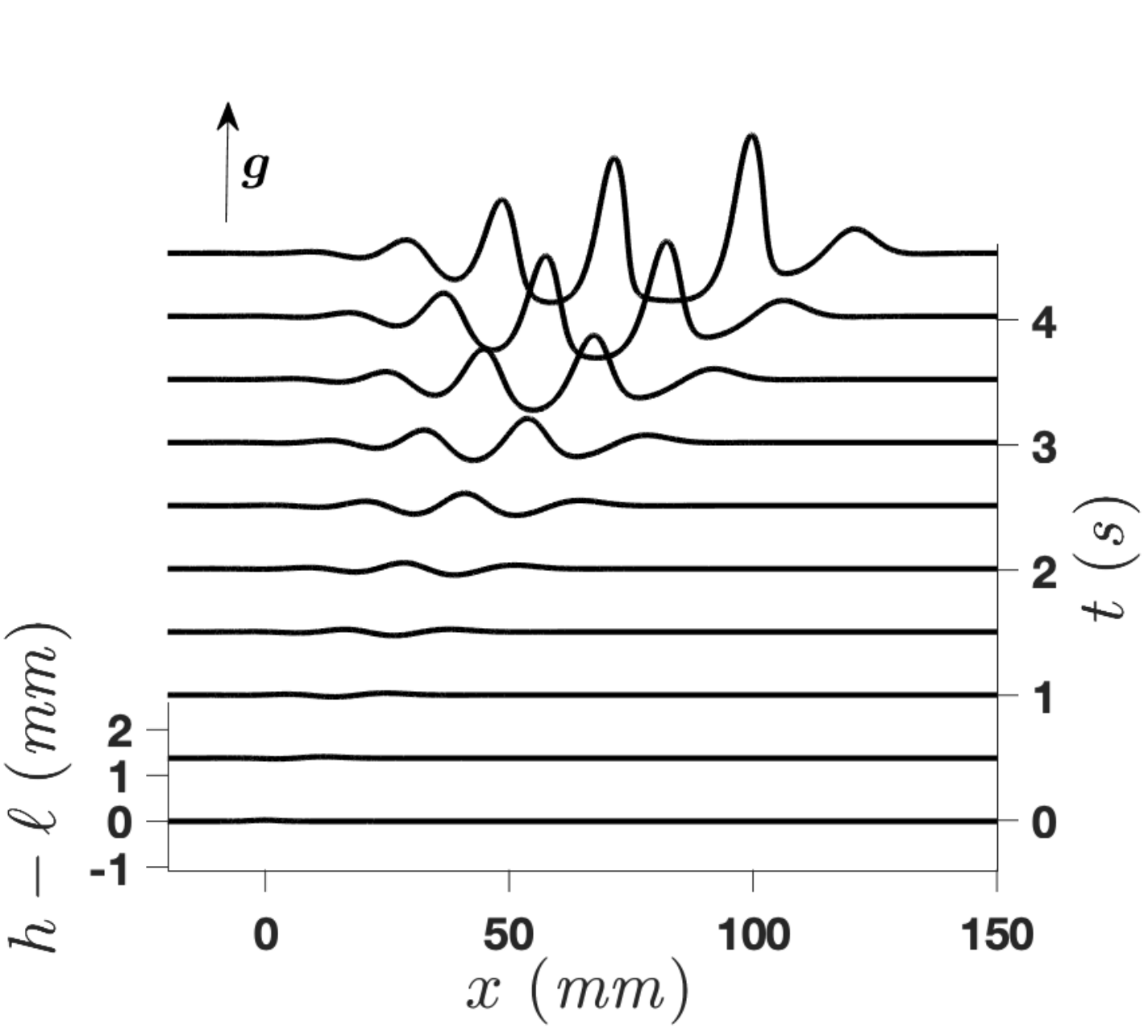}
\end{subfigure}
\begin{subfigure}{2.9in}
\caption{Evolution of wavepacket for $E_0 = 6 \times 10^5$ \emph{V}/\emph{m}.} 
\includegraphics[width=2.8in]{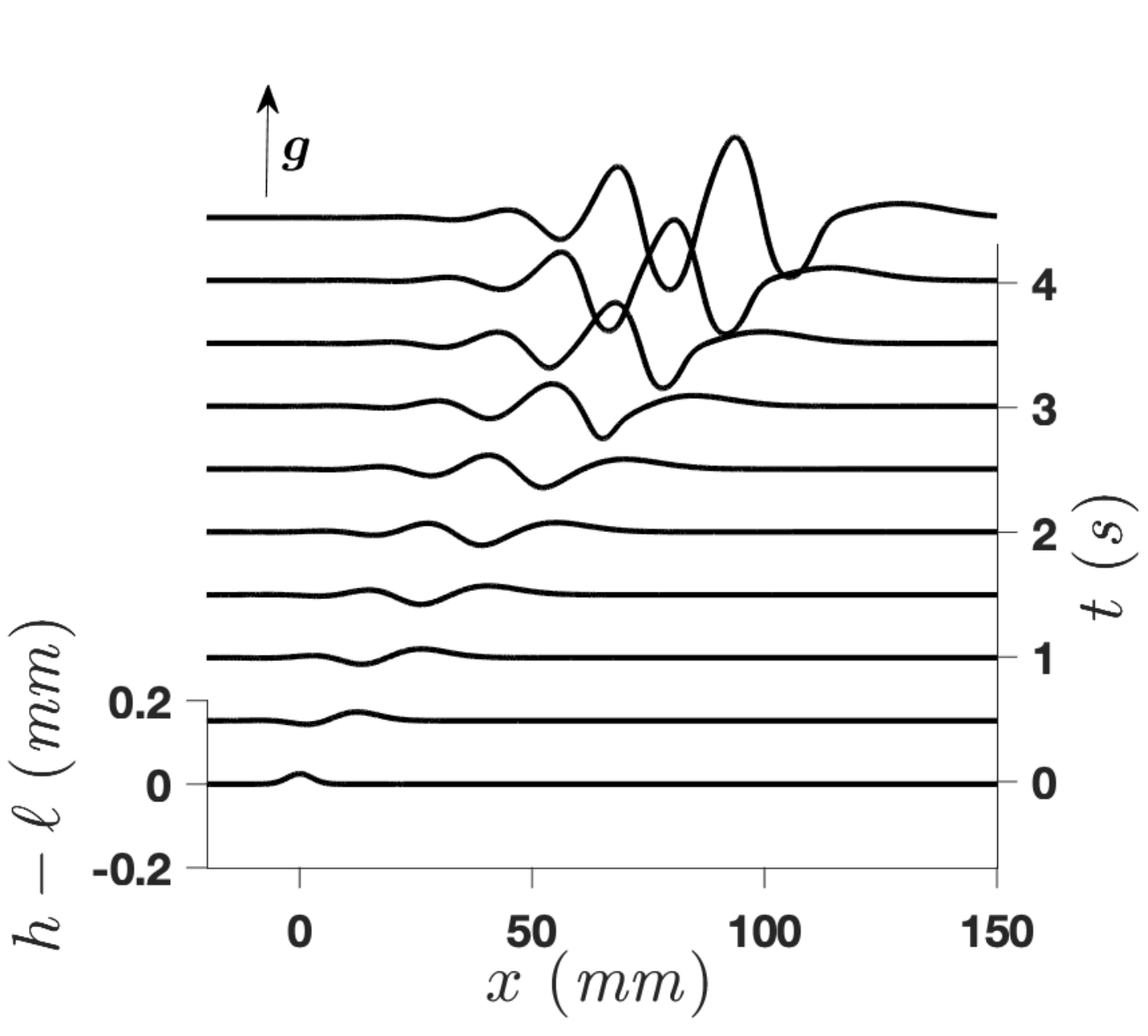}
\end{subfigure}
\begin{subfigure}{2.9in}
\caption{Upstream edge of wavepacket.} 
\includegraphics[width=2.8in]{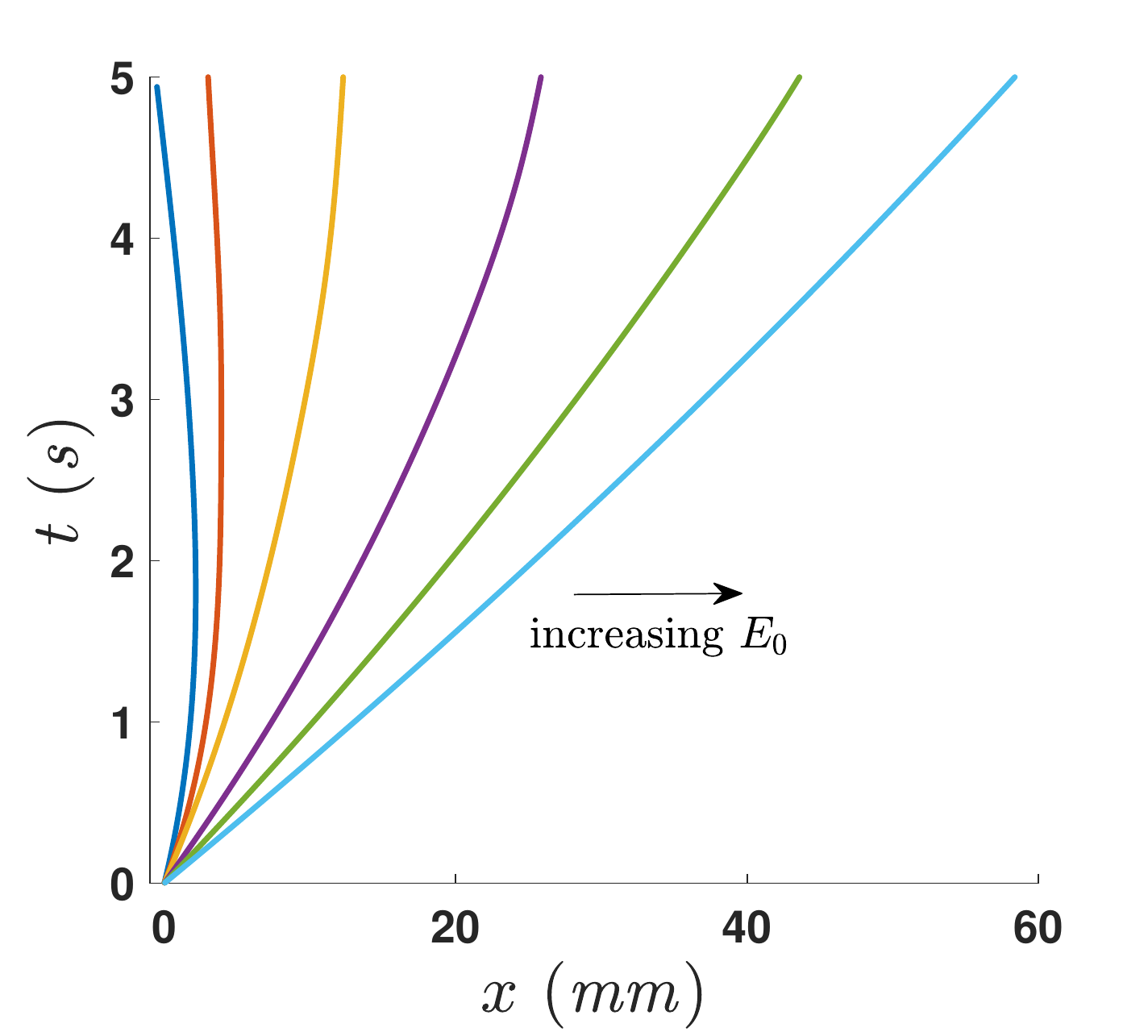}
\end{subfigure}
\caption{Panels (a,b) plot interfacial profiles starting from pulse initial data for $E_0=0,6\times 10^5$ \emph{V}/\emph{m}, respectively. For the height axes on the left hand sides, the substrate is at $-2.5$. The flow is from left to right, and we include an (off-vertical) arrow indicating the direction of gravity. Panel (c) shows the evolution of the upstream edge of the wavepacket, from left to right, the lines correspond to $E_0 = 0,1.5\times 10^5,3\times 10^5,4.5\times 10^5,6\times 10^5,7.5 \times 10^5$ \emph{V}/\emph{m}. In all panels, the profiles are shifted so that the initial Gaussian pulse is located at the origin.}
\label{dnspulse}
\end{figure}

For the DNS, we employ the open-source volume-of-fluid solver $\mathpzc{Gerris}$ \cite{popinet1,popinet2,lopez1}. 
We use viscous--gravity length and time scales as described by Samanta et al.~\cite{samanta2011falling} in order to non-dimensionalize the system, 
as it provides a convenient unit effective Reynolds number, embedding the parameter variation inside the dimensionless film thickness and surface tension coefficient instead. We consider a finite computational domain of length $L_0 = 400 \ell$ in the streamwise variable, with an extent of $50\ell$ in the perpendicular $z$-direction. For an undisturbed interface, the region $\ell < z < 50\ell$ of the computational domain is occupied by the second fluid, in our case we take air (at room temperature). For DNS, Region II is hydrodynamically active; its fluid properties are such that this has a negligible effect, and a comparison with long-wave models derived assuming that Region II is hydrodynamically passive remains valid. We set the density in Region II to $1.17$ \emph{kg}/\emph{m$^3$} and take a dynamic viscosity of
$1.83 \times 10^{-3}$ \emph{kg}/\emph{m}$\cdot$\emph{s}, 
both of which are significantly lower than their liquid film counterparts given in Table \ref{physproptable}. A spatial filtering technique 
smoothing out the interface is hence employed to improve the performance of the underlying projection solver. The no-slip and impermeability conditions are applied at the solid substrate boundary ($z=0$), and a no-stress condition is imposed on the opposite side of the computational domain at $z = 50\ell$. Dirichlet boundary conditions for the voltage potential are prescribed at the inlet and outlet, chosen to give the desired field strength $E_0$. Neumann conditions for the voltage potential are imposed at the upper and lower boundaries of the computation domain -- these are found to be an effective equivalent of the far-field conditions since we have chosen $\epsilon^{\textrm{S}} = \epsilon^{\textrm{I}}$. The Nusselt base profile, given by \eqref{basestatesdimensional} in dimensional variables, is prescribed as the initial condition, on top of which 
we add a Gaussian pulse whose amplitude is $1\%$ of the liquid film thickness, with center located $150 \ell$ from the inlet (far enough to prevent unwanted boundary noise/effects). Guided by the linear theory of the long-wave models, we choose the pulse width to be approximately $4\ell$ to induce the desired instability and minimize transient effects.
% Ruben: the exact formula for the bump is h_0*(1.0 + 0.01*exp(3600*(cos((2*pi/(400*h_0))*(x-150*h_0))-1))), it felt like too much detail to put in the text
The Nusselt solution is also used as the inlet condition, while standard outflow conditions are considered at the downstream boundary. We prescribe the maximum level of refinement around the interfacial shape, as well as close to the substrate in order to capture the details of the underlying boundary layer. Changes in the velocity field otherwise dictate the adaptive mesh refinement strategy. Considering the above, the discretized domain is described by approximately $10^5$ computational grid cells, with the target evolution to the first dripping events requiring several days of runtime on local high performance computing facilities. The code is executed in parallel on up to $16$ CPUs, with load balancing capabilities resulting in reasonably good scalability properties.

The results of the DNS are shown in figure \ref{dnspulse}. Panels (a,b) show the evolution of the pulse initial data for the cases of $E_0 = 0, 6\times 10^5$ \emph{V}/\emph{m}, respectively, with the flow from left to right. The profiles are shifted so that the initial pulse is centred at $x=0$. Note also the vector showing the direction in which gravity is acting for these hanging films. From the height scales on the left hand sides of each panel, there is a clear difference in growth rate across these two cases, and the non-electrified dynamics reaches a nonlinear phase of evolution at around $t = 4$ \emph{s}. In order to determine the absolute or convective nature of the instability, we monitor the upstream edge of the wavepacket -- plotted in figure \ref{dnspulse}(c) for a range of $E_0$. These curves were computed using the region of $(x,t)$ space for which $|h(x,t) - \ell|$ was above a given tolerance. The boundary of this region is wavy, and so the curves were computed by fitting through the farthest upstream points (similar to figure 8 in \cite{avitabile2017ducks}). There is an initial transient phase in which the wavepacket is convected downstream in all cases -- we found this is difficult to avoid by choosing better initial data due to DNS constraints. For the non-electrified case, the wavepacket edge begins to move back upstream at $t \approx 2$ \emph{s}, and similarly for $E_0 = 1.5 \times 10^5$ \emph{V}/\emph{m} at $t \approx 3$ \emph{s}. We are confident that these cases show linear (and nonlinear) absolute instability after a transient phase, as predicted by the low-dimensional models (see Table \ref{criticalE0}), rather than a linear convective instability followed by a nonlinear absolute instability as discussed by Delbende and Chomaz \cite{delbende1998nonlinear}. The choices of $E_0 \geq 3 \times 10^5$ \emph{V}/\emph{m} all showed convective instabilities (in both the linear and nonlinear phase), in agreement with the predictions of WIBL2 and the Stokes dispersion relation.

\section{Direct Numerical Simulations: Dripping}\label{sec:DNSdripping}

In this section, we perform DNS of the full Navier--Stokes equations coupled with electrostatics on a spatial domain of length $L_0$, and determine parameters for which dripping occurs. We consider inflow and outflow boundary conditions to mimic an experimental set-up, unlike the 
spatially periodic study of Kofman et al.~\cite{kofman2018prediction}. The majority of the details of the DNS set-up are exactly as in the previous section. We initialize our simulations with the Nusselt solution, and apply a time-periodic forcing at the inlet (located at $x=0$) of the form
\begin{equation} 
u(0,z,t) =  (1+ \alpha \sin(2\pi \mathfrak{f} t))\overline{u}(z), 
\end{equation}
where $\overline{u}(z)$ is the base Nusselt velocity defined in \eqref{basestatesdimensional}, and $\mathfrak{f}$ is a frequency chosen to excite the most unstable interfacial waves using predictions from the long-wave models. The strength of the time-periodic inlet perturbation, $\alpha$, was varied to reduce the transient phase of the dynamics and ensure the development of a clean wave-train close to the inlet. This form of inlet forcing was used effectively by Denner et al.~\cite{denner2018solitary} in their computation of solitary wave profiles on overlying falling films. Furthermore, we find that the resulting solutions are robust to variations in the forcing frequency $\mathfrak{f}$. For all numerical experiments, we fix the aspect ratio by taking the system length (between inlet and outlet) 
to be $L_0 = 300  \ell$. While the domain imposed to capture the dripping dynamics is marginally smaller than in the case of the pulse evolution, 
the timescale required to comprehensively investigate the instability is at least one order of magnitude larger due to a significant transient period, thus resulting in a considerably increased computational effort. The computational framework described in the previous section naturally allows for interfacial break-up, however we have opted to remove resulting drops as soon as they detach 
from the main body of fluid to avoid interactions with the opposite boundary, as well as to concentrate computational resources at the level of the investigated liquid film.

\subsection{Non-electrified case}

For non-electrified flows we considered three choices of the inclination angle $\theta$, and varied the film thickness $\ell$. These 
are given in Table \ref{tableAC} along with absolute instability regions in $\ell$ for all of the models discussed in the present paper. 
\begin{table}[H]
\begin{center}
\begin{tabular}{|c|c|c|c|c|c|c|c|c|}
  \hline
  $\theta$ & $\theta - 90(^\circ)$  &  Inertialess Benney & Benney  & WIBL1 & WIBL2 & Stokes flow & $\ell^{\textrm{drip}}$ (DNS) \\
    \hline
  $7\pi/8$ & $67.5$ & $ 1.52 \leq \ell $  & $ 1.52 \leq \ell $ & $ 1.52 \leq \ell \leq 20.1 $ & $ 1.35 \leq \ell \leq  8.76 $ &  $ 1.36 \leq \ell $  & $1.40$--$1.45$ \\
  \hline
  $5\pi/6$ & $60$ & $2.19  \leq \ell$  & $2.15  \leq \ell$ & $2.16  \leq \ell \leq 15.8$ & $2.18 \leq \ell \leq 5.21 $ &  $2.25 \leq \ell \leq 13.1$ & $2.30$--$2.50$ \\
    \hline
  $49\pi/60$ & $57$ & $ 2.51  \leq \ell $  & $ 2.42  \leq \ell $ & $ 2.44  \leq \ell \leq 14.5 $ &  ---  & --- & $2.40$--$2.60$ \\
  \hline
\end{tabular}
\end{center}
\caption{Absolute instability regions and dripping limits for the non-electrified case (lengths are given in \emph{mm}).}\label{tableAC}
\end{table}
\noindent In the case of the WIBL2 model, for example, the values are obtained by taking a constant-Kapitza number slice of the surface shown in figure \ref{ACplot2}(a); the resulting curve is then considered for the three values of $\theta$. Both WIBL2 and the Stokes flow models exhibit a non-monotonic A/C threshold, in agreement with DNS of the Navier--Stokes equations. The dashes in the last row indicate that the models are convectively unstable for all choices of $\ell$. The other models do not include viscous extensional stresses.

For each case of $\theta$, we performed DNS for a range of values of $\ell$, guided by the results of the spatial stability analysis. We do not present simulation results for large values of $\ell$ near the upper bounds of the absolute instability regions; we found immediate and extensive dripping at 
such parameters. In agreement with the results of Kofman et al.~\cite{kofman2018prediction} (see their figure 3), we are confident that only the lower A/C threshold can be used as a predictor of dripping; in this regime, viscous extensional stresses are not important and the critical angle for absolute instability depends monotonically on $\ell$. It is expected that the dripping angle should also be monotonic in $\ell$. It is worth noting
that in our DNS we varied non-dimensional parameters by an order of magnitude (although ${\Rey}$ still remains small); they range
from ${\Rey} = 0.005$ and ${\Capil} = 0.0885$ for $\ell = 1.3$ \emph{mm} and $\theta = 7\pi/8$, to ${\Rey} = 0.05$ and ${\Capil} = 0.466$ for $\ell = 2.5$ \emph{mm} and $\theta = 49\pi/60$, thus affording a fairly complete view of the physical system.

\begin{figure}
\centering
\begin{subfigure}{5.8in}
\caption{Bounded wave-train for $\ell = 1.35$ \emph{mm}.} 
\includegraphics[width=5.7in]{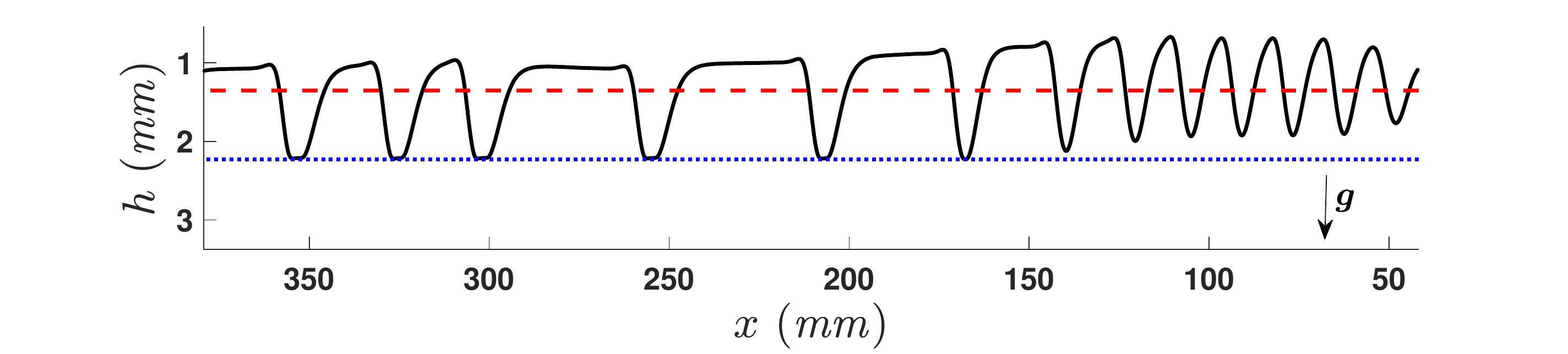}
\end{subfigure}
\begin{subfigure}{5.8in}
\caption{Wave-train with dripping due to coalescence (transient) for $\ell = 1.40$ \emph{mm}.} 
\includegraphics[width=5.7in]{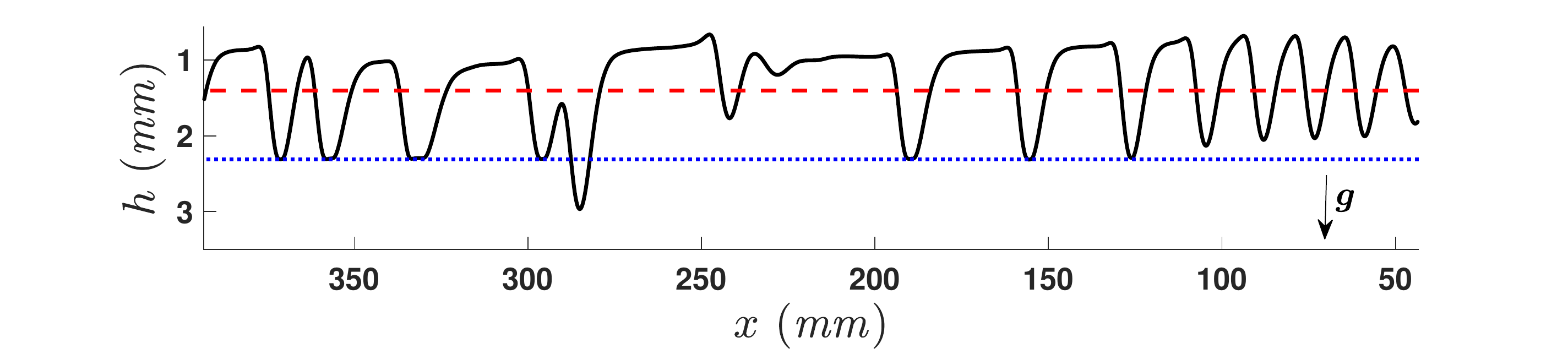}
\end{subfigure}
\begin{subfigure}{5.8in}
\caption{Dripping due to instability of solitary waves (not coalescence) for $\ell = 1.45$ \emph{mm}.} 
\includegraphics[width=5.7in]{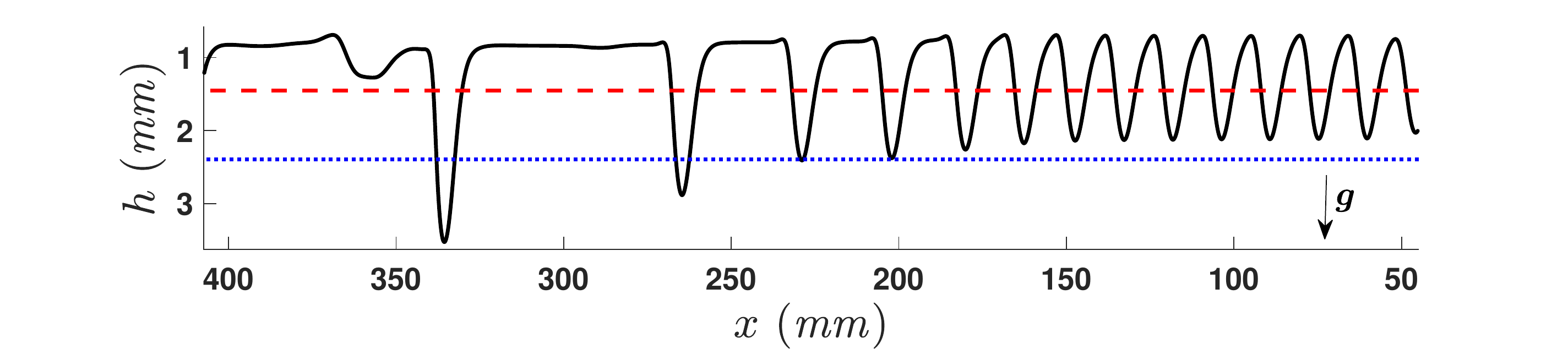}
\end{subfigure}
\caption{Snapshots of the fluid interface for $\theta = 7\pi/8$ and a range of inflow film thicknesses. The flow is from right to left, and the direction of gravity is indicated in the panels. The dashed line denotes the inflow film thicknesses $\ell$, and the dotted line is given by $h_{\textrm{max}} = 1.649 \ell$ (see the text). The waves with peaks above the dotted line in panels (b,c) eventually yield dripping events.}
\label{pi8cases}
\end{figure}

For $\theta = 7\pi/8$, we find a very clear separation between three cases which are plotted in figure \ref{pi8cases}. From simulations with $\ell = 1.25, 1.30, 1.35$ \emph{mm}, we find a linear relationship between the mean film thickness and maximum wave height, approximately $h_{\textrm{max}} = 1.649 \ell$. This linear relationship is expected and in agreement with the results in \cite{denner2016self,denner2018solitary} for overlying films. No drips are observed for these cases, even after the coalescence of neighbouring pulses. In figure \ref{pi8cases}(a) we plot a snapshot of such a bounded wave train for $\ell = 1.35$ \emph{mm}; some of the pulses are wider as they result from the coalescence of two smaller pulses, but all waves very clearly attain the maximum bound sufficiently far from the inlet. Increasing the inflow film thickness to $\ell = 1.40$ \emph{mm}, we find similar bounded wave trains (the dotted line in figure \ref{pi8cases}(b) indicates the estimate for $h_{\textrm{max}}$), but occasionally in the transient phase of the dynamics, dripping occurs due to the coalescence of two adjacent pulses. Interactions and coalescence between pulses is common for smaller values of $\ell$, however not leading to dripping events. Further increasing $\ell$ to $1.45$ \emph{mm}, we see dripping due to a different mechanism as displayed in the snapshot in figure \ref{pi8cases}(c). The individual solitary waves destabilize and drip, rather than dripping due to coalescence or pulse interaction (however, the latter is still possible in this case). The pulses appear to attain the predicted $h_{\textrm{max}}$ briefly, before further increasing in amplitude until pendant drips form. We thus define $\ell^{\textrm{drip}}$ to be the length beyond which dripping occurs due to instability of individual pulses far from the inlet (not transient effects or pulse interactions/coalescence). We believe this to be a sensible definition for this phenomenon. We give predicted ranges for $\ell^{\textrm{drip}}$ in Table \ref{tableAC} based on our DNS results -- much longer computations would be required to refine these further.

In all cases we found that the transient dynamics were significantly long, often requiring prohibitively large timescales to observe convergence to a regular state. For $\theta = 7 \pi/8$, a shallow angle close to the horizontal set-up, the dynamics are relatively well-behaved and the prevailing regime is clearly apparent, with wave-trains developing not too far downstream as shown in figure \ref{pi8cases}. On the other hand, the inclinations further from horizontal exhibited more complex pulse interactions with wave-trains developing far from the inlet, and hence a more challenging detection of features at the level of the individual pulses is required to inform our previously defined metric for dripping. The predicted intervals of $\ell^{\textrm{drip}}$ for $\theta = 5\pi/6$ and $\theta = 49\pi/60$ in Table \ref{tableAC} are correspondingly much wider. We note that the regularity of dripping is a useful indicator; for $\ell > \ell^{\textrm{drip}}$, drips often occur at regular time intervals, at a fixed spatial location. For the transient dripping with $\ell < \ell^{\textrm{drip}}$, there is very little regularity in the temporal or spatial location of the dripping events. We see from Table \ref{tableAC} that the predicted $\ell^{\textrm{drip}}$ appears to be monotonic in $\theta$, as expected.

\subsection{Electrified case}

\begin{figure}
\centering
\begin{subfigure}{2.9in}
\caption{$E_0 = 0.0\times 10^5$ \emph{V}/\emph{m}.} 
\includegraphics[width=2.8in]{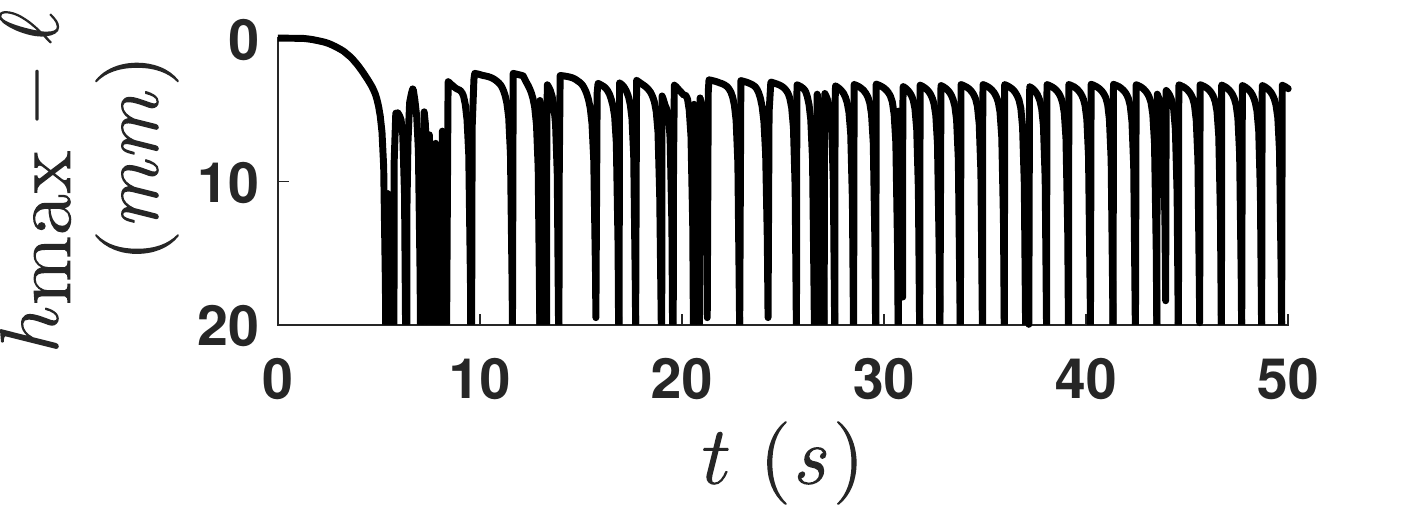}
\end{subfigure}
\begin{subfigure}{2.9in}
\caption{$E_0 = 1.5\times 10^5$ \emph{V}/\emph{m}.} 
\includegraphics[width=2.8in]{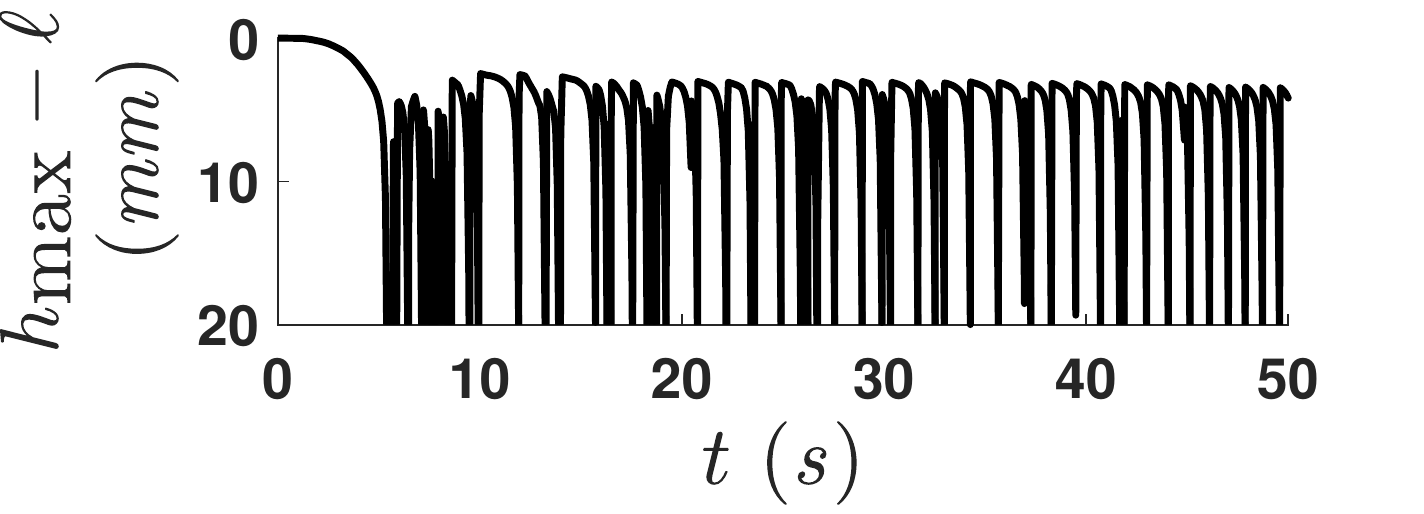}
\end{subfigure}
\begin{subfigure}{2.9in}
\caption{$E_0 = 3.0\times 10^5$ \emph{V}/\emph{m}.} 
\includegraphics[width=2.8in]{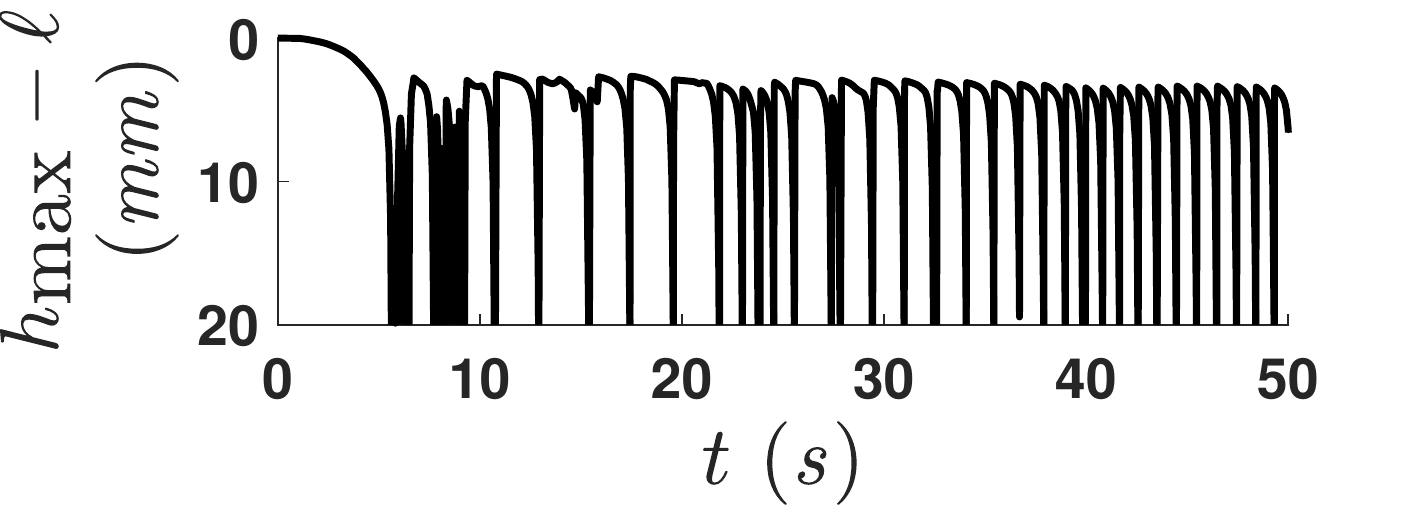}
\end{subfigure}
\begin{subfigure}{2.9in}
\caption{$E_0 = 4.5\times 10^5$ \emph{V}/\emph{m}.} 
\includegraphics[width=2.8in]{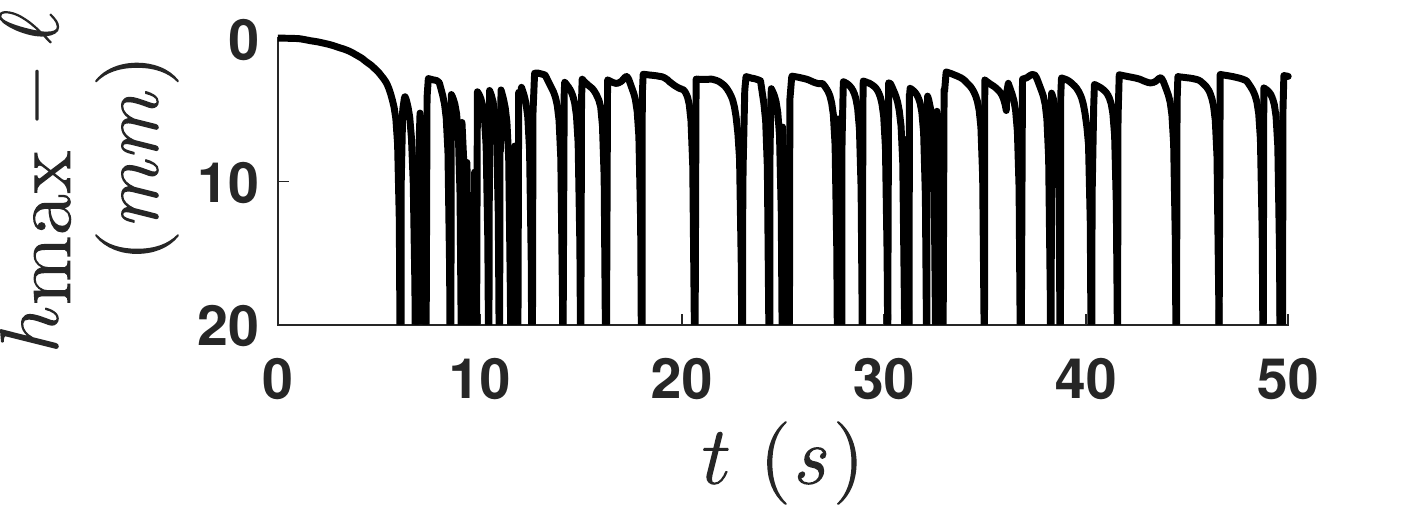}
\end{subfigure}
\begin{subfigure}{2.9in}
\caption{$E_0 = 6.0\times 10^5$ \emph{V}/\emph{m}.} 
\includegraphics[width=2.8in]{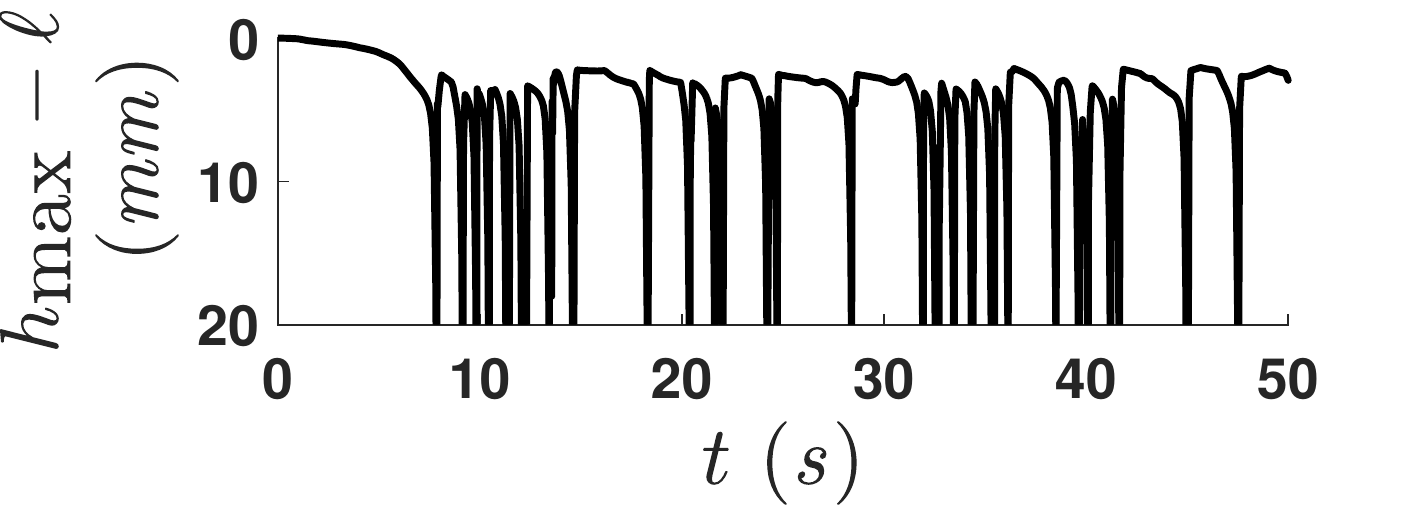}
\end{subfigure}
\begin{subfigure}{2.9in}
\caption{$E_0 = 7.5\times 10^5$ \emph{V}/\emph{m}.} 
\includegraphics[width=2.8in]{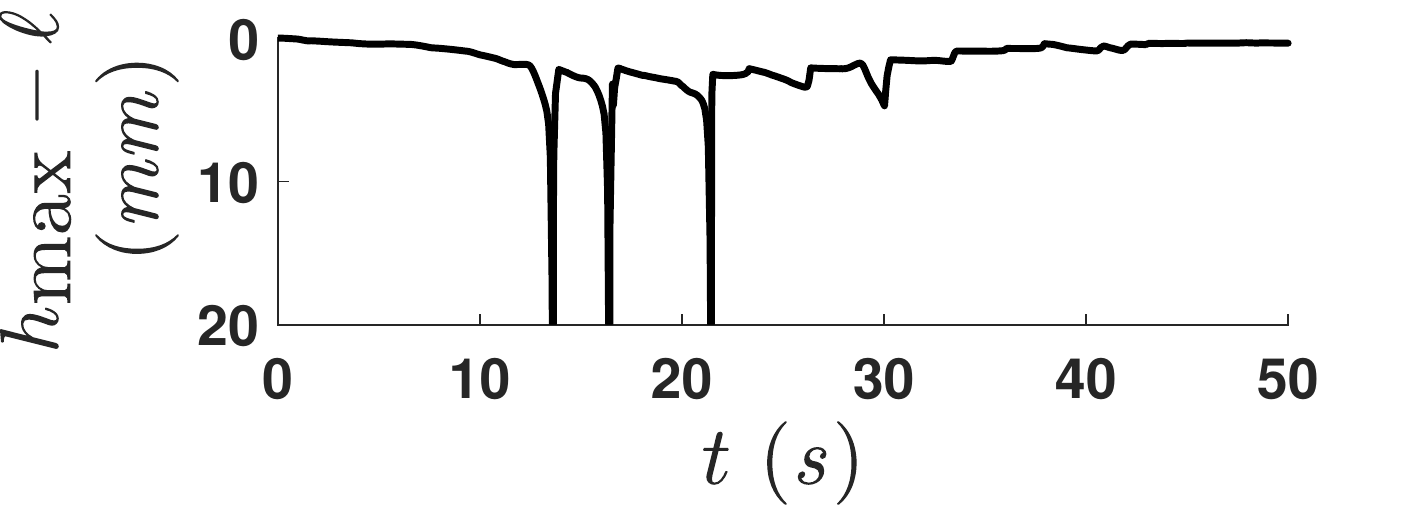}
\end{subfigure}
\begin{subfigure}{2.9in}
\caption{$E_0 = 9.0\times 10^5$ \emph{V}/\emph{m}.} 
\includegraphics[width=2.8in]{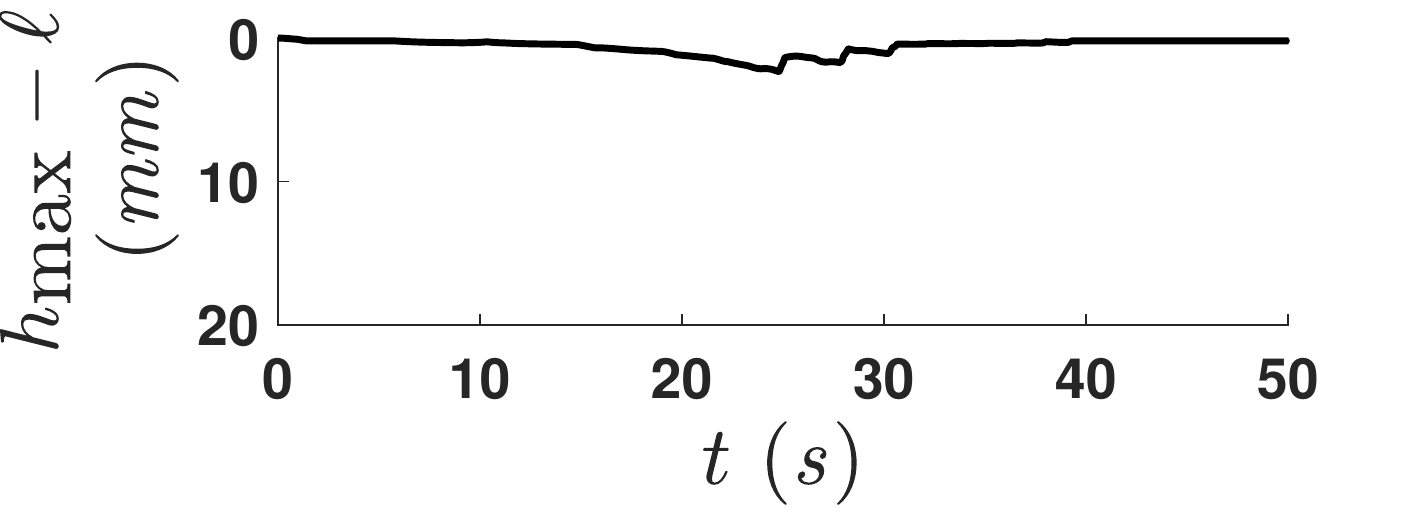}
\end{subfigure}
\caption{Time series of $h_\textrm{max}-\ell$ for a range of $E_0$. The angle of inclination is $\theta=5\pi/6$, the undisturbed film thickness $\ell=2.5$ \emph{mm}, and the domain length is $L_0=300\ell$.}
\label{electrifiedcases}
\end{figure}

\begin{figure}
\centering
\begin{subfigure}{5.8in}
\caption{$E_0 = 3\times 10^5$ \emph{V}/\emph{m}.} 
\includegraphics[width=5.7in]{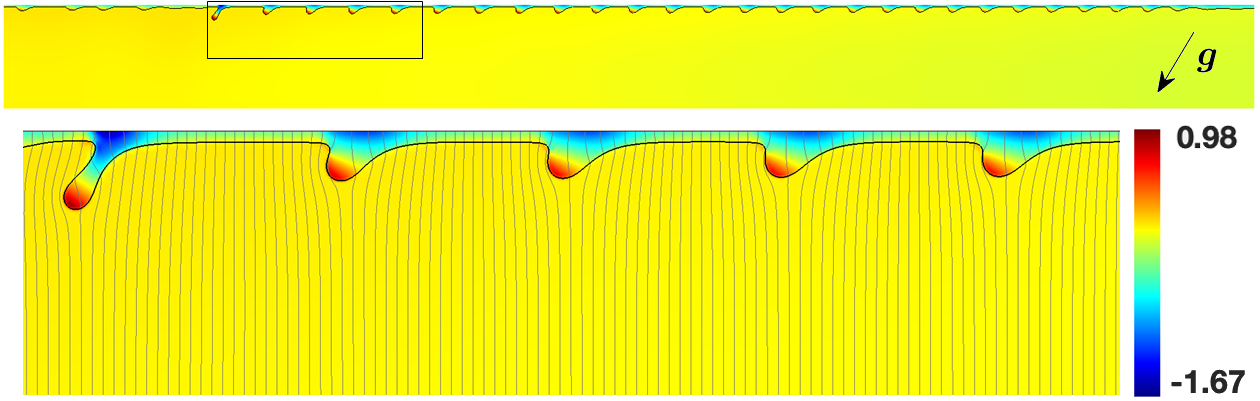}
\end{subfigure}
\begin{subfigure}{5.8in}
\caption{$E_0 = 6\times 10^5$ \emph{V}/\emph{m}.} 
\includegraphics[width=5.7in]{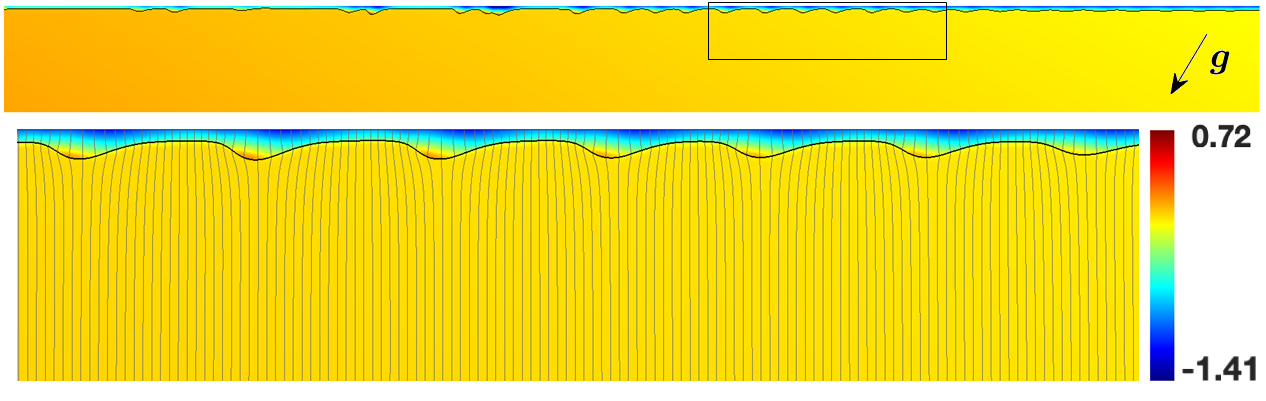}
\end{subfigure}
\caption{Snapshots of waves and dripping for electrified films. Two cases of $E_0$ are shown as given in the captions. The angle of inclination is $\theta=5\pi/6$, the undisturbed film thickness $\ell=2.5$ \emph{mm}, and the domain length is $L_0=300\ell$. The flow is from right to left, and the direction of gravity is indicated in the figure. The interfacial profile at $t \approx 67$ \emph{s} is represented with a black solid curve. The lower plots are
enlargements of the outlined rectangular regions in the upper plots of each panel. The colour bar indicates pressure values and the faint solid curves in the lower plots are voltage equipotentials.}
\label{electrified-Summary}
\end{figure}

{\color{black}
In order to study the effect of the electric field on the dripping dynamics, we focus on the case of $\theta = 5\pi/6$ and $\ell = 2.5$ \emph{mm} as considered in the previous section -- see Table \ref{criticalE0} for the critical values of $E_0$ for which the various models transition from convective to absolute instability. Since we fix the aspect ratio to $L_0 = 300 \ell$, we may express the required voltage in terms of the electric field strength as $V_0 = 300\ell E_0$. We note from Table \ref{tableAC} that the film drips without an electric field. For the chosen parameters, it can be calculated that the electric field strength required to linearly stabilize the flat film solution is approximately $6 \times 10^6$ $\emph{V}/\emph{m}$, which is double the dielectric breakdown of air and thus physically unfeasible. However, we show that dripping suppression may be attained for much smaller and physically feasible field strengths.

}

In figure \ref{electrifiedcases}, we present time series for the maximum deviation of the interface from the mean thickness, $h_{\textrm{max}} - \ell$, for a range of $E_0$. We show the first 50 \emph{s} of longer simulations in the figure; note that the axes are reversed so that downward spikes indicate dripping events. For $E_0 = 0,1.5\times 10^5, 3\times 10^5$ \emph{V}/\emph{m}, we see very regular dripping events, both temporally and spatially (from inspection of the interfacial evolution). Delay of the first dripping event, and temporal and spatial irregularity of dripping events is clear for 
$E_0 = 4.5\times 10^5, 6 \times 10^5$ \emph{V}/\emph{m}. This appears to be due to time-periodic disturbances that travel from the inlet downstream, disrupting the regular structure of the pulses and causing considerable pulse interaction. This behavior does not appear to be transient and thus our previously defined metric of dripping is invalid for the electrified problem. However, this transition between dripping due to loss of stability of pulses and dripping due to pulse interaction does agree well with the A/C threshold. For $E_0 = 7.5\times 10^5, 9 \times 10^5$ \emph{V}/\emph{m}, bounded wave trains emerge (with only a few transient drips). Dripping suppression is obtained for $E_0$ at least double the convective instability threshold -- we expect that the increased disparity between the linear and nonlinear phenomena is due to the nonlocality of the electric field effect. For nonlocal systems, long-range interactions between pulses that are not immediate neighbours are important, with interactions between neighbouring pulses strengthened \citep{lin2015coherent,blyth2018two}. Because of this, it may be possible for inlet dynamics to propagate downstream, impeding the formation of a bounded and stable pulse-train. 

{\color{black}

Finally, in figure \ref{electrified-Summary} we present snapshots of the flow with $E_0=3\times 10^5$ \emph{V}/\emph{m} in panel (a), and $E_0=6\times 10^5$ \emph{V}/\emph{m} in panel (b). 
%in a wave regime having $E_0=6\times 10^5$ V/m
%and a dripping regime at $E_0=3\times 10^5$ V/m. 
%The flow is from left to right and the image has been rotated by $30^{\circ}$ to a horizontal
%position purely for visualization reasons. 
These were obtained from the same DNS simulations used to produce figure \ref{electrifiedcases}, with the snapshots taken at $t \approx 67$ \emph{s} (beyond the limits of the time series presented in figure \ref{electrifiedcases}). The plots depict the film position, along with a pressure colour map and voltage equipotential curves in the fluid and air regions. The lower plots of each panel (these include the equipotential lines) are enlargements of the smaller rectangular regions in the upper plots, and contain five and seven
waves for panels (a) and (b), respectively.
%All other computational parameters are as in figure \ref{electrifiedcases} and the plots depict
%the film position along with a pressure color map and the voltage equipotential curves in the fluid and air regions, after $2\times 10^3$
%time units.
For the upper panel, the field is not strong enough to suppress pinching and the leftmost
finger depicted is about to break off due to a Rayleigh--Taylor instability rather than the wave merging mechanism.  As expected, we see large values of the pressure near the tips of the fingers due to capillary effects. The enlargement in figure \ref{electrified-Summary}(b) shows a developed wave train of pulses with amplitudes much smaller than those in panel (a). Dripping is still found in this case -- see the time series in figure \ref{electrifiedcases}(e) -- but is due to pulse coalescence as is visible in the upper zoomed out plot to the left of the enlarged region.

}

We note that, although the A/C threshold does not agree as closely with the dripping limit as in the non-electrified case, it provides a good order-of-magnitude approximation of the required $E_0$ for dripping stabilization. Furthermore, a temporal linear stability does not provide a good estimate of the dripping stabilization threshold since the required $E_0$ for linear stability of the system is dependent on $L_0$.

\section{Conclusions and discussion}\label{sec:Conclusions}

Absolute linear instability of the flat interface solution and the highly nonlinear process of dripping are not equivalent, however, it has been found that their thresholds are close in parameter space for non-electrified film flows, but not identical \cite{Brun1,kofman2018prediction}. In the present work, we considered the addition of a stabilizing electric field to the problem, and performed a linear stability analysis for different long-wave models and in the Stokes flow limit. 
In agreement with \cite{cimpeanu2014control,RaduAnder1}, we found that a sufficiently strong electric field may be employed to ensure linear and nonlinear stability of the fluid interface. We also found that an absolutely unstable system will become convectively unstable if the imposed electric field is sufficiently strong. Employing DNS to study the full physical system (Navier--Stokes coupled with electrostatics) with pulse initial conditions, we showed that the A/C predictions of the long-wave models are closely aligned with the A/C behavior of the full Navier-Stokes system.

Following an extensive DNS study on long domains using inflow/outflow boundary conditions that mimic experiments, we observed a separation of the fully nonlinear dynamics of hanging film flows into three types: no dripping, dripping due to coalescence and pulse interaction, and dripping due to instability of individual pulses. For both non-electrified and electrified flows, we found that the threshold of absolute instability was a good predictor for the onset of the latter of these three phenomena. For the non-electrified problem, dripping due to coalescence appears to be transient, and thus it is reasonable to define the dripping threshold as the loss of stability of pulses. However, for electrified flows, the regime of dripping due to coalescence and pulse interaction is large in parameter space, and the behavior appears to be persistent -- accordingly, we cannot define the dripping limit as in the non-electrified case. The A/C threshold remains a good order-of-magnitude estimate for the critical electric field strength required to prevent dripping (the temporal linear theory provides no such prediction). We expect this to be true for hanging film flows with a range of external effects imposed, e.g.~magnetic fields, thermal effects, surfactants. This study shows that long-wave modelling approaches and spatial linear theory can provide valuable predictions for very nonlinear processes, so that computational and experimental resources can be more appropriately targeted. There are no details in the experimental work of Brun et al.~\cite{Brun1} whether the dripping events observed are due to coalescence of waves, but the increased number of drips deeper into the absolute instability regime is congruent with our findings. The inclination angles considered in the experiments of Charogiannis and Markides \cite{markidesexperimental} and Charogiannis et al.~\cite{PhysRevFluids.3.114002} are well below the values for which we predict absolute instability or dripping from our 2D study -- their results are in agreement since no dripping is observed. Further experimental work at more extreme inclinations, such as those considered in \cite{Brun1}, would be of great interest. For the electrified flow, extension of the models to allow for weakly conducting fluid phases is warranted for comparison with future experiments and also for relevance to applications  -- Papageorgiou and Petropoulos \cite{papageorgiou2004generation} observed that a parallel field can even become destabilizing for some choices of conductivities and permittivities.

The threshold of absolute instability has been found to successfully predict the onset of nonlinear phenomena in other interfacial flow problems. Rietz et al.~\cite{rietz2017dynamics} performed an experimental study of a liquid film on the exterior of a vertical rotating cylinder. The problem is parameterized by a Reynolds number and a ratio of body forces. Reduced-order modelling approaches are viable, and the authors obtained regions of absolute and convective instability for a WIBL model. Similarly to the present study, they found three regimes of the nonlinear dynamics. No dripping was observed for convectively unstable systems. For flows just beyond the A/C threshold, they found that 2D wave-fronts destabilized into rivulets which emitted drips -- this is due to wave coalescence on the rivulets. The onsets of the linear and nonlinear phenomena agree most closely for $O(1)$ Reynolds numbers. Going further into the absolute instability regime decreases the inception length of the dripping process, until a regime is reached in which drips form immediately at the inlet. Vellingiri et al.~\cite{vellingiri2015absolute} studied the problem of a gravity-driven thin liquid film sheared by a counter-current turbulent gas flow. For sufficiently large gas flow rates, the usually downward falling film begins to flow back upstream towards the inlet, a phenomenon known as ``flooding". They find that the ``flooding point", the critical value of the gas flow rate at which standing waves form on the interface, is close to the upper limit of absolute instability.

It would also be of interest to compare the A/C predictions and DNS results with the matched asymptotics theory of Indeikina et al.~\cite{indeikina1997drop} for individual static (pinned) rivulets. The fully wetting rivulet structures observed in the experiments of Charogiannis and Markides \cite{markidesexperimental} and Charogiannis et al.~\cite{PhysRevFluids.3.114002} at moderate inclination angles (not close to vertical) have wavelength given by the most unstable spanwise mode, resulting from the balance of destabilizing cross-stream gravity and stabilizing surface tension; this is $\sqrt{2}$ larger than the wavelength of the static rivulets considered in \cite{indeikina1997drop} which correspond to the neutrally stable spanwise mode. Lin et al.~\cite{lin2012thin} considered both the dynamics on a static pinned (neutral-mode) rivulet as in \cite{indeikina1997drop}, and the development of a front on a thin precursor film (no dewetting or contact lines), finding in the latter case the most unstable mode dominates. The collection of experimental and numerical studies indicate a shift to shorter (stable) wavelengths when the film dewets. However, we also note the hysteresis effect observed in \cite{indeikina1997drop}, whereby the pinning points of a rivulet with an imposed fluid flux contract as the inclination is increased, but remain fixed as the angle is decreased. This indicates that the rivulets may exist for a range of base widths and static contact angles for a given flow rate. Can fully wetting rivulet structures survive into the dripping regime, or do they give way to one or more static rivulets as observed in \cite{rothrock1968study,indeikina1997drop}? The edges of the fluid film are observed to contract towards the centre in experiments in \cite{markidesexperimental,PhysRevFluids.3.114002}. If dripping occurs in the wedge shaped transition between the inlet and the single rivulet, the mass and fluid flux of the latter are unknown. Thus, it is unclear how to extract a useful prediction from the work of Indeikina et al.~\cite{indeikina1997drop} for the present 2D setting, since it is not obvious what the dimensions of an effective underlying static rivulet should be, in either the wetted or dewetted regime. We thus leave such comparisons to future work on the 3D problem.

The electrified aspect of the fully 3D problem is also an interesting extension. For this, we allow the electric field to be skewed at an angle $\phi$ to the streamwise flow direction, yet still parallel to the substrate surface. The far-field boundary condition for the voltage potentials in this problem is then
\begin{equation}\label{farfieldV13D} \bm{\nabla} V  \rightarrow -(E_0\cos\phi,E_0\sin\phi,0) , \quad \textrm{as } z \rightarrow \pm \infty.\end{equation}
If $\phi = 0$, we recover the arrangement considered in the present work, and if $\phi = \pi/2$ the field is directed purely in the transverse direction. Following through with the long wave analysis, we find that the electric field contribution, which has Fourier symbol $|\xi|^3$ in the 2D formulation, now appears as the operator with symbol
\begin{equation}(\xi_1^2 + \xi_2^2)^{1/2} (\xi_1\cos\phi + \xi_2 \sin \phi)^2,\end{equation}
for wavenumber vector $(\xi_1,\xi_2)$. From this expression, it is apparent that the electric field has the strongest stabilizing effect on waves which are in the same direction as the undisturbed field lines, and has no influence on interfacial waves which are perpendicular, with a smooth variation in-between. Thus, a sufficiently strong electric field ensures that an interface can only develop instabilities of a select few modes. In particular, a field set up transverse to the flow direction may be employed in experiments to preserve the 2D phase of the dynamics for overlying films, i.e.~impede secondary 3D instabilities, or to prevent the formation of rivulets for hanging films without affecting the streamwise dynamics. A detailed study of the fully 3D problem will be presented in future work. Additionally, work is underway on the full 3D problem with an electric field set up normal to the substrate, investigating the electrostatically-induced rivulet formation predicted in \cite{tomlin_papageorgiou_pavliotis_2017} for overlying liquid films.

\appendix

\section{Orr--Sommerfeld system\label{OrrSommerfeldAppendix}}

Here we provide details of the Orr--Sommerfeld system we solved to make comparisons between the full and long-wave linear theories. 
Linearizing \eqref{Navierstokesfinal1} about the exact 
Nusselt solution \eqref{basestatesdimensional2}, with perturbations denoted by tildes, we find
\begin{equation} \label{Navierstokesfinal1Orr1} {\Rey} ( \tilde{u}_t + \overline{u}\tilde{u}_x + \tilde{w}\overline{u}_z ) =  -\tilde{p}_x  + {\nabla}^2 \tilde{u} , \qquad
{\Rey} ( \tilde{w}_t + \overline{u}\tilde{w}_x ) =  -\tilde{p}_z  + {\nabla}^2 \tilde{w} , \qquad
\tilde{u}_x + \tilde{w}_z = 0,
\end{equation}
with the no-slip and impermeability conditions at the substrate surface unchanged. At $z=1$, the kinematic condition, tangential stress balance, and normal stress balance become 
%\begin{align} \tilde{w} = \tilde{h}_t + \tilde{h}_x, \qquad  \tilde{u}_z - 2 \tilde{h} + \tilde{w}_x = 0,\label{eq:kinLin}\\
%\tilde{w}_z +(1 + \tilde{h})\cot\theta -  \frac{1}{2 } \tilde{p}  +  {\Weber}\left[  \epsilon^{\iota} \tilde{V}^{\iota}_x \right]_
%{\textrm{II}}^{\textrm{I}}  =  \frac{1}{2 {\Capil}} \tilde{h}_{xx},\label{NSBlin}
%\end{align}
\begin{equation} \tilde{w} = \tilde{h}_t + \tilde{h}_x, \qquad  \tilde{u}_z - 2 \tilde{h} + \tilde{w}_x = 0, \qquad 
\tilde{w}_z +(1 + \tilde{h})\cot\theta -  \frac{1}{2 } \tilde{p}  +  {\Weber}\left[  \epsilon^{\iota} \tilde{V}^{\iota}_x \right]_
{\textrm{II}}^{\textrm{I}}  =  \frac{1}{2 {\Capil}} \tilde{h}_{xx},\label{NSBlin}
\end{equation}
where we have used (\ref{Navierstokesfinal1Orr1}c) in deriving (\ref{NSBlin}c). The Laplace equations \eqref{eq:Laplace}, far-field conditions \eqref{farfieldV2nondim}, and the solid--fluid boundary conditions \eqref{subsurfVconds1} remain unchanged, whereas the electrostatic boundary conditions at the free surface \eqref{surfacevoltcond1} linearize to
\begin{equation}\label{Vbcfluiflui2}\left[\epsilon^{\iota} ( \tilde{h}_x + \tilde{V}^{\iota}_z) \right]_{\textrm{II}}^{\textrm{I}}=0, \quad \left[\tilde{V}^{\iota} \right]_{\textrm{II}}^{\textrm{I}} = 0 \quad \textrm{at } z=1.\end{equation} 
The problem is rearranged to
obtain a system for $\tilde{h}$, $\tilde{w}$, and $\tilde{V}^{\iota}$ alone. This is achieved by eliminating the velocity $\tilde{u}$ and pressure $\tilde{p}$
from \eqref{Navierstokesfinal1Orr1} to find
\begin{equation} \label{Orrgovern1} 
\nabla^2 \left({\Rey}\tilde{w}_{t} - {\nabla}^2 \tilde{w} \right) + 
{\Rey}( 2 + \overline{u} \nabla^2 ) \tilde{w}_x = 0,
\end{equation}
and in turn eliminating $\tilde{p}$ in the normal stress balance (\ref{NSBlin}c) gives
\begin{equation} \label{eq:NSBlinear}
3\tilde{w}_{xxz}   +   \tilde{w}_{zzz} + 2\tilde{h}_{xx}\cot\theta  -  {\Rey}  (\tilde{w}_{tz} +\tilde{w}_{xz})  + 2 {\Weber}\left[  \epsilon^{\iota} \tilde{V}^{\iota}_{xxx} \right]_{\textrm{II}}^{\textrm{I}}  -  \frac{1}{{\Capil}} \tilde{h}_{xxxx} = 0,  \quad \textrm{at } z=1. 
\end{equation}
Lastly, a derivative of the tangential stress boundary condition with respect to $x$ yields
\begin{equation} 
\tilde{w}_{xx} -  \tilde{w}_{zz} = 2 \tilde{h}_x ,  \quad \textrm{at } z=1.\label{eq:TSBlinear}
\end{equation}

%By taking the divergence of the momentum equation, we obtain
%\begin{equation}\label{pexpressionlapl1}{\nabla}^2\tilde{p} = - 2{\Rey}  \overline{u}_z \tilde{w}_x = 4 {\Rey}  (z-1) \tilde{w}_x, \qquad {\nabla}^2\tilde{p}_z = 4{\Rey} \tilde{w}_x - 2{\Rey}  \overline{u}_z \tilde{w}_{xz} = 4 {\Rey}  \tilde{w}_x + 4 {\Rey}  (z-1) \tilde{w}_{xz}.\end{equation}
%Next, by taking the Laplacian of the $z$-momentum equation and using (\ref{pexpressionlapl1}b) we obtain
%\begin{equation} \label{Orrgovern1} \nabla^2 \left({\Rey}\tilde{w}_{t} - {\nabla}^2 \tilde{w} \right) + {\Rey}( 2 + \overline{u} \nabla^2 ) \tilde{w}_x = 0.\end{equation}
%Noting from (\ref{pexpressionlapl1}a) that $\nabla^2\tilde{p} = 0$ at $z=1$, we take the derivative of the $z$-momentum equation with respect to $z$ and evaluate at $z=1$ to obtain
%\begin{equation}\tilde{p}_{xx} =  {\Rey}  (\tilde{w}_{tz} +\tilde{w}_{xz}  ) -  \nabla^2 \tilde{w}_z, \quad \textrm{at } z = 1.\end{equation}
%Then, by applying $\partial_x^2$ to the normal stress balance and using the previous equation, we find
%\begin{equation} 3\tilde{w}_{xxz}   +   \tilde{w}_{zzz} + 2\tilde{h}_{xx}\cot\theta  -  {\Rey}  (\tilde{w}_{tz} +\tilde{w}_{xz})  + 2 {\Weber}\left[  \epsilon^{\iota} \tilde{V}^{\iota}_{xxx} \right]_{\textrm{II}}^{\textrm{I}}  -  \frac{1}{{\Capil}} \tilde{h}_{xxxx} = 0,  \quad \textrm{at } z=1. \end{equation}
%Lastly, by taking a derivative of the tangential stress boundary condition with respect to $x$ we obtain
%\begin{equation} \tilde{w}_{xx} -  \tilde{w}_{zz} = 2 \tilde{h}_x ,  \quad \textrm{at } z=1.\end{equation}
%

We consider perturbations of the form
\begin{equation}\label{Orrexact1} \tilde{h} = e^{i \xi x + \omega t}+\textrm{c.c.},\qquad \tilde{w} = W(z) e^{i \xi x + \omega t}+\textrm{c.c.}, 
\qquad \tilde{V}^{\iota} =  \tilde{A}^{\iota}(z) e^{i \xi x + \omega t}+\textrm{c.c.} \quad \iota= \textrm{S}, \textrm{I}, \textrm{II},
\end{equation}
where $\textrm{c.c.}$ denotes the complex conjugate; the wavenumber $\xi$ is allowed to be complex.
Defining $D\equiv \mathrm{d}/\mathrm{d}z$ casts equation \eqref{Orrgovern1} and the Laplace equations for the voltages into
\begin{align} 
& (D^2 -  \xi^2) \left[ D^2 -  \xi^2 -  {\Rey}\, \omega \right] W - i \xi {\Rey} \left[ \overline{u} (D^2 - \xi^2) +2 \right] W  = 0, \label{OrrSmaineqn1} \\
& (D^2 - \xi^2)\tilde{A}^{\iota} = 0, \qquad \iota=\textrm{S}, \textrm{I}, \textrm{II}. \label{VoltagesLinear}
\end{align}
The boundary conditions at the substrate $z=0$ are
\begin{equation}\label{substrateOrrS1}
W = 0, \qquad DW = 0,  \qquad \epsilon^{\textrm{S}} D\tilde{A}^{\textrm{S}} = 
\epsilon^{\textrm{I}} D\tilde{A}^{\textrm{I}}, \qquad \tilde{A}^{\textrm{S}} = \tilde{A}^{\textrm{I}}.
\end{equation}
The far-field conditions become
\begin{equation}\label{electroOrrS2} \tilde{A}^{\textrm{S}}(z)  \rightarrow 0 , \quad \textrm{as } z \rightarrow - \infty, \qquad \tilde{A}^{\textrm{II}}(z)  \rightarrow 0 , \quad \textrm{as } z \rightarrow + \infty,\end{equation}
and the interfacial conditions at $z=1$ are
\begin{align}
& W = i  \xi + \omega, \qquad \left[ D^2 + \xi^2\right]W  + 2i\xi = 0, \label{OrrSomNormalBC1a}\\
&{\displaystyle \left[ D^2 - 3 \xi^2  - {\Rey}   (\omega + i \xi) \right] DW  =  2\xi^2 \cot\theta + 2 i 
{\Weber} \; \xi^3 \left[  \epsilon^{\iota}  \tilde{A}^{\iota} \right]_{\textrm{II}}^{\textrm{I}}  +  \frac{\xi^4}{{\Capil}}, }\label{OrrSomNormalBC1b}\\
& \epsilon^{\textrm{I}} ( i \xi + D\tilde{A}^{\textrm{I}})= \epsilon^{\textrm{II}} ( i \xi + D\tilde{A}^{\textrm{II}}), 
\qquad  \tilde{A}^{\textrm{I}} = \tilde{A}^{\textrm{II}}.\label{electroOrrS3}
\end{align}
The electrostatics problem comprising of (\ref{VoltagesLinear},\ref{substrateOrrS1}c,d,\ref{electroOrrS2},\ref{electroOrrS3}) 
can be solved analytically, 
\begin{align} &{\tilde{A}}^{\textrm{S}}(z;\xi) = \Gamma(\xi) e^{ \operatorname{sign}(\xi_{\textrm{r}})\xi z}, \qquad {\tilde{A}}^{\textrm{I}}
(z;\xi)= \Gamma(\xi) \left[ \cosh( \xi z) + \operatorname{sign}(\xi_{\textrm{r}}) \frac{\epsilon^{\textrm{S}}}{\epsilon^{\textrm{I}}} \sinh( \xi z) \right] ,\nonumber \\
&\label{Aexprappen2}{\tilde{A}}^{\textrm{II}}(z;\xi) = \Gamma(\xi)  \left[ \cosh( \xi) + \operatorname{sign}(\xi_{\textrm{r}}) \frac{\epsilon^{\textrm{S}}}{\epsilon^{\textrm{I}}} \sinh( \xi ) \right] e^{ \operatorname{sign}(\xi_{\textrm{r}})\xi(1-z)},
\end{align}
where
\begin{equation} \label{FappendixEpses1}
\Gamma(\xi) = \frac{(\epsilon^{\textrm{II}} - \epsilon^{\textrm{I}}) i \operatorname{sign}(\xi_{\textrm{r}})}{ \left( \epsilon^{\textrm{II}} + \epsilon^{\textrm{S}} \right)\cosh (\xi) + \operatorname{sign}(\xi_{\textrm{r}}) \left( \epsilon^{\textrm{I}} + \frac{ \epsilon^{\textrm{II}} \epsilon^{\textrm{S}}}{ \epsilon^{\textrm{I}}} \right) \sinh (\xi) }.
\end{equation}
Thus, we can analytically give the electric field contribution in \eqref{OrrSomNormalBC1b} at $z=1$,
\begin{equation}\label{EfieldAppenContribute}2 i {\Weber}\; \xi^3 \left[  \epsilon^{\iota}  \tilde{A}^{\iota} \right]_{\textrm{II}}^{\textrm{I}} =   2{\Weber}' \xi^3 \frac{  \operatorname{sign}(\xi_{\textrm{r}})\cosh(\xi) +  \frac{ \epsilon^{\textrm{S}}}{ \epsilon^{\textrm{I}}} \sinh(\xi) }{ \cosh (\xi) + \operatorname{sign}(\xi_{\textrm{r}}) \left(  \frac{ ( \epsilon^{\textrm{I}} )^2 + \epsilon^{\textrm{II}} \epsilon^{\textrm{S}}}{ \epsilon^{\textrm{I}}\left( \epsilon^{\textrm{II}} + \epsilon^{\textrm{S}} \right)} \right) \sinh (\xi) },\end{equation}
(recalling the relationship \eqref{rescaledweber1} between $\Weber$ and $\Weber'$), so that only the fluid dynamics problem in the region $0\leq z \leq 1$ remains. The leading-order term of \eqref{EfieldAppenContribute} for 
$\xi\ll 1$ is $2{\Weber}' \xi^3 \operatorname{sign}(\xi_{\textrm{r}})$ in agreement with the long-wave calculations. 
This Orr--Sommerfeld system was solved using the bifurcation and continuation software \textsc{AUTO-07P}, as discussed in \cite{kalliadasis2011falling}. 
With this, we produced the solid line in figure \ref{OSplot1} showing the dependence of the critical wavenumber $\xi_{\textrm{c}}$ on the Reynolds number and electric field strength.

%{\bf{RUBEN: there were a couple of ${\Weber}^\prime$ that I removed the prime from - OK?} No -- have been put back}

%
%{\color{red}
%
%
%
%For Reynolds numbers above critical, there is a band of real wavenumbers ($\xi \in \mathbb{R}$) corresponding to linearly unstable modes, $\xi \in (0,\xi_{\textrm{{c}}})$ with $\omega_{\textrm{r}}(\xi) > 0$. Here, the endpoint $\xi_{\textrm{{c}}}$ is dependent on the non-dimensional parameters. If surface tension forces or the electric field are strengthened, $\xi_{\textrm{{c}}}$ decreases, but does not reach zero at finite Weber numbers or non-zero capillary numbers; $\omega_{\textrm{r}}(\xi) \sim \xi^2$ locally at $\xi = 0$ when the Reynolds number is beyond critical. The value of $\xi_{\textrm{c}}$ determines the shortest unstable wavelength, and hence flows with a system length below this value are stable. 
%
%
%}

\section{Stokes flow}\label{Stokesflowappend1}

At zero Reynolds number, the Orr--Sommerfeld equation (\ref{OrrSmaineqn1}) reduces to
\begin{equation}\label{StokesOS1} (D^2 -  \xi^2)^2 W = 0,  \end{equation}
with boundary conditions (\ref{substrateOrrS1}a,b,\ref{OrrSomNormalBC1a}b) and (\ref{OrrSomNormalBC1b}) 
at ${\Rey} = 0$. Equation \eqref{StokesOS1} has the general solution
\begin{equation}W = B_1(\xi)\, z\cosh(\xi z) + B_2(\xi)\, z\sinh(\xi z) + B_3(\xi) \cosh(\xi z) + B_4(\xi) \sinh(\xi z).\end{equation}
The boundary conditions (\ref{substrateOrrS1}a,b) give $B_3(\xi) = 0$ and $B_1(\xi) + \xi B_4(\xi) = 0$, respectively, 
and conditions (\ref{OrrSomNormalBC1a}b,\ref{OrrSomNormalBC1b}) at $z=1$ imply that
\begin{align} 
& B_2(\xi)  =  \frac{ i ( \xi \tanh(\xi) - 1 ) -  \xi \cot\theta -  i {\Weber} \;  \xi^2 [  \epsilon^{\iota}  \tilde{A}^{\iota} ]_{\textrm{II}}^{\textrm{I}} - (2{\Capil})^{-1} \xi^3}{( \xi^2 + \cosh^2(\xi)) \operatorname{sech}(\xi)},\\
& B_4(\xi) = \frac{B_2(\xi) (1 + \xi \tanh (\xi)) + i \operatorname{sech}(\xi)}{\xi^2}.
\end{align}
With these, the dispersion relation in the Stokes flow limit can be obtained from (\ref{OrrSomNormalBC1a}a) as
\begin{equation} \omega =  - \frac{  i \xi ( 1 + \xi^2 + \cosh^2(\xi) )  }{ \xi^2 + \cosh^2(\xi)} - \left( \cot\theta +  i {\Weber} \;  \xi \; [  \epsilon^{\iota}  \tilde{A}^{\iota} ]_{\textrm{II}}^{\textrm{I}} + (2{\Capil})^{-1} \xi^2 \right) \frac{\sinh(\xi ) \cosh(\xi ) - \xi }{\xi( \xi^2 + \cosh^2(\xi))},
  \end{equation}
where the electric field contribution is given by \eqref{EfieldAppenContribute}. With the small wavenumber approximations
\begin{equation}
\frac{ i \xi (1 + \xi^2 + \cosh^2(\xi) )}{ \xi^2 + \cosh^2(\xi)} = 2i\xi + O(\xi^3),
\qquad  \frac{\sinh(\xi ) \cosh(\xi ) - \xi  }{\xi( \xi^2 + \cosh^2(\xi))} = \frac{2}{3}\xi^2 + O(\xi^4),  
\end{equation}
and that of the electric field term given above, we recover, to leading order, the Benney dispersion relation \eqref{lindispBenney1} at zero Reynolds number.

\begin{acknowledgments}
 RJT acknowledges a PhD studentship from the Engineering and Physical Sciences Research Council (EPSRC). 
 DTP was partly supported by EPSRC grant EP/L020564/1.
 RC acknowledges the support of the 
 Mathematical Institute of the University of Oxford, as well as the staff actively maintaining 
 the Imperial College Research Computing Service (DOI:\href{http://doi.org/10.14469/hpc/2232}{10.14469/hpc/2232}).
 
\end{acknowledgments}

\bibliography{StabilisingEfieldbib}

\begin{thebibliography}{70}%
\makeatletter
\providecommand \@ifxundefined [1]{%
 \@ifx{#1\undefined}
}%
\providecommand \@ifnum [1]{%
 \ifnum #1\expandafter \@firstoftwo
 \else \expandafter \@secondoftwo
 \fi
}%
\providecommand \@ifx [1]{%
 \ifx #1\expandafter \@firstoftwo
 \else \expandafter \@secondoftwo
 \fi
}%
\providecommand \natexlab [1]{#1}%
\providecommand \enquote  [1]{``#1''}%
\providecommand \bibnamefont  [1]{#1}%
\providecommand \bibfnamefont [1]{#1}%
\providecommand \citenamefont [1]{#1}%
\providecommand \href@noop [0]{\@secondoftwo}%
\providecommand \href [0]{\begingroup \@sanitize@url \@href}%
\providecommand \@href[1]{\@@startlink{#1}\@@href}%
\providecommand \@@href[1]{\endgroup#1\@@endlink}%
\providecommand \@sanitize@url [0]{\catcode `\\12\catcode `\$12\catcode
  `\&12\catcode `\#12\catcode `\^12\catcode `\_12\catcode `\%12\relax}%
\providecommand \@@startlink[1]{}%
\providecommand \@@endlink[0]{}%
\providecommand \url  [0]{\begingroup\@sanitize@url \@url }%
\providecommand \@url [1]{\endgroup\@href {#1}{\urlprefix }}%
\providecommand \urlprefix  [0]{URL }%
\providecommand \Eprint [0]{\href }%
\providecommand \doibase [0]{http://dx.doi.org/}%
\providecommand \selectlanguage [0]{\@gobble}%
\providecommand \bibinfo  [0]{\@secondoftwo}%
\providecommand \bibfield  [0]{\@secondoftwo}%
\providecommand \translation [1]{[#1]}%
\providecommand \BibitemOpen [0]{}%
\providecommand \bibitemStop [0]{}%
\providecommand \bibitemNoStop [0]{.\EOS\space}%
\providecommand \EOS [0]{\spacefactor3000\relax}%
\providecommand \BibitemShut  [1]{\csname bibitem#1\endcsname}%
\let\auto@bib@innerbib\@empty
%</preamble>
\bibitem [{\citenamefont {Stone}\ \emph {et~al.}(2004)\citenamefont {Stone},
  \citenamefont {Stroock},\ and\ \citenamefont
  {Ajdari}}]{doi:10.1146/annurev.fluid.36.050802.122124}%
  \BibitemOpen
  \bibfield  {author} {\bibinfo {author} {\bibfnamefont {H.~A.}\ \bibnamefont
  {Stone}}, \bibinfo {author} {\bibfnamefont {A.~D.}\ \bibnamefont {Stroock}},
  \ and\ \bibinfo {author} {\bibfnamefont {A.}~\bibnamefont {Ajdari}},\
  }\bibfield  {title} {\enquote {\bibinfo {title} {Engineering flows in small
  devices: Microfluidics toward a lab-on-a-chip},}\ }\href {\doibase
  10.1146/annurev.fluid.36.050802.122124} {\bibfield  {journal} {\bibinfo
  {journal} {Annual Review of Fluid Mechanics}\ }\textbf {\bibinfo {volume}
  {36}},\ \bibinfo {pages} {381--411} (\bibinfo {year} {2004})}\BibitemShut
  {NoStop}%
\bibitem [{\citenamefont {Miyara}(1999)}]{Miyara1999}%
  \BibitemOpen
  \bibfield  {author} {\bibinfo {author} {\bibfnamefont {A.}~\bibnamefont
  {Miyara}},\ }\bibfield  {title} {\enquote {\bibinfo {title} {Numerical
  analysis on flow dynamics and heat transfer of falling liquid films with
  interfacial waves},}\ }\href {\doibase 10.1007/s002310050328} {\bibfield
  {journal} {\bibinfo  {journal} {Heat Mass Transfer}\ }\textbf {\bibinfo
  {volume} {35}},\ \bibinfo {pages} {298--306} (\bibinfo {year}
  {1999})}\BibitemShut {NoStop}%
\bibitem [{\citenamefont {Serifi}\ \emph {et~al.}(2004)\citenamefont {Serifi},
  \citenamefont {Malamataris},\ and\ \citenamefont
  {Bontozoglou}}]{serifi2004transient}%
  \BibitemOpen
  \bibfield  {author} {\bibinfo {author} {\bibfnamefont {K.}~\bibnamefont
  {Serifi}}, \bibinfo {author} {\bibfnamefont {N.~A.}\ \bibnamefont
  {Malamataris}}, \ and\ \bibinfo {author} {\bibfnamefont {V.}~\bibnamefont
  {Bontozoglou}},\ }\bibfield  {title} {\enquote {\bibinfo {title} {Transient
  flow and heat transfer phenomena in inclined wavy films},}\ }\href@noop {}
  {\bibfield  {journal} {\bibinfo  {journal} {Int. J. Therm. Sci.}\ }\textbf
  {\bibinfo {volume} {43}},\ \bibinfo {pages} {761--767} (\bibinfo {year}
  {2004})}\BibitemShut {NoStop}%
\bibitem [{\citenamefont {Shorts}\ \emph {et~al.}(2005)\citenamefont {Shorts},
  \citenamefont {Baygents},\ and\ \citenamefont {Goldstein}}]{Shorts_et_al}%
  \BibitemOpen
  \bibfield  {author} {\bibinfo {author} {\bibfnamefont {M.~B.}\ \bibnamefont
  {Shorts}}, \bibinfo {author} {\bibfnamefont {J.~C.}\ \bibnamefont
  {Baygents}}, \ and\ \bibinfo {author} {\bibfnamefont {R.~E.}\ \bibnamefont
  {Goldstein}},\ }\bibfield  {title} {\enquote {\bibinfo {title} {Stalactite
  growth as a free-boundary problem},}\ }\href@noop {} {\bibfield  {journal}
  {\bibinfo  {journal} {Phys. Fluids}\ }\textbf {\bibinfo {volume} {17}},\
  \bibinfo {pages} {083101} (\bibinfo {year} {2005})}\BibitemShut {NoStop}%
\bibitem [{\citenamefont {Camporeale}(2017)}]{Camporeale}%
  \BibitemOpen
  \bibfield  {author} {\bibinfo {author} {\bibfnamefont {C.}~\bibnamefont
  {Camporeale}},\ }\bibfield  {title} {\enquote {\bibinfo {title} {An
  asymptotic approach to the crenulation instability},}\ }\href@noop {}
  {\bibfield  {journal} {\bibinfo  {journal} {J. Fluid Mech.}\ }\textbf
  {\bibinfo {volume} {826}},\ \bibinfo {pages} {636--652} (\bibinfo {year}
  {2017})}\BibitemShut {NoStop}%
\bibitem [{\citenamefont {Kapitza}\ and\ \citenamefont
  {Kapitza}(1949)}]{kapitza1949wave}%
  \BibitemOpen
  \bibfield  {author} {\bibinfo {author} {\bibfnamefont {P.~L.}\ \bibnamefont
  {Kapitza}}\ and\ \bibinfo {author} {\bibfnamefont {S.~P.}\ \bibnamefont
  {Kapitza}},\ }\bibfield  {title} {\enquote {\bibinfo {title} {Wave flow of
  thin layers of viscous liquids. part iii. experimental research of a wave
  flow regime},}\ }\href@noop {} {\bibfield  {journal} {\bibinfo  {journal}
  {Zhurnal Eksperimentalnoi i Teoreticheskoi Fiziki}\ }\textbf {\bibinfo
  {volume} {19}},\ \bibinfo {pages} {105--120} (\bibinfo {year}
  {1949})}\BibitemShut {NoStop}%
\bibitem [{\citenamefont {Nusselt}(1916)}]{nusselt1}%
  \BibitemOpen
  \bibfield  {author} {\bibinfo {author} {\bibfnamefont {W.}~\bibnamefont
  {Nusselt}},\ }\bibfield  {title} {\enquote {\bibinfo {title} {Die
  oberfl{Å }chenkondensation des wasserdampfe},}\ }\href@noop {} {\bibfield
  {journal} {\bibinfo  {journal} {Z. Ver. Deut. Indr.}\ }\textbf {\bibinfo
  {volume} {60}},\ \bibinfo {pages} {541--546} (\bibinfo {year}
  {1916})}\BibitemShut {NoStop}%
\bibitem [{\citenamefont {Yih}(1955)}]{yih1955proceedings}%
  \BibitemOpen
  \bibfield  {author} {\bibinfo {author} {\bibfnamefont {C.~S.}\ \bibnamefont
  {Yih}},\ }\bibfield  {title} {\enquote {\bibinfo {title} {Proceedings of the
  2nd us congress on applied mechanics},}\ \ }(\bibinfo {organization} {ASME},\
  \bibinfo {year} {1955})\BibitemShut {NoStop}%
\bibitem [{\citenamefont {Yih}(1963)}]{Yih1}%
  \BibitemOpen
  \bibfield  {author} {\bibinfo {author} {\bibfnamefont {C.~S.}\ \bibnamefont
  {Yih}},\ }\bibfield  {title} {\enquote {\bibinfo {title} {Stability of liquid
  flow down an inclined plane},}\ }\href {\doibase
  http://dx.doi.org/10.1063/1.1706737} {\bibfield  {journal} {\bibinfo
  {journal} {Phys. Fluids}\ }\textbf {\bibinfo {volume} {6}},\ \bibinfo {pages}
  {321--334} (\bibinfo {year} {1963})}\BibitemShut {NoStop}%
\bibitem [{\citenamefont {Benjamin}(1957)}]{FLM:367246}%
  \BibitemOpen
  \bibfield  {author} {\bibinfo {author} {\bibfnamefont {T.~B.}\ \bibnamefont
  {Benjamin}},\ }\bibfield  {title} {\enquote {\bibinfo {title} {Wave formation
  in laminar flow down an inclined plane},}\ }\href {\doibase
  10.1017/S0022112057000373} {\bibfield  {journal} {\bibinfo  {journal} {J.
  Fluid Mech.}\ }\textbf {\bibinfo {volume} {2}},\ \bibinfo {pages} {554--573}
  (\bibinfo {year} {1957})}\BibitemShut {NoStop}%
\bibitem [{\citenamefont {Liu}\ and\ \citenamefont
  {Gollub}(1994)}]{liu1994solitary}%
  \BibitemOpen
  \bibfield  {author} {\bibinfo {author} {\bibfnamefont {J.}~\bibnamefont
  {Liu}}\ and\ \bibinfo {author} {\bibfnamefont {J.~P.}\ \bibnamefont
  {Gollub}},\ }\bibfield  {title} {\enquote {\bibinfo {title} {Solitary wave
  dynamics of film flows},}\ }\href@noop {} {\bibfield  {journal} {\bibinfo
  {journal} {Physics of Fluids}\ }\textbf {\bibinfo {volume} {6}},\ \bibinfo
  {pages} {1702--1712} (\bibinfo {year} {1994})}\BibitemShut {NoStop}%
\bibitem [{\citenamefont {Kharlamov}\ \emph {et~al.}(2015)\citenamefont
  {Kharlamov}, \citenamefont {Guzanov}, \citenamefont {Bobylev}, \citenamefont
  {Alekseenko},\ and\ \citenamefont {Markovich}}]{kharlamov2015transition}%
  \BibitemOpen
  \bibfield  {author} {\bibinfo {author} {\bibfnamefont {S.~M.}\ \bibnamefont
  {Kharlamov}}, \bibinfo {author} {\bibfnamefont {V.~V.}\ \bibnamefont
  {Guzanov}}, \bibinfo {author} {\bibfnamefont {A.~V.}\ \bibnamefont
  {Bobylev}}, \bibinfo {author} {\bibfnamefont {S.~V.}\ \bibnamefont
  {Alekseenko}}, \ and\ \bibinfo {author} {\bibfnamefont {D.~M.}\ \bibnamefont
  {Markovich}},\ }\bibfield  {title} {\enquote {\bibinfo {title} {The
  transition from two-dimensional to three-dimensional waves in falling liquid
  films: Wave patterns and transverse redistribution of local flow rates},}\
  }\href@noop {} {\bibfield  {journal} {\bibinfo  {journal} {Physics of
  Fluids}\ }\textbf {\bibinfo {volume} {27}},\ \bibinfo {pages} {114106}
  (\bibinfo {year} {2015})}\BibitemShut {NoStop}%
\bibitem [{\citenamefont {Alekseenko}\ \emph {et~al.}(1994)\citenamefont
  {Alekseenko}, \citenamefont {Nakoryakov},\ and\ \citenamefont
  {Pokusaev}}]{alekseenko1994wave}%
  \BibitemOpen
  \bibfield  {author} {\bibinfo {author} {\bibfnamefont {S.~V.}\ \bibnamefont
  {Alekseenko}}, \bibinfo {author} {\bibfnamefont {V.~E.}\ \bibnamefont
  {Nakoryakov}}, \ and\ \bibinfo {author} {\bibfnamefont {B.~G.}\ \bibnamefont
  {Pokusaev}},\ }\href@noop {} {\emph {\bibinfo {title} {Wave flow of liquid
  films}}}\ (\bibinfo  {publisher} {Begell House New York},\ \bibinfo {year}
  {1994})\BibitemShut {NoStop}%
\bibitem [{\citenamefont {Park}\ and\ \citenamefont
  {Nosoko}(2003)}]{park2003three}%
  \BibitemOpen
  \bibfield  {author} {\bibinfo {author} {\bibfnamefont {C.~D.}\ \bibnamefont
  {Park}}\ and\ \bibinfo {author} {\bibfnamefont {T.}~\bibnamefont {Nosoko}},\
  }\bibfield  {title} {\enquote {\bibinfo {title} {Three-dimensional wave
  dynamics on a falling film and associated mass transfer},}\ }\href@noop {}
  {\bibfield  {journal} {\bibinfo  {journal} {AIChE Journal}\ }\textbf
  {\bibinfo {volume} {49}},\ \bibinfo {pages} {2715--2727} (\bibinfo {year}
  {2003})}\BibitemShut {NoStop}%
\bibitem [{\citenamefont {Rothrock}(1968)}]{rothrock1968study}%
  \BibitemOpen
  \bibfield  {author} {\bibinfo {author} {\bibfnamefont {D.~A.}\ \bibnamefont
  {Rothrock}},\ }\emph {\bibinfo {title} {A study of flows down the underside
  of an inclined plane}},\ \href@noop {} {Ph.D. thesis},\ \bibinfo  {school}
  {University of Cambridge} (\bibinfo {year} {1968})\BibitemShut {NoStop}%
\bibitem [{\citenamefont {Charogiannis}\ and\ \citenamefont
  {Markides}(2016)}]{markidesexperimental}%
  \BibitemOpen
  \bibfield  {author} {\bibinfo {author} {\bibfnamefont {A.}~\bibnamefont
  {Charogiannis}}\ and\ \bibinfo {author} {\bibfnamefont {C.~N.}\ \bibnamefont
  {Markides}},\ }\bibfield  {title} {\enquote {\bibinfo {title} {Application of
  planar laser-induced fluorescence for the investigation of interfacial waves
  and rivulet structures in liquid films flowing down inverted substrates},}\
  }\href {\doibase 10.1615/InterfacPhenomHeatTransfer.2017019587} {\bibfield
  {journal} {\bibinfo  {journal} {Interfacial phenomena and heat transfer}\
  }\textbf {\bibinfo {volume} {4}},\ \bibinfo {pages} {234--251} (\bibinfo
  {year} {2016})}\BibitemShut {NoStop}%
\bibitem [{\citenamefont {Charogiannis}\ \emph {et~al.}(2018)\citenamefont
  {Charogiannis}, \citenamefont {Denner}, \citenamefont {van Wachem},
  \citenamefont {Kalliadasis}, \citenamefont {Scheid},\ and\ \citenamefont
  {Markides}}]{PhysRevFluids.3.114002}%
  \BibitemOpen
  \bibfield  {author} {\bibinfo {author} {\bibfnamefont {A.}~\bibnamefont
  {Charogiannis}}, \bibinfo {author} {\bibfnamefont {F.}~\bibnamefont
  {Denner}}, \bibinfo {author} {\bibfnamefont {B.~G.~M.}\ \bibnamefont {van
  Wachem}}, \bibinfo {author} {\bibfnamefont {S.}~\bibnamefont {Kalliadasis}},
  \bibinfo {author} {\bibfnamefont {B.}~\bibnamefont {Scheid}}, \ and\ \bibinfo
  {author} {\bibfnamefont {C.~N.}\ \bibnamefont {Markides}},\ }\bibfield
  {title} {\enquote {\bibinfo {title} {Experimental investigations of liquid
  falling films flowing under an inclined planar substrate},}\ }\href {\doibase
  10.1103/PhysRevFluids.3.114002} {\bibfield  {journal} {\bibinfo  {journal}
  {Phys. Rev. Fluids}\ }\textbf {\bibinfo {volume} {3}},\ \bibinfo {pages}
  {114002} (\bibinfo {year} {2018})}\BibitemShut {NoStop}%
\bibitem [{\citenamefont {Indeikina}\ \emph {et~al.}(1997)\citenamefont
  {Indeikina}, \citenamefont {Veretennikov},\ and\ \citenamefont
  {Chang}}]{indeikina1997drop}%
  \BibitemOpen
  \bibfield  {author} {\bibinfo {author} {\bibfnamefont {A.}~\bibnamefont
  {Indeikina}}, \bibinfo {author} {\bibfnamefont {I.}~\bibnamefont
  {Veretennikov}}, \ and\ \bibinfo {author} {\bibfnamefont {H.-C.}\
  \bibnamefont {Chang}},\ }\bibfield  {title} {\enquote {\bibinfo {title} {Drop
  fall-off from pendent rivulets},}\ }\href@noop {} {\bibfield  {journal}
  {\bibinfo  {journal} {Journal of Fluid Mechanics}\ }\textbf {\bibinfo
  {volume} {338}},\ \bibinfo {pages} {173--201} (\bibinfo {year}
  {1997})}\BibitemShut {NoStop}%
\bibitem [{\citenamefont {Brun}\ \emph {et~al.}(2015)\citenamefont {Brun},
  \citenamefont {Damiano}, \citenamefont {Rieu}, \citenamefont {Balestra},\
  and\ \citenamefont {Gallaire}}]{Brun1}%
  \BibitemOpen
  \bibfield  {author} {\bibinfo {author} {\bibfnamefont {P.-T.}\ \bibnamefont
  {Brun}}, \bibinfo {author} {\bibfnamefont {A.}~\bibnamefont {Damiano}},
  \bibinfo {author} {\bibfnamefont {P.}~\bibnamefont {Rieu}}, \bibinfo {author}
  {\bibfnamefont {G.}~\bibnamefont {Balestra}}, \ and\ \bibinfo {author}
  {\bibfnamefont {F.}~\bibnamefont {Gallaire}},\ }\bibfield  {title} {\enquote
  {\bibinfo {title} {{R}ayleigh-{T}aylor instability under an inclined
  plane},}\ }\href {\doibase 10.1063/1.4927857} {\bibfield  {journal} {\bibinfo
   {journal} {Physics of Fluids}\ }\textbf {\bibinfo {volume} {27}},\ \bibinfo
  {pages} {084107} (\bibinfo {year} {2015})}\BibitemShut {NoStop}%
\bibitem [{\citenamefont {Rohlfs}\ \emph {et~al.}(2017)\citenamefont {Rohlfs},
  \citenamefont {Pischke},\ and\ \citenamefont {Scheid}}]{Rohlfs}%
  \BibitemOpen
  \bibfield  {author} {\bibinfo {author} {\bibfnamefont {W.}~\bibnamefont
  {Rohlfs}}, \bibinfo {author} {\bibfnamefont {P.}~\bibnamefont {Pischke}}, \
  and\ \bibinfo {author} {\bibfnamefont {B.}~\bibnamefont {Scheid}},\
  }\bibfield  {title} {\enquote {\bibinfo {title} {Hydrodynamic waves in films
  flowing under an inclined plane},}\ }\href@noop {} {\bibfield  {journal}
  {\bibinfo  {journal} {Phys. Rev. Fluids}\ }\textbf {\bibinfo {volume} {2}},\
  \bibinfo {pages} {044003} (\bibinfo {year} {2017})}\BibitemShut {NoStop}%
\bibitem [{\citenamefont {Benney}(1966)}]{Benney}%
  \BibitemOpen
  \bibfield  {author} {\bibinfo {author} {\bibfnamefont {D.~J.}\ \bibnamefont
  {Benney}},\ }\bibfield  {title} {\enquote {\bibinfo {title} {Long waves on
  liquid films},}\ }\href@noop {} {\bibfield  {journal} {\bibinfo  {journal}
  {J. Math. Phys.}\ }\textbf {\bibinfo {volume} {45}},\ \bibinfo {pages}
  {150--155} (\bibinfo {year} {1966})}\BibitemShut {NoStop}%
\bibitem [{\citenamefont {Gjevik}(1970)}]{gjevik1970occurrence}%
  \BibitemOpen
  \bibfield  {author} {\bibinfo {author} {\bibfnamefont {B.}~\bibnamefont
  {Gjevik}},\ }\bibfield  {title} {\enquote {\bibinfo {title} {Occurrence of
  finite-amplitude surface waves on falling liquid films},}\ }\href@noop {}
  {\bibfield  {journal} {\bibinfo  {journal} {The Physics of Fluids}\ }\textbf
  {\bibinfo {volume} {13}},\ \bibinfo {pages} {1918--1925} (\bibinfo {year}
  {1970})}\BibitemShut {NoStop}%
\bibitem [{\citenamefont {Pumir}\ \emph {et~al.}(1983)\citenamefont {Pumir},
  \citenamefont {Manneville},\ and\ \citenamefont
  {Pomeau}}]{pumir1983solitary}%
  \BibitemOpen
  \bibfield  {author} {\bibinfo {author} {\bibfnamefont {A.}~\bibnamefont
  {Pumir}}, \bibinfo {author} {\bibfnamefont {P.}~\bibnamefont {Manneville}}, \
  and\ \bibinfo {author} {\bibfnamefont {Y.}~\bibnamefont {Pomeau}},\
  }\bibfield  {title} {\enquote {\bibinfo {title} {On solitary waves running
  down an inclined plane},}\ }\href@noop {} {\bibfield  {journal} {\bibinfo
  {journal} {J. Fluid Mech.}\ }\textbf {\bibinfo {volume} {135}},\ \bibinfo
  {pages} {27--50} (\bibinfo {year} {1983})}\BibitemShut {NoStop}%
\bibitem [{\citenamefont {Rosenau}\ \emph {et~al.}(1992)\citenamefont
  {Rosenau}, \citenamefont {Oron},\ and\ \citenamefont
  {Hyman}}]{rosenau1992bounded}%
  \BibitemOpen
  \bibfield  {author} {\bibinfo {author} {\bibfnamefont {P.}~\bibnamefont
  {Rosenau}}, \bibinfo {author} {\bibfnamefont {A.}~\bibnamefont {Oron}}, \
  and\ \bibinfo {author} {\bibfnamefont {J.~M.}\ \bibnamefont {Hyman}},\
  }\bibfield  {title} {\enquote {\bibinfo {title} {Bounded and unbounded
  patterns of the {B}enney equation},}\ }\href@noop {} {\bibfield  {journal}
  {\bibinfo  {journal} {Phys. Fluid. Fluid Dynam.}\ }\textbf {\bibinfo {volume}
  {4}},\ \bibinfo {pages} {1102--1104} (\bibinfo {year} {1992})}\BibitemShut
  {NoStop}%
\bibitem [{\citenamefont {Salamon}\ \emph {et~al.}(1994)\citenamefont
  {Salamon}, \citenamefont {Armstrong},\ and\ \citenamefont {Brown}}]{Salamon}%
  \BibitemOpen
  \bibfield  {author} {\bibinfo {author} {\bibfnamefont {T.~R.}\ \bibnamefont
  {Salamon}}, \bibinfo {author} {\bibfnamefont {R.~C.}\ \bibnamefont
  {Armstrong}}, \ and\ \bibinfo {author} {\bibfnamefont {R.~A.}\ \bibnamefont
  {Brown}},\ }\bibfield  {title} {\enquote {\bibinfo {title} {Traveling waves
  on vertical films: Numerical analysis using the finite element method},}\
  }\href {\doibase 10.1063/1.868222} {\bibfield  {journal} {\bibinfo  {journal}
  {Physics of Fluids}\ }\textbf {\bibinfo {volume} {6}},\ \bibinfo {pages}
  {2202--2220} (\bibinfo {year} {1994})}\BibitemShut {NoStop}%
\bibitem [{\citenamefont {Oron}\ and\ \citenamefont
  {Gottlieb}(2002)}]{oron2002nonlinear}%
  \BibitemOpen
  \bibfield  {author} {\bibinfo {author} {\bibfnamefont {Alexander}\
  \bibnamefont {Oron}}\ and\ \bibinfo {author} {\bibfnamefont {O}~\bibnamefont
  {Gottlieb}},\ }\bibfield  {title} {\enquote {\bibinfo {title} {Nonlinear
  dynamics of temporally excited falling liquid films},}\ }\href@noop {}
  {\bibfield  {journal} {\bibinfo  {journal} {Physics of Fluids}\ }\textbf
  {\bibinfo {volume} {14}},\ \bibinfo {pages} {2622--2636} (\bibinfo {year}
  {2002})}\BibitemShut {NoStop}%
\bibitem [{\citenamefont {Gottlieb}\ and\ \citenamefont
  {Oron}(2004)}]{gottlieb2004stability}%
  \BibitemOpen
  \bibfield  {author} {\bibinfo {author} {\bibfnamefont {O.}~\bibnamefont
  {Gottlieb}}\ and\ \bibinfo {author} {\bibfnamefont {A.}~\bibnamefont
  {Oron}},\ }\bibfield  {title} {\enquote {\bibinfo {title} {Stability and
  bifurcations of parametrically excited thin liquid films},}\ }\href@noop {}
  {\bibfield  {journal} {\bibinfo  {journal} {International Journal of
  Bifurcation and Chaos}\ }\textbf {\bibinfo {volume} {14}},\ \bibinfo {pages}
  {4117--4141} (\bibinfo {year} {2004})}\BibitemShut {NoStop}%
\bibitem [{\citenamefont {Oron}\ and\ \citenamefont
  {Gottlieb}(2004)}]{Oron2004}%
  \BibitemOpen
  \bibfield  {author} {\bibinfo {author} {\bibfnamefont {A.}~\bibnamefont
  {Oron}}\ and\ \bibinfo {author} {\bibfnamefont {O.}~\bibnamefont
  {Gottlieb}},\ }\bibfield  {title} {\enquote {\bibinfo {title} {Subcritical
  and supercritical bifurcations of the first- and second-order benney
  equations},}\ }\href {\doibase 10.1007/s10665-004-2760-7} {\bibfield
  {journal} {\bibinfo  {journal} {Journal of Engineering Mathematics}\ }\textbf
  {\bibinfo {volume} {50}},\ \bibinfo {pages} {121--140} (\bibinfo {year}
  {2004})}\BibitemShut {NoStop}%
\bibitem [{\citenamefont {Scheid}\ \emph {et~al.}(2005)\citenamefont {Scheid},
  \citenamefont {Ruyer-Quil}, \citenamefont {Thiele}, \citenamefont {Kabov},
  \citenamefont {Legros},\ and\ \citenamefont {Colinet}}]{scheid2005validity}%
  \BibitemOpen
  \bibfield  {author} {\bibinfo {author} {\bibfnamefont {B.}~\bibnamefont
  {Scheid}}, \bibinfo {author} {\bibfnamefont {C.}~\bibnamefont {Ruyer-Quil}},
  \bibinfo {author} {\bibfnamefont {U.}~\bibnamefont {Thiele}}, \bibinfo
  {author} {\bibfnamefont {O.~A.}\ \bibnamefont {Kabov}}, \bibinfo {author}
  {\bibfnamefont {J.~C.}\ \bibnamefont {Legros}}, \ and\ \bibinfo {author}
  {\bibfnamefont {P.}~\bibnamefont {Colinet}},\ }\bibfield  {title} {\enquote
  {\bibinfo {title} {Validity domain of the benney equation including the
  marangoni effect for closed and open flows},}\ }\href@noop {} {\bibfield
  {journal} {\bibinfo  {journal} {Journal of Fluid Mechanics}\ }\textbf
  {\bibinfo {volume} {527}},\ \bibinfo {pages} {303--335} (\bibinfo {year}
  {2005})}\BibitemShut {NoStop}%
\bibitem [{\citenamefont {Kapitza}(1948{\natexlab{a}})}]{kapitza1948wave1}%
  \BibitemOpen
  \bibfield  {author} {\bibinfo {author} {\bibfnamefont {P.~L.}\ \bibnamefont
  {Kapitza}},\ }\bibfield  {title} {\enquote {\bibinfo {title} {Wave flow of
  thin layers of viscous liquids. part i. free flow},}\ }\href@noop {}
  {\bibfield  {journal} {\bibinfo  {journal} {Zhurnal Eksperimentalnoi i
  Teoreticheskoi Fiziki}\ }\textbf {\bibinfo {volume} {18}},\ \bibinfo {pages}
  {3--18} (\bibinfo {year} {1948}{\natexlab{a}})}\BibitemShut {NoStop}%
\bibitem [{\citenamefont {Kapitza}(1948{\natexlab{b}})}]{kapitza1948wave2}%
  \BibitemOpen
  \bibfield  {author} {\bibinfo {author} {\bibfnamefont {P.~L.}\ \bibnamefont
  {Kapitza}},\ }\bibfield  {title} {\enquote {\bibinfo {title} {Wave flow of
  thin layers of viscous liquids. part ii. fluid flow in the presence of
  continuous gas flow and heat transfer},}\ }\href@noop {} {\bibfield
  {journal} {\bibinfo  {journal} {Zhurnal Eksperimentalnoi i Teoreticheskoi
  Fiziki}\ }\textbf {\bibinfo {volume} {18}},\ \bibinfo {pages} {19--28}
  (\bibinfo {year} {1948}{\natexlab{b}})}\BibitemShut {NoStop}%
\bibitem [{\citenamefont {Shkadov}(1967)}]{shkadov1967wave}%
  \BibitemOpen
  \bibfield  {author} {\bibinfo {author} {\bibfnamefont {V.~Y.}\ \bibnamefont
  {Shkadov}},\ }\bibfield  {title} {\enquote {\bibinfo {title} {Wave flow
  regimes of a thin layer of viscous fluid subject to gravity},}\ }\href@noop
  {} {\bibfield  {journal} {\bibinfo  {journal} {Fluid Dynamics}\ }\textbf
  {\bibinfo {volume} {2}},\ \bibinfo {pages} {29--34} (\bibinfo {year}
  {1967})}\BibitemShut {NoStop}%
\bibitem [{\citenamefont {Ruyer-Quil}\ and\ \citenamefont
  {Manneville}(1998)}]{ruyer1998modeling}%
  \BibitemOpen
  \bibfield  {author} {\bibinfo {author} {\bibfnamefont {C.}~\bibnamefont
  {Ruyer-Quil}}\ and\ \bibinfo {author} {\bibfnamefont {P.}~\bibnamefont
  {Manneville}},\ }\bibfield  {title} {\enquote {\bibinfo {title} {Modeling
  film flows down inclined planes},}\ }\href@noop {} {\bibfield  {journal}
  {\bibinfo  {journal} {The European Physical Journal B-Condensed Matter and
  Complex Systems}\ }\textbf {\bibinfo {volume} {6}},\ \bibinfo {pages}
  {277--292} (\bibinfo {year} {1998})}\BibitemShut {NoStop}%
\bibitem [{\citenamefont {Ruyer-Quil}\ and\ \citenamefont
  {Manneville}(2000)}]{ruyer2000improved}%
  \BibitemOpen
  \bibfield  {author} {\bibinfo {author} {\bibfnamefont {C.}~\bibnamefont
  {Ruyer-Quil}}\ and\ \bibinfo {author} {\bibfnamefont {P.}~\bibnamefont
  {Manneville}},\ }\bibfield  {title} {\enquote {\bibinfo {title} {Improved
  modeling of flows down inclined planes},}\ }\href@noop {} {\bibfield
  {journal} {\bibinfo  {journal} {The European Physical Journal B-Condensed
  Matter and Complex Systems}\ }\textbf {\bibinfo {volume} {15}},\ \bibinfo
  {pages} {357--369} (\bibinfo {year} {2000})}\BibitemShut {NoStop}%
\bibitem [{\citenamefont {Ruyer-Quil}\ and\ \citenamefont
  {Manneville}(2002)}]{ruyer2002further}%
  \BibitemOpen
  \bibfield  {author} {\bibinfo {author} {\bibfnamefont {C.}~\bibnamefont
  {Ruyer-Quil}}\ and\ \bibinfo {author} {\bibfnamefont {P.}~\bibnamefont
  {Manneville}},\ }\bibfield  {title} {\enquote {\bibinfo {title} {Further
  accuracy and convergence results on the modeling of flows down inclined
  planes by weighted-residual approximations},}\ }\href@noop {} {\bibfield
  {journal} {\bibinfo  {journal} {Physics of Fluids}\ }\textbf {\bibinfo
  {volume} {14}},\ \bibinfo {pages} {170--183} (\bibinfo {year}
  {2002})}\BibitemShut {NoStop}%
\bibitem [{\citenamefont {Denner}\ \emph {et~al.}(2018)\citenamefont {Denner},
  \citenamefont {Charogiannis}, \citenamefont {Pradas}, \citenamefont
  {Markides}, \citenamefont {van Wachem},\ and\ \citenamefont
  {Kalliadasis}}]{denner2018solitary}%
  \BibitemOpen
  \bibfield  {author} {\bibinfo {author} {\bibfnamefont {F.}~\bibnamefont
  {Denner}}, \bibinfo {author} {\bibfnamefont {A.}~\bibnamefont
  {Charogiannis}}, \bibinfo {author} {\bibfnamefont {M.}~\bibnamefont
  {Pradas}}, \bibinfo {author} {\bibfnamefont {C.~N.}\ \bibnamefont
  {Markides}}, \bibinfo {author} {\bibfnamefont {B.~G.~M.}\ \bibnamefont {van
  Wachem}}, \ and\ \bibinfo {author} {\bibfnamefont {S.}~\bibnamefont
  {Kalliadasis}},\ }\bibfield  {title} {\enquote {\bibinfo {title} {Solitary
  waves on falling liquid films in the inertia-dominated regime},}\ }\href@noop
  {} {\bibfield  {journal} {\bibinfo  {journal} {Journal of Fluid Mechanics}\
  }\textbf {\bibinfo {volume} {837}},\ \bibinfo {pages} {491--519} (\bibinfo
  {year} {2018})}\BibitemShut {NoStop}%
\bibitem [{\citenamefont {Kalliadasis}\ \emph {et~al.}(2012)\citenamefont
  {Kalliadasis}, \citenamefont {Ruyer-Quil}, \citenamefont {Scheid},\ and\
  \citenamefont {Velarde}}]{kalliadasis2011falling}%
  \BibitemOpen
  \bibfield  {author} {\bibinfo {author} {\bibfnamefont {S.}~\bibnamefont
  {Kalliadasis}}, \bibinfo {author} {\bibfnamefont {C.}~\bibnamefont
  {Ruyer-Quil}}, \bibinfo {author} {\bibfnamefont {B.}~\bibnamefont {Scheid}},
  \ and\ \bibinfo {author} {\bibfnamefont {M.~G.}\ \bibnamefont {Velarde}},\
  }\href@noop {} {\emph {\bibinfo {title} {Falling liquid films}}},\ Vol.\
  \bibinfo {volume} {176}\ (\bibinfo  {publisher} {Springer Science \& Business
  Media},\ \bibinfo {year} {2012})\BibitemShut {NoStop}%
\bibitem [{\citenamefont {Scheid}\ \emph {et~al.}(2016)\citenamefont {Scheid},
  \citenamefont {Kofman},\ and\ \citenamefont {Rohlfs}}]{ScheidKofman1}%
  \BibitemOpen
  \bibfield  {author} {\bibinfo {author} {\bibfnamefont {B.}~\bibnamefont
  {Scheid}}, \bibinfo {author} {\bibfnamefont {N.}~\bibnamefont {Kofman}}, \
  and\ \bibinfo {author} {\bibfnamefont {W.}~\bibnamefont {Rohlfs}},\
  }\bibfield  {title} {\enquote {\bibinfo {title} {Critical inclination for
  absolute/convective instability transition in inverted falling films},}\
  }\href {\doibase 10.1063/1.4946827} {\bibfield  {journal} {\bibinfo
  {journal} {Physics of Fluids}\ }\textbf {\bibinfo {volume} {28}},\ \bibinfo
  {pages} {044107} (\bibinfo {year} {2016})}\BibitemShut {NoStop}%
\bibitem [{\citenamefont {Kofman}\ \emph {et~al.}(2018)\citenamefont {Kofman},
  \citenamefont {Rohlfs}, \citenamefont {Gallaire}, \citenamefont {Scheid},\
  and\ \citenamefont {Ruyer-Quil}}]{kofman2018prediction}%
  \BibitemOpen
  \bibfield  {author} {\bibinfo {author} {\bibfnamefont {N.}~\bibnamefont
  {Kofman}}, \bibinfo {author} {\bibfnamefont {W.}~\bibnamefont {Rohlfs}},
  \bibinfo {author} {\bibfnamefont {F.}~\bibnamefont {Gallaire}}, \bibinfo
  {author} {\bibfnamefont {B.}~\bibnamefont {Scheid}}, \ and\ \bibinfo {author}
  {\bibfnamefont {C.}~\bibnamefont {Ruyer-Quil}},\ }\bibfield  {title}
  {\enquote {\bibinfo {title} {Prediction of two-dimensional dripping onset of
  a liquid film under an inclined plane},}\ }\href@noop {} {\bibfield
  {journal} {\bibinfo  {journal} {International Journal of Multiphase Flow}\
  }\textbf {\bibinfo {volume} {104}},\ \bibinfo {pages} {286--293} (\bibinfo
  {year} {2018})}\BibitemShut {NoStop}%
\bibitem [{\citenamefont {Lin}\ \emph {et~al.}(2012)\citenamefont {Lin},
  \citenamefont {Kondic},\ and\ \citenamefont {Filippov}}]{lin2012thin}%
  \BibitemOpen
  \bibfield  {author} {\bibinfo {author} {\bibfnamefont {T.-S.}\ \bibnamefont
  {Lin}}, \bibinfo {author} {\bibfnamefont {L.}~\bibnamefont {Kondic}}, \ and\
  \bibinfo {author} {\bibfnamefont {A.}~\bibnamefont {Filippov}},\ }\bibfield
  {title} {\enquote {\bibinfo {title} {Thin films flowing down inverted
  substrates: Three-dimensional flow},}\ }\href@noop {} {\bibfield  {journal}
  {\bibinfo  {journal} {Physics of Fluids}\ }\textbf {\bibinfo {volume} {24}},\
  \bibinfo {pages} {022105} (\bibinfo {year} {2012})}\BibitemShut {NoStop}%
\bibitem [{\citenamefont {Conroy}\ \emph {et~al.}(2019)\citenamefont {Conroy},
  \citenamefont {Esp\'{\i}n}, \citenamefont {Matar},\ and\ \citenamefont
  {Kumar}}]{PhysRevFluids.4.034001}%
  \BibitemOpen
  \bibfield  {author} {\bibinfo {author} {\bibfnamefont {D.~T.}\ \bibnamefont
  {Conroy}}, \bibinfo {author} {\bibfnamefont {L.}~\bibnamefont {Esp\'{\i}n}},
  \bibinfo {author} {\bibfnamefont {O.~K.}\ \bibnamefont {Matar}}, \ and\
  \bibinfo {author} {\bibfnamefont {S.}~\bibnamefont {Kumar}},\ }\bibfield
  {title} {\enquote {\bibinfo {title} {Thermocapillary and electrohydrodynamic
  effects on the stability of dynamic contact lines},}\ }\href {\doibase
  10.1103/PhysRevFluids.4.034001} {\bibfield  {journal} {\bibinfo  {journal}
  {Phys. Rev. Fluids}\ }\textbf {\bibinfo {volume} {4}},\ \bibinfo {pages}
  {034001} (\bibinfo {year} {2019})}\BibitemShut {NoStop}%
\bibitem [{\citenamefont {Cimpeanu}\ \emph {et~al.}(2014)\citenamefont
  {Cimpeanu}, \citenamefont {Papageorgiou},\ and\ \citenamefont
  {Petropoulos}}]{cimpeanu2014control}%
  \BibitemOpen
  \bibfield  {author} {\bibinfo {author} {\bibfnamefont {R.}~\bibnamefont
  {Cimpeanu}}, \bibinfo {author} {\bibfnamefont {D.~T.}\ \bibnamefont
  {Papageorgiou}}, \ and\ \bibinfo {author} {\bibfnamefont {P.~G.}\
  \bibnamefont {Petropoulos}},\ }\bibfield  {title} {\enquote {\bibinfo {title}
  {On the control and suppression of the rayleigh-taylor instability using
  electric fields},}\ }\href@noop {} {\bibfield  {journal} {\bibinfo  {journal}
  {Physics of Fluids}\ }\textbf {\bibinfo {volume} {26}},\ \bibinfo {pages}
  {022105} (\bibinfo {year} {2014})}\BibitemShut {NoStop}%
\bibitem [{\citenamefont {Anderson}\ \emph {et~al.}(2017)\citenamefont
  {Anderson}, \citenamefont {Cimpeanu}, \citenamefont {Papageorgiou},\ and\
  \citenamefont {Petropoulos}}]{RaduAnder1}%
  \BibitemOpen
  \bibfield  {author} {\bibinfo {author} {\bibfnamefont {T.~G.}\ \bibnamefont
  {Anderson}}, \bibinfo {author} {\bibfnamefont {R.}~\bibnamefont {Cimpeanu}},
  \bibinfo {author} {\bibfnamefont {D.~T.}\ \bibnamefont {Papageorgiou}}, \
  and\ \bibinfo {author} {\bibfnamefont {P.~G.}\ \bibnamefont {Petropoulos}},\
  }\bibfield  {title} {\enquote {\bibinfo {title} {Electric field stabilization
  of viscous liquid layers coating the underside of a surface},}\ }\href
  {\doibase 10.1103/PhysRevFluids.2.054001} {\bibfield  {journal} {\bibinfo
  {journal} {Phys. Rev. Fluids}\ }\textbf {\bibinfo {volume} {2}},\ \bibinfo
  {pages} {054001} (\bibinfo {year} {2017})}\BibitemShut {NoStop}%
\bibitem [{\citenamefont {Cimpeanu}\ and\ \citenamefont
  {Papageorgiou}(2015)}]{cimpeanu2015electrostatically}%
  \BibitemOpen
  \bibfield  {author} {\bibinfo {author} {\bibfnamefont {R.}~\bibnamefont
  {Cimpeanu}}\ and\ \bibinfo {author} {\bibfnamefont {D.~T.}\ \bibnamefont
  {Papageorgiou}},\ }\bibfield  {title} {\enquote {\bibinfo {title}
  {Electrostatically induced mixing in confined stratified multi-fluid
  systems},}\ }\href@noop {} {\bibfield  {journal} {\bibinfo  {journal}
  {International Journal of Multiphase Flow}\ }\textbf {\bibinfo {volume}
  {75}},\ \bibinfo {pages} {194--204} (\bibinfo {year} {2015})}\BibitemShut
  {NoStop}%
\bibitem [{\citenamefont {Kord}\ and\ \citenamefont
  {Capecelatro}(2019)}]{Kord}%
  \BibitemOpen
  \bibfield  {author} {\bibinfo {author} {\bibfnamefont {A.}~\bibnamefont
  {Kord}}\ and\ \bibinfo {author} {\bibfnamefont {J.}~\bibnamefont
  {Capecelatro}},\ }\bibfield  {title} {\enquote {\bibinfo {title} {Optimal
  perturbations for controlling the growth of a rayleighÐtaylor instability},}\
  }\href@noop {} {\bibfield  {journal} {\bibinfo  {journal} {J. Fluid Mech.}\
  }\textbf {\bibinfo {volume} {876}},\ \bibinfo {pages} {150--185} (\bibinfo
  {year} {2019})}\BibitemShut {NoStop}%
\bibitem [{\citenamefont {Blyth}\ \emph {et~al.}(2018)\citenamefont {Blyth},
  \citenamefont {Tseluiko}, \citenamefont {Lin},\ and\ \citenamefont
  {Kalliadasis}}]{blyth2018two}%
  \BibitemOpen
  \bibfield  {author} {\bibinfo {author} {\bibfnamefont {M.~G.}\ \bibnamefont
  {Blyth}}, \bibinfo {author} {\bibfnamefont {D.}~\bibnamefont {Tseluiko}},
  \bibinfo {author} {\bibfnamefont {T.-S.}\ \bibnamefont {Lin}}, \ and\
  \bibinfo {author} {\bibfnamefont {S.}~\bibnamefont {Kalliadasis}},\
  }\bibfield  {title} {\enquote {\bibinfo {title} {Two-dimensional pulse
  dynamics and the formation of bound states on electrified falling films},}\
  }\href@noop {} {\bibfield  {journal} {\bibinfo  {journal} {Journal of Fluid
  Mechanics}\ }\textbf {\bibinfo {volume} {855}},\ \bibinfo {pages} {210--235}
  (\bibinfo {year} {2018})}\BibitemShut {NoStop}%
\bibitem [{\citenamefont {Papageorgiou}(2019)}]{Papageorgiou2019}%
  \BibitemOpen
  \bibfield  {author} {\bibinfo {author} {\bibfnamefont {D.~T.}\ \bibnamefont
  {Papageorgiou}},\ }\bibfield  {title} {\enquote {\bibinfo {title} {Film flows
  in the presence of electric fields},}\ }\href@noop {} {\bibfield  {journal}
  {\bibinfo  {journal} {Annual Review of Fluid Mechanics}\ }\textbf {\bibinfo
  {volume} {51}},\ \bibinfo {pages} {155--187} (\bibinfo {year}
  {2019})}\BibitemShut {NoStop}%
\bibitem [{\citenamefont {Papageorgiou}\ and\ \citenamefont
  {Petropoulos}(2004)}]{papageorgiou2004generation}%
  \BibitemOpen
  \bibfield  {author} {\bibinfo {author} {\bibfnamefont {D.~T.}\ \bibnamefont
  {Papageorgiou}}\ and\ \bibinfo {author} {\bibfnamefont {P.~G.}\ \bibnamefont
  {Petropoulos}},\ }\bibfield  {title} {\enquote {\bibinfo {title} {Generation
  of interfacial instabilities in charged electrified viscous liquid films},}\
  }\href@noop {} {\bibfield  {journal} {\bibinfo  {journal} {J. Eng. Math.}\
  }\textbf {\bibinfo {volume} {50}},\ \bibinfo {pages} {223--240} (\bibinfo
  {year} {2004})}\BibitemShut {NoStop}%
\bibitem [{\citenamefont {Melcher}\ and\ \citenamefont
  {Taylor}(1969)}]{MelcherTaylor}%
  \BibitemOpen
  \bibfield  {author} {\bibinfo {author} {\bibfnamefont {J.~R.}\ \bibnamefont
  {Melcher}}\ and\ \bibinfo {author} {\bibfnamefont {G.~I.}\ \bibnamefont
  {Taylor}},\ }\bibfield  {title} {\enquote {\bibinfo {title}
  {Electrohydrodynamics: A review of the role of interfacial shear stresses},}\
  }\href {\doibase 10.1146/annurev.fl.01.010169.000551} {\bibfield  {journal}
  {\bibinfo  {journal} {Annu. Rev. Fluid Mech.}\ }\textbf {\bibinfo {volume}
  {1}},\ \bibinfo {pages} {111--146} (\bibinfo {year} {1969})}\BibitemShut
  {NoStop}%
\bibitem [{\citenamefont {Saville}(1997)}]{doi:10.1146/annurev.fluid.29.1.27}%
  \BibitemOpen
  \bibfield  {author} {\bibinfo {author} {\bibfnamefont {D.~A.}\ \bibnamefont
  {Saville}},\ }\bibfield  {title} {\enquote {\bibinfo {title}
  {Electrohydrodynamics: {T}he {T}aylor{--}{M}elcher leaky dielectric model},}\
  }\href {\doibase 10.1146/annurev.fluid.29.1.27} {\bibfield  {journal}
  {\bibinfo  {journal} {Annu. Rev. Fluid Mech.}\ }\textbf {\bibinfo {volume}
  {29}},\ \bibinfo {pages} {27--64} (\bibinfo {year} {1997})}\BibitemShut
  {NoStop}%
\bibitem [{\citenamefont {Pease}\ and\ \citenamefont
  {Russel}(2002)}]{pease2002linear}%
  \BibitemOpen
  \bibfield  {author} {\bibinfo {author} {\bibfnamefont {L.~F.}\ \bibnamefont
  {Pease}}\ and\ \bibinfo {author} {\bibfnamefont {W.~B.}\ \bibnamefont
  {Russel}},\ }\bibfield  {title} {\enquote {\bibinfo {title} {Linear stability
  analysis of thin leaky dielectric films subjected to electric fields},}\
  }\href@noop {} {\bibfield  {journal} {\bibinfo  {journal} {J. Non-Newtonian
  Fluid Mech.}\ }\textbf {\bibinfo {volume} {102}},\ \bibinfo {pages}
  {233--250} (\bibinfo {year} {2002})}\BibitemShut {NoStop}%
\bibitem [{\citenamefont {Tseluiko}\ and\ \citenamefont
  {Papageorgiou}(2006)}]{tseluiko2006wave}%
  \BibitemOpen
  \bibfield  {author} {\bibinfo {author} {\bibfnamefont {D.}~\bibnamefont
  {Tseluiko}}\ and\ \bibinfo {author} {\bibfnamefont {D.~T.}\ \bibnamefont
  {Papageorgiou}},\ }\bibfield  {title} {\enquote {\bibinfo {title} {Wave
  evolution on electrified falling films},}\ }\href@noop {} {\bibfield
  {journal} {\bibinfo  {journal} {J. Fluid Mech.}\ }\textbf {\bibinfo {volume}
  {556}},\ \bibinfo {pages} {361--386} (\bibinfo {year} {2006})}\BibitemShut
  {NoStop}%
\bibitem [{\citenamefont {Tomlin}\ \emph {et~al.}(2017)\citenamefont {Tomlin},
  \citenamefont {Papageorgiou},\ and\ \citenamefont
  {Pavliotis}}]{tomlin_papageorgiou_pavliotis_2017}%
  \BibitemOpen
  \bibfield  {author} {\bibinfo {author} {\bibfnamefont {R.~J.}\ \bibnamefont
  {Tomlin}}, \bibinfo {author} {\bibfnamefont {D.~T.}\ \bibnamefont
  {Papageorgiou}}, \ and\ \bibinfo {author} {\bibfnamefont {G.~A.}\
  \bibnamefont {Pavliotis}},\ }\bibfield  {title} {\enquote {\bibinfo {title}
  {Three-dimensional wave evolution on electrified falling films},}\ }\href
  {\doibase 10.1017/jfm.2017.250} {\bibfield  {journal} {\bibinfo  {journal}
  {Journal of Fluid Mechanics}\ }\textbf {\bibinfo {volume} {822}},\ \bibinfo
  {pages} {54–79} (\bibinfo {year} {2017})}\BibitemShut {NoStop}%
\bibitem [{\citenamefont {Huerre}\ and\ \citenamefont
  {Monkewitz}(1990)}]{huerre1990local}%
  \BibitemOpen
  \bibfield  {author} {\bibinfo {author} {\bibfnamefont {P.}~\bibnamefont
  {Huerre}}\ and\ \bibinfo {author} {\bibfnamefont {P.~A.}\ \bibnamefont
  {Monkewitz}},\ }\bibfield  {title} {\enquote {\bibinfo {title} {Local and
  global instabilities in spatially developing flows},}\ }\href@noop {}
  {\bibfield  {journal} {\bibinfo  {journal} {Annual review of fluid
  mechanics}\ }\textbf {\bibinfo {volume} {22}},\ \bibinfo {pages} {473--537}
  (\bibinfo {year} {1990})}\BibitemShut {NoStop}%
\bibitem [{\citenamefont {Fokas}\ and\ \citenamefont
  {Papageorgiou}(2005)}]{fokas2005absolute}%
  \BibitemOpen
  \bibfield  {author} {\bibinfo {author} {\bibfnamefont {A.~S.}\ \bibnamefont
  {Fokas}}\ and\ \bibinfo {author} {\bibfnamefont {D.~T.}\ \bibnamefont
  {Papageorgiou}},\ }\bibfield  {title} {\enquote {\bibinfo {title} {Absolute
  and convective instability for evolution pdes on the half-line},}\
  }\href@noop {} {\bibfield  {journal} {\bibinfo  {journal} {Studies in Applied
  Mathematics}\ }\textbf {\bibinfo {volume} {114}},\ \bibinfo {pages} {95--114}
  (\bibinfo {year} {2005})}\BibitemShut {NoStop}%
\bibitem [{\citenamefont {Doedel}\ \emph {et~al.}(2007)\citenamefont {Doedel},
  \citenamefont {Fairgrieve}, \citenamefont {Sandstede}, \citenamefont
  {Champneys}, \citenamefont {Kuznetsov},\ and\ \citenamefont
  {Wang}}]{doedel2007auto}%
  \BibitemOpen
  \bibfield  {author} {\bibinfo {author} {\bibfnamefont {E.~J.}\ \bibnamefont
  {Doedel}}, \bibinfo {author} {\bibfnamefont {T.~F.}\ \bibnamefont
  {Fairgrieve}}, \bibinfo {author} {\bibfnamefont {B.}~\bibnamefont
  {Sandstede}}, \bibinfo {author} {\bibfnamefont {A.~R.}\ \bibnamefont
  {Champneys}}, \bibinfo {author} {\bibfnamefont {Y.~A.}\ \bibnamefont
  {Kuznetsov}}, \ and\ \bibinfo {author} {\bibfnamefont {X.}~\bibnamefont
  {Wang}},\ }\bibfield  {title} {\enquote {\bibinfo {title} {Auto-07p:
  Continuation and bifurcation software for ordinary differential equations},}\
  }\href@noop {} {\  (\bibinfo {year} {2007})}\BibitemShut {NoStop}%
\bibitem [{\citenamefont {Burcham}\ and\ \citenamefont
  {Saville}(2000)}]{burcham2000electrohydrodynamic}%
  \BibitemOpen
  \bibfield  {author} {\bibinfo {author} {\bibfnamefont {C.~L.}\ \bibnamefont
  {Burcham}}\ and\ \bibinfo {author} {\bibfnamefont {D.~A.}\ \bibnamefont
  {Saville}},\ }\bibfield  {title} {\enquote {\bibinfo {title} {The
  electrohydrodynamic stability of a liquid bridge: microgravity experiments on
  a bridge suspended in a dielectric gas},}\ }\href@noop {} {\bibfield
  {journal} {\bibinfo  {journal} {J. Fluid Mech.}\ }\textbf {\bibinfo {volume}
  {405}},\ \bibinfo {pages} {37--56} (\bibinfo {year} {2000})}\BibitemShut
  {NoStop}%
\bibitem [{\citenamefont {Uguz}\ \emph {et~al.}(2008)\citenamefont {Uguz},
  \citenamefont {Ozen},\ and\ \citenamefont {Aubry}}]{uguz2008electric}%
  \BibitemOpen
  \bibfield  {author} {\bibinfo {author} {\bibfnamefont {A.~K.}\ \bibnamefont
  {Uguz}}, \bibinfo {author} {\bibfnamefont {O.}~\bibnamefont {Ozen}}, \ and\
  \bibinfo {author} {\bibfnamefont {N.}~\bibnamefont {Aubry}},\ }\bibfield
  {title} {\enquote {\bibinfo {title} {Electric field effect on a two-fluid
  interface instability in channel flow for fast electric times},}\ }\href@noop
  {} {\bibfield  {journal} {\bibinfo  {journal} {Physics of Fluids}\ }\textbf
  {\bibinfo {volume} {20}},\ \bibinfo {pages} {031702} (\bibinfo {year}
  {2008})}\BibitemShut {NoStop}%
\bibitem [{\citenamefont {Craster}\ and\ \citenamefont
  {Matar}(2005)}]{:/content/aip/journal/pof2/17/3/10.1063/1.1852459}%
  \BibitemOpen
  \bibfield  {author} {\bibinfo {author} {\bibfnamefont {R.~V.}\ \bibnamefont
  {Craster}}\ and\ \bibinfo {author} {\bibfnamefont {O.~K.}\ \bibnamefont
  {Matar}},\ }\bibfield  {title} {\enquote {\bibinfo {title} {Electrically
  induced pattern formation in thin leaky dielectric films},}\ }\href {\doibase
  http://dx.doi.org/10.1063/1.1852459} {\bibfield  {journal} {\bibinfo
  {journal} {Phys. Fluids}\ }\textbf {\bibinfo {volume} {17}} (\bibinfo {year}
  {2005}),\ http://dx.doi.org/10.1063/1.1852459}\BibitemShut {NoStop}%
\bibitem [{\citenamefont {Tipler}(1987)}]{tipler1987college}%
  \BibitemOpen
  \bibfield  {author} {\bibinfo {author} {\bibfnamefont {P.~A.}\ \bibnamefont
  {Tipler}},\ }\href@noop {} {\emph {\bibinfo {title} {College Physics}}}\
  (\bibinfo  {publisher} {Worth Publishers},\ \bibinfo {year}
  {1987})\BibitemShut {NoStop}%
\bibitem [{\citenamefont {Popinet}(2003)}]{popinet1}%
  \BibitemOpen
  \bibfield  {author} {\bibinfo {author} {\bibfnamefont {S.}~\bibnamefont
  {Popinet}},\ }\bibfield  {title} {\enquote {\bibinfo {title} {Gerris: A
  tree-based adaptive solver for the incompressible {Euler} equations in
  complex geometries},}\ }\href {\doibase 10.1016/S0021-9991(03)00298-5}
  {\bibfield  {journal} {\bibinfo  {journal} {J. Comput. Phys.}\ }\textbf
  {\bibinfo {volume} {190}},\ \bibinfo {pages} {572--600} (\bibinfo {year}
  {2003})}\BibitemShut {NoStop}%
\bibitem [{\citenamefont {Popinet}(2009)}]{popinet2}%
  \BibitemOpen
  \bibfield  {author} {\bibinfo {author} {\bibfnamefont {S.}~\bibnamefont
  {Popinet}},\ }\bibfield  {title} {\enquote {\bibinfo {title} {An accurate
  adaptive solver for surface-tension-driven interfacial flows},}\ }\href
  {\doibase 10.1016/j.jcp.2009.04.042} {\bibfield  {journal} {\bibinfo
  {journal} {J. Comput. Phys.}\ }\textbf {\bibinfo {volume} {228}},\ \bibinfo
  {pages} {5838--5866} (\bibinfo {year} {2009})}\BibitemShut {NoStop}%
\bibitem [{\citenamefont {L\'opez-Herrera}\ \emph {et~al.}(2011)\citenamefont
  {L\'opez-Herrera}, \citenamefont {Popinet},\ and\ \citenamefont
  {Herrada}}]{lopez1}%
  \BibitemOpen
  \bibfield  {author} {\bibinfo {author} {\bibfnamefont {J.~M.}\ \bibnamefont
  {L\'opez-Herrera}}, \bibinfo {author} {\bibfnamefont {S.}~\bibnamefont
  {Popinet}}, \ and\ \bibinfo {author} {\bibfnamefont {M.~A.}\ \bibnamefont
  {Herrada}},\ }\bibfield  {title} {\enquote {\bibinfo {title} {A
  charge-conservative approach for simulating electrohydrodynamic two-phase
  flows using volume-of-fluid},}\ }\href@noop {} {\bibfield  {journal}
  {\bibinfo  {journal} {J. Comput. Phys.}\ }\textbf {\bibinfo {volume} {230}},\
  \bibinfo {pages} {1939--1955} (\bibinfo {year} {2011})}\BibitemShut {NoStop}%
\bibitem [{\citenamefont {Samanta}\ \emph {et~al.}(2011)\citenamefont
  {Samanta}, \citenamefont {Ruyer-Quil},\ and\ \citenamefont
  {Goyeau}}]{samanta2011falling}%
  \BibitemOpen
  \bibfield  {author} {\bibinfo {author} {\bibfnamefont {A.}~\bibnamefont
  {Samanta}}, \bibinfo {author} {\bibfnamefont {C.}~\bibnamefont {Ruyer-Quil}},
  \ and\ \bibinfo {author} {\bibfnamefont {B.}~\bibnamefont {Goyeau}},\
  }\bibfield  {title} {\enquote {\bibinfo {title} {A falling film down a
  slippery inclined plane},}\ }\href@noop {} {\bibfield  {journal} {\bibinfo
  {journal} {Journal of Fluid Mechanics}\ }\textbf {\bibinfo {volume} {684}},\
  \bibinfo {pages} {353--383} (\bibinfo {year} {2011})}\BibitemShut {NoStop}%
\bibitem [{\citenamefont {Avitabile}\ \emph {et~al.}(2017)\citenamefont
  {Avitabile}, \citenamefont {Desroches}, \citenamefont {Knobloch},\ and\
  \citenamefont {Krupa}}]{avitabile2017ducks}%
  \BibitemOpen
  \bibfield  {author} {\bibinfo {author} {\bibfnamefont {D.}~\bibnamefont
  {Avitabile}}, \bibinfo {author} {\bibfnamefont {M.}~\bibnamefont
  {Desroches}}, \bibinfo {author} {\bibfnamefont {E.}~\bibnamefont {Knobloch}},
  \ and\ \bibinfo {author} {\bibfnamefont {M.}~\bibnamefont {Krupa}},\
  }\bibfield  {title} {\enquote {\bibinfo {title} {Ducks in space: from
  nonlinear absolute instability to noise-sustained structures in a
  pattern-forming system},}\ }\href@noop {} {\bibfield  {journal} {\bibinfo
  {journal} {Proc. R. Soc. A}\ }\textbf {\bibinfo {volume} {473}},\ \bibinfo
  {pages} {20170018} (\bibinfo {year} {2017})}\BibitemShut {NoStop}%
\bibitem [{\citenamefont {Delbende}\ and\ \citenamefont
  {Chomaz}(1998)}]{delbende1998nonlinear}%
  \BibitemOpen
  \bibfield  {author} {\bibinfo {author} {\bibfnamefont {I.}~\bibnamefont
  {Delbende}}\ and\ \bibinfo {author} {\bibfnamefont {J.-M.}\ \bibnamefont
  {Chomaz}},\ }\bibfield  {title} {\enquote {\bibinfo {title} {Nonlinear
  convective/absolute instabilities in parallel two-dimensional wakes},}\
  }\href@noop {} {\bibfield  {journal} {\bibinfo  {journal} {Physics of
  Fluids}\ }\textbf {\bibinfo {volume} {10}},\ \bibinfo {pages} {2724--2736}
  (\bibinfo {year} {1998})}\BibitemShut {NoStop}%
\bibitem [{\citenamefont {Denner}\ \emph {et~al.}(2016)\citenamefont {Denner},
  \citenamefont {Pradas}, \citenamefont {Charogiannis}, \citenamefont
  {Markides}, \citenamefont {van Wachem},\ and\ \citenamefont
  {Kalliadasis}}]{denner2016self}%
  \BibitemOpen
  \bibfield  {author} {\bibinfo {author} {\bibfnamefont {F.}~\bibnamefont
  {Denner}}, \bibinfo {author} {\bibfnamefont {M.}~\bibnamefont {Pradas}},
  \bibinfo {author} {\bibfnamefont {A.}~\bibnamefont {Charogiannis}}, \bibinfo
  {author} {\bibfnamefont {C.~N.}\ \bibnamefont {Markides}}, \bibinfo {author}
  {\bibfnamefont {B.~G.~M.}\ \bibnamefont {van Wachem}}, \ and\ \bibinfo
  {author} {\bibfnamefont {S.}~\bibnamefont {Kalliadasis}},\ }\bibfield
  {title} {\enquote {\bibinfo {title} {Self-similarity of solitary waves on
  inertia-dominated falling liquid films},}\ }\href@noop {} {\bibfield
  {journal} {\bibinfo  {journal} {Physical Review E}\ }\textbf {\bibinfo
  {volume} {93}},\ \bibinfo {pages} {033121} (\bibinfo {year}
  {2016})}\BibitemShut {NoStop}%
\bibitem [{\citenamefont {Lin}\ \emph {et~al.}(2015)\citenamefont {Lin},
  \citenamefont {Pradas}, \citenamefont {Kalliadasis}, \citenamefont
  {Papageorgiou},\ and\ \citenamefont {Tseluiko}}]{lin2015coherent}%
  \BibitemOpen
  \bibfield  {author} {\bibinfo {author} {\bibfnamefont {T.-S.}\ \bibnamefont
  {Lin}}, \bibinfo {author} {\bibfnamefont {M.}~\bibnamefont {Pradas}},
  \bibinfo {author} {\bibfnamefont {S.}~\bibnamefont {Kalliadasis}}, \bibinfo
  {author} {\bibfnamefont {D.~T.}\ \bibnamefont {Papageorgiou}}, \ and\
  \bibinfo {author} {\bibfnamefont {D.}~\bibnamefont {Tseluiko}},\ }\bibfield
  {title} {\enquote {\bibinfo {title} {Coherent structures in nonlocal
  dispersive active-dissipative systems},}\ }\href@noop {} {\bibfield
  {journal} {\bibinfo  {journal} {SIAM Journal on Applied Mathematics}\
  }\textbf {\bibinfo {volume} {75}},\ \bibinfo {pages} {538--563} (\bibinfo
  {year} {2015})}\BibitemShut {NoStop}%
\bibitem [{\citenamefont {Rietz}\ \emph {et~al.}(2017)\citenamefont {Rietz},
  \citenamefont {Scheid}, \citenamefont {Gallaire}, \citenamefont {Kofman},
  \citenamefont {Kneer},\ and\ \citenamefont {Rohlfs}}]{rietz2017dynamics}%
  \BibitemOpen
  \bibfield  {author} {\bibinfo {author} {\bibfnamefont {M.}~\bibnamefont
  {Rietz}}, \bibinfo {author} {\bibfnamefont {B.}~\bibnamefont {Scheid}},
  \bibinfo {author} {\bibfnamefont {F.}~\bibnamefont {Gallaire}}, \bibinfo
  {author} {\bibfnamefont {N.}~\bibnamefont {Kofman}}, \bibinfo {author}
  {\bibfnamefont {R.}~\bibnamefont {Kneer}}, \ and\ \bibinfo {author}
  {\bibfnamefont {W.}~\bibnamefont {Rohlfs}},\ }\bibfield  {title} {\enquote
  {\bibinfo {title} {Dynamics of falling films on the outside of a vertical
  rotating cylinder: waves, rivulets and dripping transitions},}\ }\href@noop
  {} {\bibfield  {journal} {\bibinfo  {journal} {Journal of Fluid Mechanics}\
  }\textbf {\bibinfo {volume} {832}},\ \bibinfo {pages} {189--211} (\bibinfo
  {year} {2017})}\BibitemShut {NoStop}%
\bibitem [{\citenamefont {Vellingiri}\ \emph {et~al.}(2015)\citenamefont
  {Vellingiri}, \citenamefont {Tseluiko},\ and\ \citenamefont
  {Kalliadasis}}]{vellingiri2015absolute}%
  \BibitemOpen
  \bibfield  {author} {\bibinfo {author} {\bibfnamefont {R.}~\bibnamefont
  {Vellingiri}}, \bibinfo {author} {\bibfnamefont {D.}~\bibnamefont
  {Tseluiko}}, \ and\ \bibinfo {author} {\bibfnamefont {S.}~\bibnamefont
  {Kalliadasis}},\ }\bibfield  {title} {\enquote {\bibinfo {title} {Absolute
  and convective instabilities in counter-current gas--liquid film flows},}\
  }\href@noop {} {\bibfield  {journal} {\bibinfo  {journal} {Journal of Fluid
  Mechanics}\ }\textbf {\bibinfo {volume} {763}},\ \bibinfo {pages} {166--201}
  (\bibinfo {year} {2015})}\BibitemShut {NoStop}%
\end{thebibliography}%

\end{document}